\documentclass{article}

\usepackage{arxiv}

\usepackage{amssymb}
\usepackage{epsfig}
\usepackage{graphicx,graphbox}
\usepackage[labelformat=simple]{subcaption}

\usepackage{amsmath}
\usepackage{amssymb}
\usepackage{amsthm} 
\usepackage{epigraph}
\usepackage{color}
\usepackage{soul}
\usepackage{ulem}

\newtheorem{theorem}{Theorem}

\newtheorem{proposition}[theorem]{Proposition}

\begin{document}

\title{\normalfont \bf A geometrical Green-Naghdi  type system for   dispersive-like waves in  prismatic channels}
\date{\today }
\author{Sergey Gavrilyuk\thanks{Aix Marseille Univ, CNRS, IUSTI,   UMR 7343,  Marseille, France,   {\tt sergey.gavrilyuk@univ-amu.fr}}  $\,$ 
 and 
Mario Ricchiuto\thanks{INRIA, Univ. Bordeaux, CNRS, Bordeaux INP, IMB, UMR 5251, 200 Avenue de la Vieille Tour, 33405 Talence cedex, France,   {\tt mario.ricchiuto@inria.fr}} }

\maketitle

\begin{abstract}	
We consider 2D  free surface gravity  waves  in  prismatic channels   over topography  
with   bathymetric variations uniquely in the transverse direction.  
Starting from the Saint-Venant equations (shallow water equations) we 
  derive a 1D transverse averaged model  describing  dispersive effects solely related  to   variations of the channel topography. 
These effects have been demonstrated in (Chassagne  et al.,   {\it J. Fluid Mech.}  vol. 870, 2019, 595-616) 
to be predominant in the propagation of bores with Froude numbers below a critical value of about 1.15.
The model proposed is fully nonlinear,   Galilean invariant, and  admits a variational formulation under natural assumptions about the channel geometry. 
It is endowed with an  exact energy conservation law, and admits exact  travelling wave solutions.   
Our model  generalizes and improves the linear equations  proposed by   (Chassagne  et al.,    {\it J. Fluid Mech.}  vol. 870, 2019, 595-616), as well as   
in  (Quezada de Luna and Ketcheson, {\it  J. Fluid Mech.} vol. 917, 2021, A45).
The system is recast in two useful forms appropriate for its numerical approximations,
whose properties are discussed. Numerical results allow to verify   against analytical solutions 
the implementation of these formulations, 
and validate our model against fully 2D nonlinear shallow  water simulations, as well as the famous experiments by (Treske, {\it J. Hyd. Res.} vol. 32, 1994, 355-370). 
\end{abstract}



\tableofcontents

\section{Introduction}

{\color{black}This paper presents a new fully nonlinear 
one--dimensional  dispersive model 
for gravity waves over topography  in channels with prismatic sections.
The model   accounts  for  geometrical dispersion processes  appearing in  hydrostatic flows, and 
originated {\color{black} solely} by bathymetric variations. }\\

The physical background  of this work is  {\color{black} the propagation of undular bores in prismatic channels, and can
be related to the laboratory experiments}   by  \cite{Treske}, as well as 
to the {\it in situ} measurements  in the Garonne and Seine rivers, reported in \cite{Bonneton_CRG,Bonneton_JGR}. 
Before   the   Froude number  is high enough for wave  breaking   to occur ($\sim$1.3), two different dispersive propagation regimes are clearly observable
both in  laboratory and  river experiments.  One of them, occurring for Froude {\color{black} numbers} higher than $\sim$ 1.15, corresponds to the usual dispersive undular bore
known since   the experiments by  \cite{Favre}. These waves are well understood, and can be related theoretically to the classical  dispersion relation
from Airy theory using the analogy by
 \cite{Lemoine}. For lower Froude numbers,  longer waves are observed which do not correspond to classical
dispersive propagation. This phenomenon has been studied  in \cite{Chassagne18,Chassagne}
 where  it has been shown  numerically and theoretically that these low Froude {\color{black} number} undular bores are generated by a purely geometrical process related 
 to refraction in the transverse direction. These waves  {\color{black} are unaffected by  vertical kinematics, and appear} also  in hydrostatic flows. {\color{black} So, despite of their dispersive nature,  these waves  can be simulated using the 
 hyperbolic shallow water equations as shown in \cite{Chassagne18,Chassagne}.} 
 {\color{black} Such a dispersive behaviour of  waves in  materials with micro-structure    is well known, see e.g. {\color{black} \cite{Berezovski11}}  and \cite{BEREZOVSKI20131981}. Following the asymptotic analysis of \cite{ketcheson2015}  for  anisotropic media,
  in  \cite{Chassagne18,Chassagne} the authors introduce the idea of a  scale separation
 between the characteristic transverse length  ($\it{l}$, running across the section), and the  characteristic longitudinal  length  ($L$, in the main propagation direction along the channel axis). Using {\color{black} the  smallness of   $\it{l}/L$  the authors derive an asymptotic} 
{\color{black}  
 dispersive linear wave equation providing 
 an  asymptotic section averaged approximation  of the  linearized shallow water system. The dispersion effects, and the dispersion coefficient, 
 only  depend  on the channel's geometry. Using the theory  by  \cite{Lemoine}, the authors
use the dispersion relation of  their linear model  to obtain  a good}    quantitative prediction  of the wavelengths  
{ \color{black} of the low Froude number  waves.}
  These  waves are  purely related to  {\color{black} geometrical effects, and in particular to bathymetric  variations}. 
  They have been baptised ``dispersive-like'' 
 in  \cite{Chassagne}.  Later on, \cite{Quesada}  used the same formal development to identify solitary waves associated to geometrical dispersion.  {\color{black} They also
 introduce ad-hoc non-linear terms to obtain heuristic Boussinesq and KdV-like models. }
Recently, the same averaging procedure has been applied in \cite{ketcheson2024dispersiveeffectiveequationtransverse}
 to    the Saint-Venant equations assuming  periodic   topography and small amplitude waves.

{\color{black} From the modelling point of view,   the first rigorous Boussinesq  model for channels with non-rectangular sections  dates  to the early work
by \cite{peregrine68}, focusing on prismatic channels. First generalizations allowing small section variations
along the channel  axis were proposed by {\color{black} \cite{TengWu92,TengWu97}.}   
Equations accounting for arbitrary non-uniform cross sections     have been developed in \cite{Winckler}.
All the above models are obtained in a very general setting which starts  from the  Euler equations and then goes through classical
scaling assumptions on  length and velocity scales to arrive to  models including weak dispersive effects. 
These scaling hypotheses  differentiate between horizontal and vertical motions, and
do not account for the differences between  transverse and longitudinal horizontal processes. 
Despite of this general setting, it has been shown in   \cite{Jouy}  that   e.g.  the  model by \cite{Winckler}  only accounts
for waves associated to geometrical variations, as the linear model by \cite{Chassagne}.
  There is thus some ambiguity on which type of dispersion is  
modelled by existing equations, and which range of applications they can be used for.}

Here we propose a 1D fully nonlinear model for  dispersive-like waves originated from 2D bathymetric variations.
 In this respect,  the present work is inherently different from previous ones on section averaged dispersive models.
 { \color{black} There is no ambiguity  on the dispersive process   modelled, which is fully relying on   horizontal kinematics.}
The new  geometrical Green-Naghdi model (gGN) model  proposed  is derived from the nonlinear hyperbolic shallow water equations.
 We call this model the Green-Naghdi type (and not, as conventionally, the Serre-Green-Naghdi type) to emphasize that
it is based on the model taking into account the effects of 2D bathymetry.
 {\color{black} The physical assumptions made   are somewhat similar to those 
 invoked in \cite{Chassagne}, albeit retaining  the  full hydrodynamic  non-linearity. 
{\color{black} The new section averaged  gGN  model   is thus   fully non-linear}, it has a {\color{black} Lagrangian}  structure, and, {\color{black} as a consequence,}
 an exact  energy  conservation law.  We study the propagation  properties of the model, and
show the existence of several types of exact travelling wave solutions including solitary waves and periodic waves.
 We  propose two reformulations to solve the PDE system numerically: the first based
a elliptic-hyperbolic decomposition   \cite{Filippini}, the second using   hyperbolic relaxation {\color{black} approach} \cite{Favrie}.
We discuss the numerical approximation of both formulations, and provide a thorough investigation or the model,
which is validated both using comparisons with transverse averages of full 2D  shallow water simulations,
and against the experiments by \cite{Treske} in the low Froude {\color{black} number} range.}

 The structure of the paper is the following. In section \ref{2} we introduce the asymptotic hypothesis and discuss the {\color{black} formal} derivation of the model.
 Section \ref{3} is devoted to the study of some of its properties: energy conservation law, dispersion relation, travelling wave solutions.
 In section \ref{4} we propose some re-formulations of the system more suitable for its numerical approximation. 
 In section \ref{5} we  validate the model by  comparing it with  averaged 2D shallow water computations, {\color{black} as well as  with experiments  in prismatic channels with a trapezoidal cross-section}. {\color{black} 
 Additional technical details  related to the model and numerical methods are given in three Appendices}.

%
%
%
%
%
\section{Equations averaged in  $y$-direction}
\label{2}
We consider    shallow water flows over topography described by the  classical 2D Saint-Venant equations :
\begin{subequations}
\label{dimensional}
\begin{align}
&h_t +(hu)_x+(hv)_y =0, \label{eq:1a}\\
&(hu)_t +(hu^2+gh^2/2)_x+(huv)_y =-gh b_x,\label{eq:1b}\\ 
&(hv)_t +(huv)_x+(hu^2+gh^2/2)_y =-gh b_y.\label{eq:1c}
\end{align}
\end{subequations}
 Here $t$ is  time, $(x,y)$ are the Cartesian  coordinates {\color{black} respectively running in the longitudinal (channel axis, $x$), 
 and transverse (across the sections, $y$) directions  {\color{black} (see the sketch in figure \ref{fig:periodic structure-a})}.
  We denote by {\color{black} $h$ the fluid depth},  $(u,v) $ the $x$-and $y$ - velocity components,  
  $b(x,y)$ is the bottom topography}, and $g$ is the  acceleration of gravity.

  {\color{black}  We assume that the transverse and longitudinal propagation processes are
  characterized by two different length scales ${\it l} $ and $L$, respectively,   {\color{black} such} that ${\it l} \ll L$. We thus introduce
  the small parameter
\begin{equation}\label{eq:eps}
\varepsilon := \dfrac{{\it l}}{L}.
  \end{equation}
{\color{black} Let $H$ be the  characteristic vertical length,  usually taken as the characteristic undisturbed water depth, see e.g.  \cite{peregrine68}.  
We introduce the dimensionless variables (denoted with the `tilde' sign) }
  } 
\begin{equation*}
x=L\tilde x, \; y=\varepsilon L\; \tilde y, \; t=L/\sqrt{gH} \; \tilde t, \; u=\sqrt{gH}\; \tilde u, \; v=\varepsilon\sqrt{gH}\; \tilde v, 
\end{equation*}
\begin{equation*}
 h=H\; \tilde h, \;  b=H\; {\tilde b}.
\end{equation*}
{\color{black} Introducing the above scaling in the equations,   and suppressing for simplicity   the    `tilde', we can write  the  dimensionless system:}
\begin{subequations}
	\label{dimensionless}
	\begin{align}
	&h_t +(hu)_x+(hv)_y =0, \label{eq:2a}\\
	&(hu)_t +\left(hu^2+\frac{h^2}{2}\right)_x +(huv)_y+hb_x =0,\label{eq:2b}\\ 
	&\varepsilon^2(v_t +uv_x+vv_y)+(h+b)_y=0,\label{eq:2c}
\end{align}
\end{subequations} 
{\color{black} Note that   the equation for $v$ is  written  in  non-conservative form for later convenience.} 
 We now introduce some of our main hypotheses. 
 
 The first is that the channel has a prismatic bathymetry defined as 
$b=b(y)$.  A slow variation of $b$ with $x$ could also be included,  for example $\partial_x b=\mathcal{O}(\varepsilon^\beta)$ for some $\beta >0$. 
We will leave this out for simplicity. 

The second important hypothesis is that $(hv)\big|_{y=0}=(hv)\big|_{y=l}$ { \color{black} and $(hvu)\big|_{y=0}=(hvu)\big|_{y=l}$.  Note that these two
conditions are equivalent  for flows in channels with banks ($h=0$ on both sides), as well as in channels with reflective straight walls ($v=0$ on both sides). 
They are obviously  satisfied for flows which are $l$-periodic in $y$-direction, with an     $l$-periodic bathymetry:  $b(y)=b(l+y)$. }

{\color{black} Our aim is to replace the 2D equations with an equivalent system of averaged 1D equations, which exhibit dispersive behavior similar to that of the original model.    
To this end we will average   \eqref{dimensionless} in the transverse direction  across a  length $\it{l}$,  which
we assume to be  constant. This length can be a given channel width (for channels with vertical walls), or the  width  of a reference/initial wet section, or the period in the case of periodic topography. }
Figures \ref{fig:periodic structure-b} and \ref{fig:periodic structure-c}  show typical  geometries   encountered in practice, and gives some initial  notation.

	\begin{center}\begin{figure}
	\begin{subfigure}{\textwidth}
	\centering\includegraphics[width=0.4\textwidth]{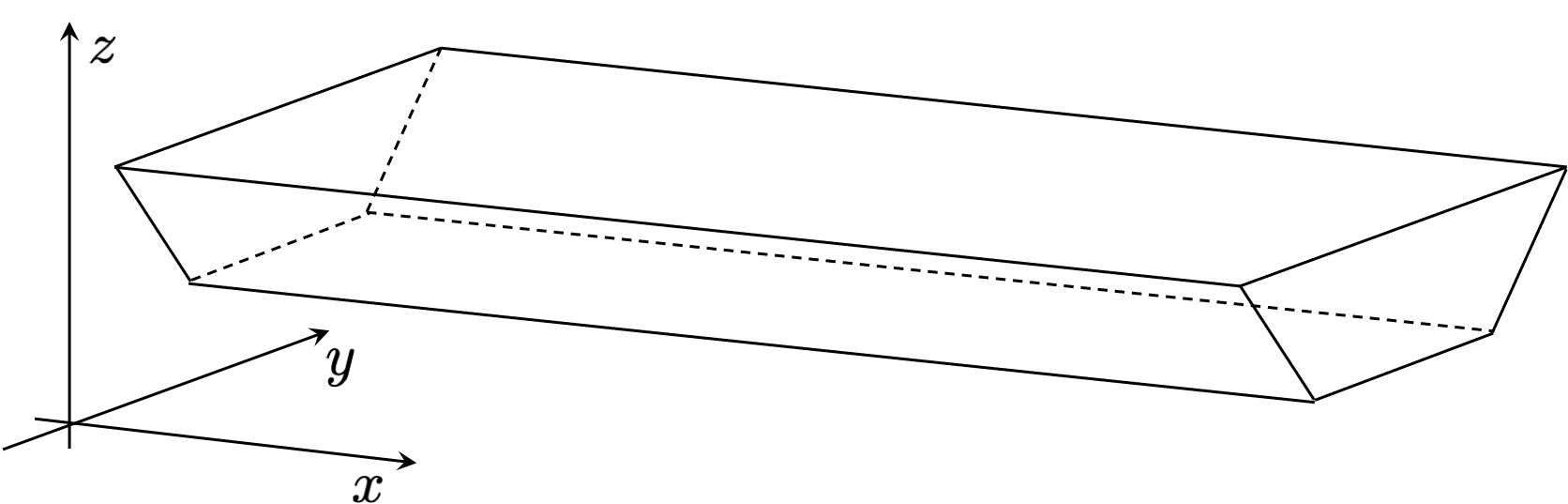}
	\caption{}
	\label{fig:periodic structure-a}
	\end{subfigure}
			\begin{subfigure}{0.5\textwidth}
				\centering\includegraphics[width=0.8\textwidth]{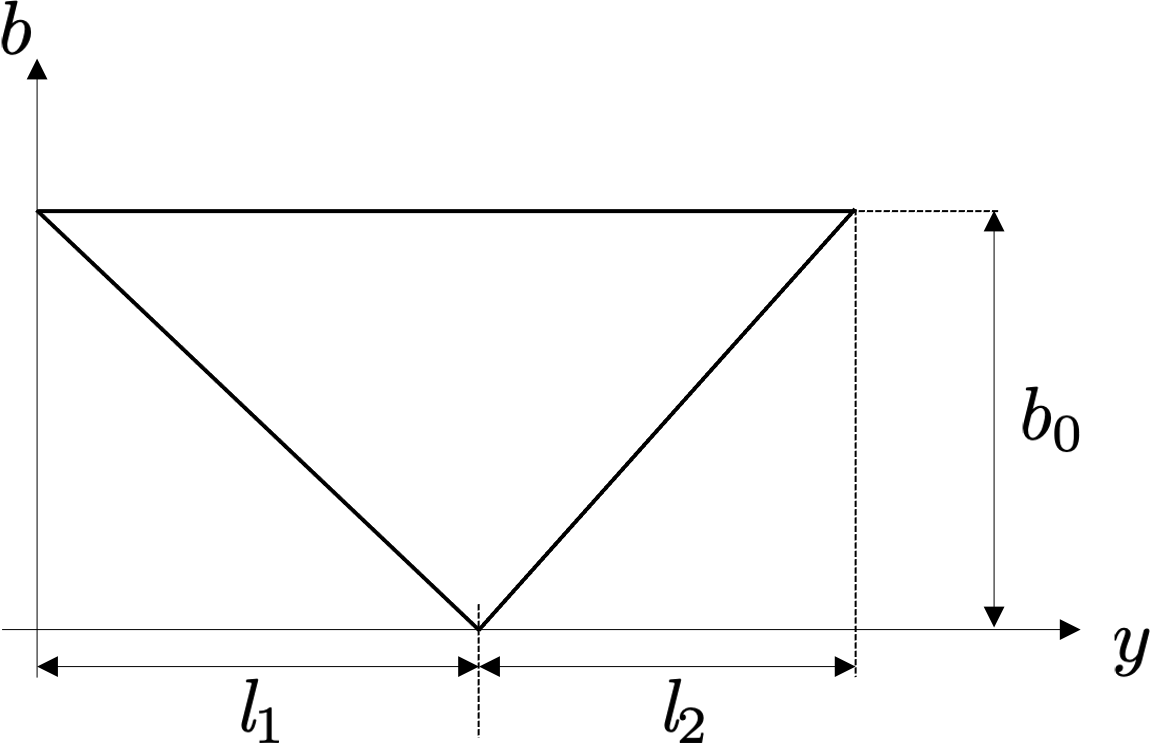}
				\caption{}
				\label{fig:periodic structure-b}
			\end{subfigure}\hfill
						\begin{subfigure}{0.5\textwidth}
								\centering\includegraphics[width=0.8\textwidth]{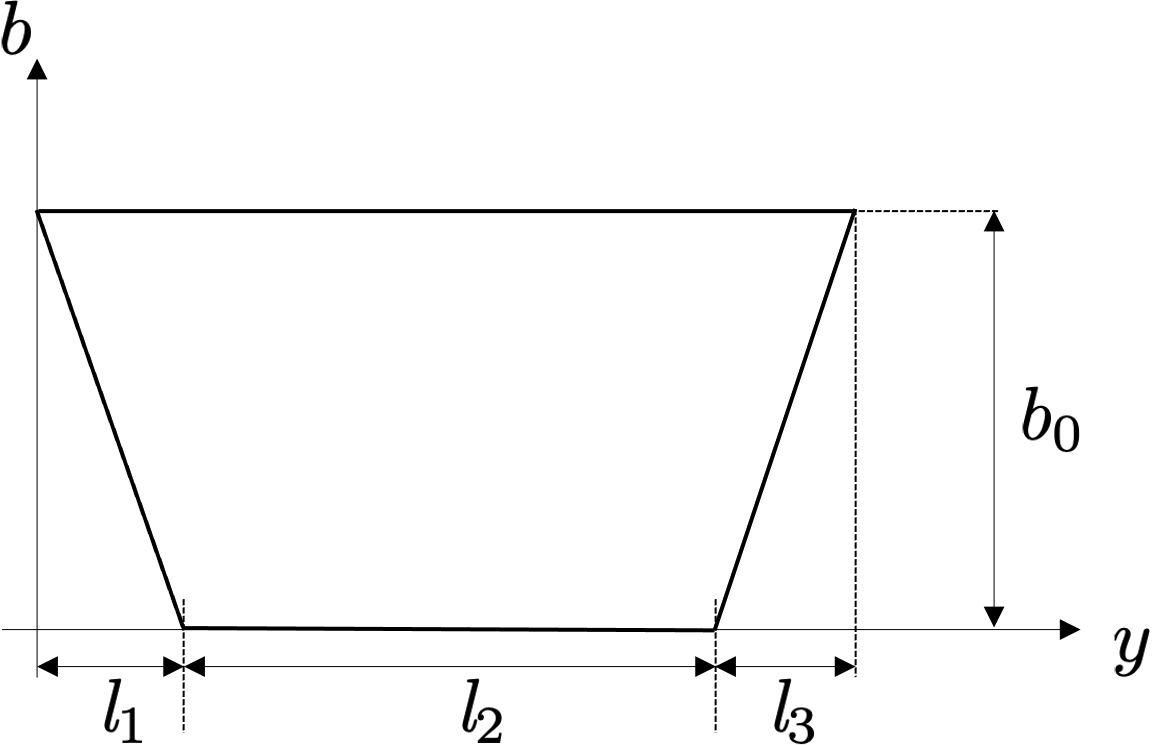}
								\caption{}
								\label{fig:periodic structure-c}
			\end{subfigure}
		\caption{{\color{black}Channel geometry. (a)   longitudinal  axis $x$, transverse  axis  $y$,  vertical  axis  $z$. (b)    triangular section with ${\it l}={\it l}_1+{\it l}_2$, and $b(0)=b({\it l})=b_0, \; b({\it l}_1)=0$. (c)   trapezoidal  section with ${\it l}={\it l}_1 + {\it l}_2 +{\it l}_3$ and 
		  $b(0)=b({\it l})=b_0, \; b({\it l}_1)= b({\it l}_1+{\it l}_2)=0$. 
			\label{fig:periodic structure}}}
\end{figure}
	\end{center}

\subsection{Generalities and  asymptotic expansion}
{\color{black} For any function $f(t,x,y)$}, we will use  two types of averaging operators in $y$-direction.  One is standard  : 
\begin{equation}
\overline{f}(t,x)=\frac{1}{\it l}\int_0^{\it l}f(t,x,y)dy
\end{equation} 
The second one defines  the {\color{black} weighted} averaged value
\begin{equation}
\langle f\rangle=\frac{1}{\it l\overline{h}}\int_0^{\it l}h(t,x,y)f(t,x,y)dy.
\end{equation} 
Sometimes the last averaging is called {\it Favre averaging}. The advantage of the last  definition is that the mass conservation law is exactly satisfied in the usual sense, with the  velocity $\langle u\rangle$.  The dimensionless averaged conservative form of equations \eqref{eq:1a}, \eqref{eq:1b}  in the case $b=b(y)$ becomes :  
\begin{subequations}
	\label{averaging}
	\begin{align}
	&\overline{h}_t +(\overline{h}\langle u\rangle)_x =0, \label{eq:3a}\\
	&(\overline{h}\langle u\rangle)_t +(\overline{hu^2}+\frac{1}{2}\overline{h^2})_x=0.\label{eq:3b} 
\end{align}
\end{subequations}
We need  to find $\overline{hu^2}$ and $\overline{h^2}$. Equation \eqref{eq:2c} implies :
\begin{equation}
h+b=\overline{h}+\overline{b}+{\cal{O}}(\varepsilon^2).
\label{h_expression}
\end{equation}
We want to find this $ {\cal{O}}(\varepsilon^2)$ term in \eqref{h_expression} in explicit form. 
{\color{black} For this,  we make  the following {\it  ansatz} for the velocity $u$ in $x$ -direction  :
\begin{equation}
\label{velocity_ansatz}
u(t,x,y)= \langle u\rangle(t,x)+{\cal{O}}(\varepsilon^2  ).
\end{equation}
This {\it  ansatz} means that the shear effects {\color{black} of $u$} in $y$-direction are negligible, i.e. the velocity in  $x$-direction is almost $y$ - independent  (see \cite{Michel-Dansac} for the derivation of a model for fluid flows in prismatic channels with shear effects, but without dispersive effects).   
The {\it ansatz} \eqref{velocity_ansatz} can be considered  as a  potential  flow assumption commonly  used for the derivation of the Saint-Venant equations. Indeed, the equations  \eqref{eq:2a}--\eqref{eq:2c} imply the dimensionless vorticity equation in the form 
\begin{equation*}
\frac{D}{Dt}\left(\frac{\omega}{h}\right) =0, \quad \frac{D}{Dt}=\frac{\partial}{\partial t} +u\frac{\partial}{\partial x} +v\frac{\partial}{\partial y}, \quad \omega = \varepsilon ^2 v_x-u_y.
\end{equation*}
If  $\omega=0$ initially, it stays zero for any time. Integrating the equation $\omega=0$ with respect to $y$, we have
\begin{equation*}
u(t,x,y)=c(t,x)+\varepsilon^2 \int_0^y v_x(t,x,s)ds. 
\end{equation*}
Multiplying the expression for $u$ by $h$, one obtains
\begin{equation*}
h(t,x,y)\, u(t,x,y)=h(t,x,y)\,c(t,x)+\varepsilon^2 h(t,x,y)\,\int_0^y v_x(t,x,s)ds.
\end{equation*}
The averaging  with respect to $y$ gives then 
\begin{equation*}
c(t,x)= \langle u \rangle(t,x)+{\cal O}(\varepsilon^2),
\end{equation*}
and
\begin{equation*}
u- \langle u \rangle ={\cal O}(\varepsilon^2). 
\end{equation*}
Also, 
\begin{equation}
\label{estimation}
\overline{hu^2}=\overline{h(u- \langle u \rangle+ \langle u \rangle)^2}=\overline{h(u- \langle u \rangle)^2}+ 2\overline{h(u- \langle u \rangle)}\, \langle u \rangle+\overline{h}\langle u \rangle^2
\end{equation} 
\begin{equation*}
=\overline{h(u- \langle u \rangle)^2}+\overline{h} \langle u \rangle^2=\overline{h} \langle u \rangle^2+{\cal{O}}(\varepsilon^4).
\end{equation*} }
An additional hypothesis about the behavior of $v$ will be now specified. 
Replacing {\it  ansatz} \eqref{velocity_ansatz} into the mass conservation law \eqref{eq:2a}, one obtains for $v$ the equation :
\begin{equation}
\overline{h}_t+((\overline{h}+\overline{b}-b)\langle u \rangle )_x+\color{black} (hv)_y={\cal{O}}{(\varepsilon^2)}.
\end{equation}
Since
\begin{equation*}
\overline{h}_t+(\overline{h}\langle u \rangle )_x=0,
\end{equation*}
one gets 
\begin{equation*}
(\langle u \rangle (\overline{b}-b))_x+\color{black} (hv)_y={\cal{O}}{(\varepsilon^2)}.
\end{equation*}
{\color{black} It implies  :  
\begin{equation*}
{\color{black} hv}=C(t,x)- S(y)\; \langle u \rangle_x +{\cal{O}}{(\varepsilon^2)}, \quad S(y)= \int_0^y(\overline{b}-b(y')) dy'.
\end{equation*}
}
{\color{black} The  function  $S(y)$ plays an important role in this work. It has the   property  $S(0)=S({\it l})=0$.  Moreover we prove in appendix \ref{app:geo1} that in the   case  $b(y)=b(l-y)$ for any $0<y<l$, i.e. $b(y)$  symmetric with respect to the channel axis, we have   $\bar S=0$. 
 We will see the importance of this result shortly. }

{\color{black} Concerning the integration  `constant' $C(t,x)$,  we need to determine it using  additional hypotheses.} For  the applications we have in mind {\color{black} (flow in prismatic channels)}, we  use  the requirement of a  vanishing  average discharge  in the $y$- direction : 
\begin{equation*}
	\int_{0}^{\it l}hv dy=0,
\end{equation*}
which leads to    $C(t,x)=\overline{S}\; \langle u \rangle_x+{\cal{O}}{(\varepsilon^2)}$.

\subsection{Small-scale geometrical effects on the  section averaged flow}
Using the above results  we can now write {\color{black} the asymptotic formula for $v$  }:
\begin{equation}
{\color{black} hv 	=(\overline{S}- S(y))\; \langle u \rangle_x+{\cal{O}}{(\varepsilon^2)}}.
	\label{v0 expression}
\end{equation}
{\color{black} If $\overline{S}=0$, then, due to the property $S(0)=S(l)=0$, the  vanishing transverse flow discharge implies directly the smallness of the transverse velocities  : $v(t,x,0)={\cal{O}}{(\varepsilon^2)}$ and $v(t,x,l)={\cal{O}}{(\varepsilon^2)}$. This justifies, in particular,  the application of the averaged model to flows in  prismatic channels with vertical walls. 
Equations \eqref{v0 expression}   and mass conservation \eqref{eq:3a}  combined give the following expression  for the transverse velocity (below, ${\cal{O}}{(\varepsilon^2)}$ error will not be indicated and replace by the sign $\approx$)} :

{\color{black}
	\begin{equation}
	v(t,x,y)\approx\frac{(\overline{S}- S(y)) }{\overline{h}+S_y}\; \langle u \rangle_x=\frac{(\overline{S}- S(y))\dot \tau }{1+\tau S_y},
	\label{v0}
	\end{equation}
	with $\tau=1/{\overline{h}}$, and  the ``dot"  denotes   total derivative along the longitudinal direction  :
	\begin{equation}
	\dot{\tau} =\frac{\partial \tau}{\partial t}+\langle u \rangle\frac{\partial \tau}{\partial x}.
	\end{equation}
	We set 
	\begin{equation}
	\sigma (y) =  \overline{S}-S(y).
	\label{dispersion_coeffs1}
	\end{equation}
	The function $\sigma (y)$ has the following properties :
	\begin{equation}\label{eq.sigma}
	\sigma (0)=\sigma (l)=\overline{S}, \quad \overline{\sigma}=0,\quad \frac{d\sigma}{dy} = b(y) - \bar b,\quad
	 \quad \overline{\frac{d\sigma}{dy}} =0.
	\end{equation}
	We can  thus  write  the following approximation for $v$ :
	\begin{equation}
	v(t,x,y)\approx\frac{\sigma(y) \dot{\tau}(t,x)}{\displaystyle 1-\tau(t,x) \frac{d \sigma(y)}{dy}}.
	\end{equation}

We are now able to find the} ${\cal O}(\varepsilon^2)$ terms from \eqref{eq:2c}: 
\begin{equation}
(h+b)_y=-\varepsilon^2(v_t+uv_x+vv_y)+{\cal{O}}(\varepsilon^4)=-\varepsilon^2\left(\dot{v}  +v v _y\right)+{\cal{O}}(\varepsilon^4).
\label{non_hydrostatic}
\end{equation}
{\color{black} 
	One has
	\begin{equation}
	\dot{v}(t,x,y)\approx\frac{\sigma \ddot{\tau}}{\displaystyle 1-\tau \frac{d \sigma}{dy}}+\frac{\sigma\displaystyle\frac{d \sigma}{dy}\dot{\tau}^2}{\left(\displaystyle 1-\tau \frac{d \sigma}{dy}\right)^2},	
	\label{symplified_v0}
	\end{equation}
	and 
	\begin{equation}
	h+b= \overline h+\overline b-\varepsilon^2
	\left(
	M-\overline M\right)+\mathcal{O}(\varepsilon^4), \quad M(t,x,y)=
	\int_0^y \dot{v}\left(t,x,s \right)ds+\frac{v^2 (t,x,y)}{2}.
	\label{non_hydrostatic2}
	\end{equation}
	Or
	\begin{equation}
	h= \overline{h}-\frac{d\sigma}{dy}-\varepsilon^2
	\left(
	M-\overline M\right)+{\cal{O}}(\varepsilon^4).
	\end{equation}
	Thus 
	\begin{equation}
	\overline{h^2}=\overline{\left(\overline {h}-\frac{d\sigma}{dy}\right)^2}-2\varepsilon^2 \overline{\left(\overline {h}-\frac{d\sigma}{dy}\right)\left(M-\overline{M}\right)}+{\cal{O}}(\varepsilon^4)
	\end{equation}
	\begin{equation}
	=\overline{h}^2+2\varepsilon^2 \overline{M\, \frac{d\sigma}{dy}}+const+{\cal{O}}(\varepsilon^4), 
	\end{equation}
	because $\displaystyle\overline{\frac{d\sigma}{dy}}=0$.  From the above we can deduce that within ${\cal{O}}(\varepsilon^4)$  
	\begin{equation}
	 \left(\dfrac{\overline{h^2}}{2}\right)_x= \left(  \dfrac{\overline{h}^2}{2} \right)_x+ \varepsilon^2 \left(\overline{M\, \frac{d\sigma}{dy}}\right)_x, 
	\end{equation}
            where the general expression for $M$ is 
	\begin{equation}
	M=\int_0^y\left(\frac{\sigma(s) \ddot{\tau}}{\displaystyle 1-\tau \frac{d \sigma(s)}{ds}}+\frac{\sigma(s)\displaystyle\frac{d\sigma(s)}{ds}\dot{\tau}^2}{\left(\displaystyle 1-\tau \frac{d \sigma}{ds}\right)^2}\right)ds+\frac{\sigma^2(y) \dot{\tau}^2}{2\left(\displaystyle 1-\tau \frac{d \sigma}{dy}\right)^2}.
	\end{equation}
	Note that the denominator $ 1- \tau \frac{d \sigma}{dy}= 1 -(b(y) - \bar b)/\bar h$ can be easily shown to be strictly positive for many sections of practical relevance as e.g. trapezoidal ones. \\
	
	We introduce  two important particular cases, and then will devote the paper to the  in-depth study of the one of them. 
	We first consider channels verifying  {\color{black} the assumption $\overline{S}=0$ which is, in particular, satisfied for symmetric channel  cross-sections (see appendix \ref{app:geo1}  for  a proof). Formally,  since  only    ${\mathcal{O}}(\varepsilon ^2)$  terms are retained in
	 the equations,  this   assumption can  be weakened to $\overline{S}={\mathcal{O}}(\varepsilon ^\beta)$, with $\beta >0$. In this case one could speak of  quasi-symmetric  channel cross-sections.   This first family of models can be characterized by the following property.} 
	
	\begin{proposition}[Lagrangian structure for symmetric and quasi-symmetric channel cross-sections\label{prop1}]
		             Consider   quasi-symmetric  channel  cross-sections for which $\bar S = {\cal O}(\varepsilon^\beta)$ with $\beta >0$.
	Then, up to terms of order  $ {\cal O}(\varepsilon^{\beta})$,
	the  term $\displaystyle \overline{M\, \frac{d\sigma}{dy}}$ can be written as {\color{black} the variational  derivative } of the Lagrangian potential
	\begin{equation*}
	{\cal{L}}(\tau, \dot\tau)=\overline{N\frac{d\sigma}{dy}}
	\end{equation*}
	with
	\begin{equation}
	N(\tau,\dot\tau,y)=\frac{\dot \tau^2}{2}\int_{0}^{y}\frac{\sigma(s)ds}{1-\displaystyle \tau \frac{d\sigma}{ds}}.
	\end{equation}
	More precisely, we have 
	\begin{equation}
	\overline{M\frac{d\sigma}{dy}}=-\left(\frac{\partial {\cal{L}}}{\partial \tau }-\frac{D}{Dt}\left(\frac{\partial {\cal{L}}}{\partial \dot \tau}\right)\right)
	+{\cal O}(\varepsilon^{\beta})
	\label{variational_formulation_10}
	\end{equation}
{\color{black}	For symmetric channel  cross-sections  verifying  the condition 	$\bar S=0$, the expression  \eqref{variational_formulation_10} is exact (the  term ${\cal O}(\varepsilon^{\beta})$ identically  vanishes). }
	\end{proposition}
The proof is given in appendix  \ref{app.prop1}.

The last proposition shows that for quasi-symmetric channel cross-sections  the pressure term in our model has a Lagrangian structure : 
		\begin{equation}
\frac{\overline{h^2}}{2}\approx \frac{(\overline{h})^2}{2}-\varepsilon^2\frac{\delta {\cal L}}{\delta \tau}, \quad {\rm with} \quad \frac{\delta {\cal L}}{\delta \tau}=\frac{\partial {\cal{L}}}{\partial \tau }-\frac{D}{Dt}\left(\frac{\partial {\cal{L}}}{\partial \dot \tau}\right).
	\label{variational_formulation_1} 
	\end{equation}
We call this  model {\it symmetric geometrical Green-Naghdi equations}.  
The advantage of the formulation \eqref{variational_formulation_1} are multiple. In particular, the form \eqref{variational_formulation_1} guarantees the  variational formulation of our equations  (\cite{Gavrilyuk}), and, as a consequence, the existence of an exact energy balance law. More details are given in the next Section. \\

The derivation above provides a large family of models. Its form however does not allow   closed form expressions for 
the dispersion coefficients. To obtain some simplification we consider a different geometrical {\it  ansatz}.
{\color{black} As we have already mentioned above}, the condition $1 - \displaystyle \tau \frac{d\sigma}{dy} >0$ can be easily shown for sections of practical importance.
 {\color{black} An} interesting limit is obtained when considering  channels which are wide compared to the depth.
This typically the case in realistic applications both in natural and artificial environments.
For this case, we can assume further that  ratio $\displaystyle \tau \frac{d\sigma}{dy}$ is small, namely $\displaystyle\tau \frac{d\sigma}{dy}=\mathcal{O}(\epsilon^{\gamma})$
for some $\gamma >0$.  In this case one can easily show that 
$$
M =   \int_0^y\left(\sigma(s) \ddot{\tau}+\sigma(s)\displaystyle\frac{d\sigma(s)}{ds}\dot{\tau}^2\,\right)ds+\frac{\sigma^2(y)}{2}\dot{\tau^2} + \mathcal{O}(\epsilon^{\gamma})
$$
Neglecting the asymptotically small $\mathcal{O}(\epsilon^{\gamma})$ term, we obtain 
	\begin{equation*}
	M\approx \ddot{\tau}\int_0^y\sigma(s)ds+\sigma^2(y)\dot\tau^2-\dot\tau^2\frac{\sigma^2(0)}{2}. 
	\end{equation*}
	Hence, 
	\begin{equation}
	\overline{\frac{d\sigma}{dy} M}=\ddot{\tau}\overline{ \frac{d\sigma}{dy}\int_0^y\sigma(s)ds}=
	\ddot{\tau}\left(\overline{\frac{d}{dy}\left(\sigma(y)\int_0^y\sigma(s)ds\right)-\sigma^2}\right)
	\label{variational_formulation_2} 
	\end{equation}
	\begin{equation}\label{eq.chi}
=\ddot{\tau}\left(\sigma(l) \overline{\sigma}-\overline{\sigma^2}\right)	=-\ddot{\tau}\overline{\sigma^2}	=-\chi  \ddot{\tau}, \quad {\rm with }\quad 	\quad \chi =\overline{S^2}-\overline{S}^2>0.
	\end{equation}
This limit also enjoys   a very important structural  property,  which can be formulated in the following way:
	\begin{proposition}[Lagrangian structure for wide  channel cross-sections\label{prop2}] 
	Consider  wide channels for which
	$\displaystyle\tau \frac{d\sigma}{dy}=\mathcal{O}(\epsilon^{\gamma})$,  with $\gamma >0$. 
	Then, up to terms of order  $ {\cal O}(\varepsilon^{\gamma})$,
	the hydrodynamic term $\displaystyle\overline{M\, \frac{d\sigma}{dy}}$ can be written as  the variational derivative of  a Lagrangian potential.
	More precisely we have 
	\begin{equation}
	\overline{M\frac{d\sigma}{dy}}=-\left(\frac{\partial {\cal{L}}}{\partial \tau }-\frac{D}{Dt}\left(\frac{\partial {\cal{L}}}{\partial \dot \tau}\right)\right)
	+{\cal O}(\varepsilon^{\gamma})
	\label{variational_formulation_1_1} 
	\end{equation}
	with 	
	\begin{equation}\label{eq.Lsimplified}
	{\cal{L}}(\tau, \dot\tau)= -\chi\, \frac{\dot\tau^2}{2}.
	\end{equation}
	and $\chi$ as in  \eqref{eq.chi}.
	\end{proposition}

{\color{black} We  will refer to the last model, having a very simple form.} as   {\it simplified geometrical Green-Naghdi equations}. We will devote the remainder of the paper to  the study of its theoretical properties, and to
its applications to the propagation of undular bores in channels.

{\color{black} The  simplified geometrical Green-Naghdi model can be considered as a prototype {\color{black} averaged} model, containing only one (geometrical) parameter $\chi$. In the case of symmetric channel cross-sections, $ \chi =\overline{S^2}$.} \\

For the {\color{black} prototype average model}  equations  \eqref{eq:3a} - \eqref{eq:3b} can be written 
in dimensional variables as follows:
\begin{eqnarray}
	&\overline{h}_t +(\overline{h}\langle u\rangle)_x =0, \label{eq:4a}\\
	&
	(\overline{h}\langle u\rangle)_t +(\overline{h}\langle u\rangle^2+ g \frac{\overline{h}^2}{2}  + p )_x=0,\label{eq:4b} 
\end{eqnarray}
where we set 
\begin{equation}
\label{dispersion3}
p =  -\chi \ddot\tau, \quad \tau =\dfrac{1}{\overline h}, \quad \chi =\overline{S^2}-\overline{S}^2>0.
\end{equation}
with  $ \chi$ having  dimension $[m^4]$.  
}

\subsection{Expressions for the geometric parameter for the triangular and trapezoidal sections}
The value of the parameter $\chi$ for a triangular type  bottom (see Figure  \ref{fig:periodic structure-b}) is :
\begin{equation}\label{eq:chi_tri}
\chi=\frac{b_0^2({\it l}_1^2+4{\it l}_1{\it l}_2+{\it l}_2^2)}{720}.
\end{equation}
Here $b_0$ is the bottom height ($b_0=b(0)=b(l)$),  and ${\it l}_1$ is the position of the bottom singularity ($b({\it l}_1)=0$, and ${\it l}={\it l}_1+{\it l}_2$). 
The value of the parameter $\chi$ for a trapezoidal type bottom  is :
\begin{equation}\label{eq:chi_trap}
\chi=\frac{b_0^2({\it l}_1+{\it l}_3)({\it l}_1^3+{\it l}_1^2(6{\it l}_2+5{\it l}_3)+{\it l}_3(15{\it l}_2^2+6{\it l}_2{\it l}_3+{\it l}_3^2)+{\it l}_1(15{\it l}_2^2+24{\it l}_2{\it l}_3+5{\it l}_3^2))}{720({\it l}_1+{\it l}_2+{\it l}_3)^2}.
\end{equation}
Here $b_0$ is the bottom height ($b_0=b(0)=b(l)$),  ${\it l}_1$ is the position of the first bottom singularity ($b({\it l}_1)=0$), ${\it l}_1+{\it l}_2$  is the position of the second  bottom singularity ($b({\it l}_1+{\it l}_2)=0$), $b(y)=0$ for $y\in ({\it l}_1, {\it l}_1+{\it l}_2)$, and ${\it l}={\it l}_1+{\it l}_2+{\it l}_3$.

The formulas coincide if one takes ${\it l}_2=0$ and replaces ${\it l}_3$ by ${\it l}_2$. Moreover the two formulas are symmetric with respect   the pairs  (${\it l}_1,\,{\it l}_2$) and (${\it l}_1,\,{\it l}_3$), respectively. This implies that the dispersion effects are  insensitive 
to mirroring of asymmetric shapes, as one should expect. Similarly, one easily verifies that asymmetric configurations have a smaller dispersion coefficient than symmetric ones.  For this, one has  to look for  the  minimum of $\chi$  under  the constraint $l_1+l_2=l$ in the case of the triangular type bottom,  or under  the constraint   $l_1+l_2+l_3=l$ in the case of trapezoidal type bottom.  For the triangular  bottom topography, the maximal value of $\chi/(b_0^2 l^2)$ is achieved for $l_1=l_2=l/2$,  and is equal  approximately to $0.0020833$. For the trapezoidal bottom  the maximal value of $\chi/b_0^2l^2$ is achieved for 
$\displaystyle l_1=l_3=(21-\sqrt{41}) l/40 $,  and is equal approximately to $0.0027139$.

\section{Properties of the geometrical Green-Naghdi type system}
\label{3}
\subsection{Galilean invariance}
{\color{black} The geometrical models  obtained in the previous section can be formally  written as
$$
\begin{aligned}
\dot{\bar h} +& \bar h \langle u\rangle_x =0, \\
\bar h  \dot{ \langle u\rangle } +& g\bar h \bar h_x + p(\tau,\dot\tau,\ddot\tau)_x=0,
\end{aligned}
$$
}
{\color{black}   These equations  involve  only space derivative and  first and second 1D material time  derivatives  (denoted, as usually, by `dot'). 
 Now, consider a   Galilean  reference  frame moving with a constant  velocity.
Both  space and  material  time derivatives  are invariant with respect to this change of the reference frame.
Thus   the model  is  Galilean invariant. } 

{\color{black}
	\subsection{  {\color{black} Lagrangian} structure and energy equation}
 Consider the Hamilton action 
	\begin{equation}
	a=\int_{t_0}^{t_1}\; L\; dt,\quad L=\int_{-\infty}^{+\infty}\left(\overline{h}\frac{\langle u\rangle^2}{2}-W(\overline{h}, \dot{\overline{h}})\right)dx, \quad W(\overline{h}, \dot{\overline{h}})=\overline{h}\, w(\overline h,  \dot{\overline h}).
	\label{Lagrangian_1}
	\end{equation}
Here $t_i$, $i=0,1$,  are given time instants, and $w(\overline h,  \dot{\overline h})$  is a given potential. The mass conservation law 
\begin{equation*}
\overline{h}_t +(\overline{h}\langle u\rangle)_x =0 
\end{equation*}	
is considered as a differential constraint. 
Then, the Euler--Lagrange equation (momentum equation) for \eqref{Lagrangian_1}
can be written as (see e.g. \cite{Gavrilyuk} and \cite{Sergey}): 
\begin{equation*}
\left(\overline{h}\langle u\rangle\right)_t +\left(\overline{h}\langle u\rangle^2+\overline{h}^2\frac{\delta w}{\delta \overline{h}}\right)_x =0, \quad {\rm with}\quad \frac{\delta w}{\delta \overline{h}}=\frac{\partial w}{\partial \overline{h}}-\frac{D}{Dt}\left(\frac{\partial w}{\partial \dot{\overline h}}\right). 	
\end{equation*}
Introducing a new potential ${\tilde w}(\tau,\dot\tau)$ as :
\begin{equation*}
\tilde{w}(\tau,  \dot{\tau})=\tilde{w}\left(\frac{1}{\overline{h}}, - \frac{\dot{\overline{h}}}{\overline{h}^2}\right)=w(\overline{h},  \dot{\overline{h}}),
\end{equation*}
one can also rewrite the momentum equation in the form
\begin{equation*}
\left(\overline{h}\langle u\rangle\right)_t +\left(\overline{h}\langle u\rangle^2-\frac{\delta \tilde w}{\delta \tau}\right)_x =0, \quad  {\rm with}\quad \frac{\delta \tilde{w}}{\delta \tau}=\frac{\partial \tilde{w}}{\partial {\tau}}-\frac{D}{Dt}\left(\frac{\partial \tilde{w}}{\partial \dot{\tau}}\right).
\end{equation*}
Thus, the averaged  equations in the transverse direction both for the symmetrical case and the simplified case are obtained by taking  
\begin{equation*}
\tilde w(\tau, \dot \tau)=\frac{g}{2\tau}+ {\cal L}(\tau, \dot \tau).  
\end{equation*}
Setting $ \bar q = \overline{h}\langle u\rangle$, the system admits the energy conservation law which is a consequence of invariance of the Lagrangian under time shift :
\begin{equation}
\left(\overline{h}\left(\frac{\langle u\rangle^2}{2}+g\frac{\overline h}{2}+E\right)\right)_t +\left(\overline{q} \left(\frac{\langle u\rangle^2}{2}+g\frac{\overline h}{2}+E\right)  +
  \left(g\frac{\overline{h}^2}{2}-  \frac{\delta \cal L}{\delta \tau}\right) \langle u\rangle\right)_x=0,
\end{equation} 
with
\begin{equation*}
E=  \cal L-\dot{\tau}\frac{\partial \cal L}{\partial \dot \tau}.
\end{equation*}
In particular, for  model \eqref{eq:4a}-\eqref{eq:4b}-\eqref{dispersion3} we have 
\begin{equation}
\left(\overline{h}\left(\frac{\langle u\rangle^2}{2}+ g\frac{\overline h}{2}+\frac{ \chi\dot\tau^2}{2}\right)\right)_t +\left(\overline{q} \left(\frac{\langle u\rangle^2}{2}+g\frac{\overline h}{2}+\frac{  \chi\dot\tau^2}{2}\right)  + \left(g\frac{\overline{h}^2}{2}- \chi \ddot\tau\right) \langle u\rangle\right)_x=0.
\end{equation}
}

\subsection{Linear dispersion properties\label{sec:dispersion}}

Consider now the linearized version of \eqref{eq:4a}, \eqref{eq:4b} and  \eqref{dispersion3} {\color{black} on a flow at rest, with constant transverse average depth $ \overline h_0$} : 
\begin{eqnarray}
	&\overline{h}_t +  \overline h_0\langle u\rangle_x \!=& 0, \label{eq:4aL}\\
	&  \langle u\rangle_t   + g \overline{h}_x    =&  \frac{\chi}{\overline{h}_0^2}    \langle  u\rangle_{xxt}.\label{eq:4bL} 
\end{eqnarray}
{\color{black} Note that for simple section shapes, as those of figures \ref{fig:periodic structure} and \ref{fig:dispersion-a}, $\overline h_0$ is related to the maximum depth $b_0$ by simple geometrical 
relations (cf. also section \ref{sec:treske}, for explicit formulas in the trapezoidal case).} 

Using standard techniques 
we  show that the dispersion relation of the system is 
\begin{equation}\label{disp-rel}
\omega^2 =  k^2\dfrac{c_0^2}{ 1 +  \frac{\chi}{\overline{h}_0^2}  k^2 }
\end{equation}
with $k$ the wave-number  {\color{black} and $c_0=\sqrt{g \bar h_0}$. In figures \ref{fig:dispersion-b} and \ref{fig:dispersion-c} we show the typical behaviour of the non-dimensional phase speed $\omega/kc_0$ 
as a function of the \emph{horizontal} reduced wave number $k{\it l}$, 
for the case of a trapezoidal section. We consider the case of a deep channel with steep slopes, and of a  shallow channel with mild slope.
Using the notation in     figure  \ref{fig:dispersion-a}, these cases correspond respectively to $(\beta,w,b_0)=(1/2, 2\text{m}, 2\text{m} )$, and $(\beta,w,b_0)=(1/3,1.24\text{m},0.2\text{m})$.
The latter   is representative of the channel used in the well known experiments by  \cite{Treske}.
The coefficient $\chi$ is computed from \eqref{eq:chi_trap} with 
${\it l}_2=w$, ${\it l}_1={\it l}_3=b_0/\beta$.
For completeness, we compare the curves with those obtained using the dispersion relation of the linear model obtained in  \cite{Chassagne}.
The   curves  show comparable trends, with  some differences     for shorter waves  which are within classical modelling error. 
In particular, for reduced wave numbers  below   $k {\it l} = \pi$  are within  
 2\% in the first case, and   0.5\% for the mild slopes. }

\begin{figure}
	\begin{center}
\begin{subfigure}{0.235\textwidth}		
\includegraphics[align=c,width=\textwidth]{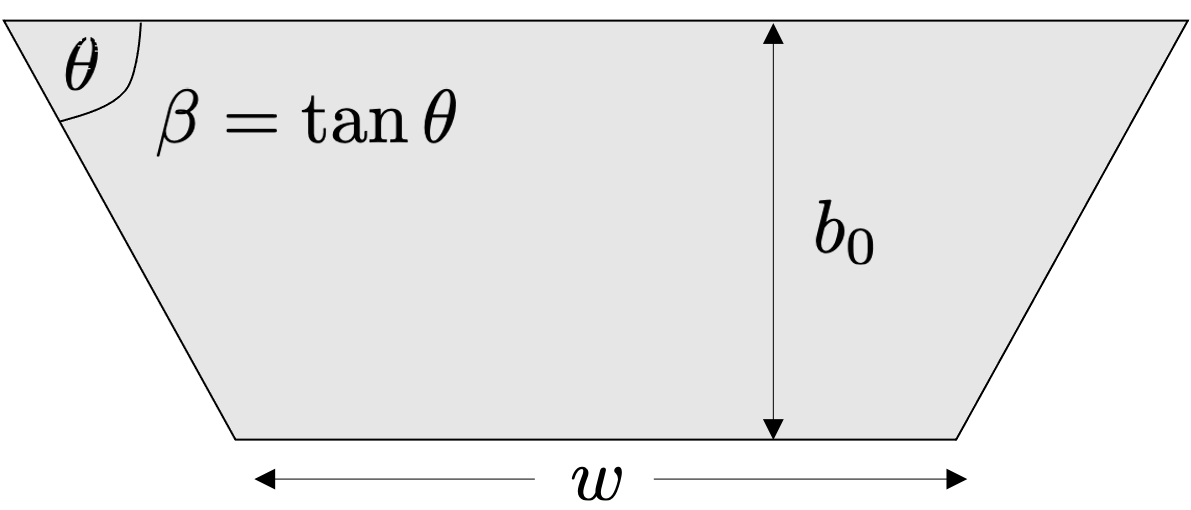}
\caption{}
\label{fig:dispersion-a}
\end{subfigure}\hfill
\begin{subfigure}{0.375\textwidth}	
\includegraphics[align=c,width=\textwidth]{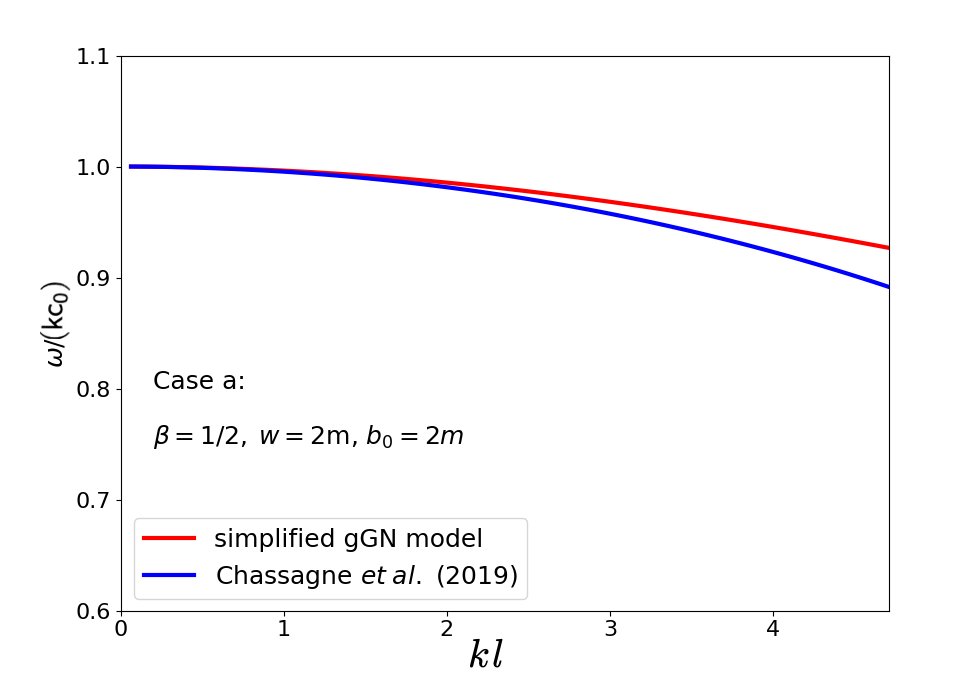}
\caption{}
\label{fig:dispersion-b}
\end{subfigure}\hfill
\begin{subfigure}{0.375\textwidth}
\includegraphics[align=c,width=\textwidth]{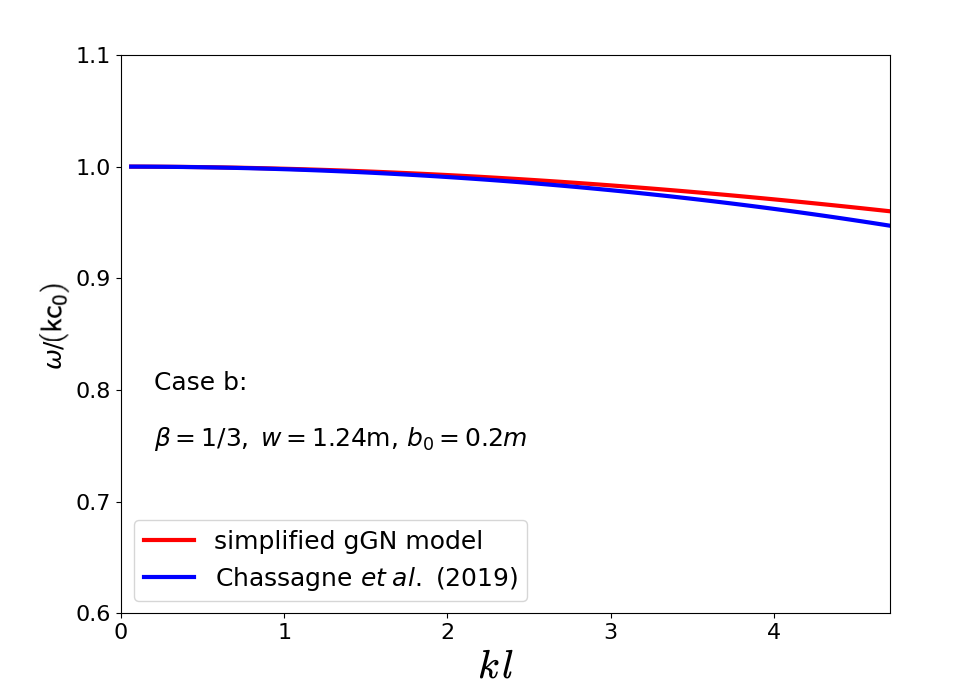}
\caption{}
\label{fig:dispersion-c}
\end{subfigure}
		\caption{\color{black}Dispersion relation for trapezoidal sections. (a)  notation. (b) phase speed for a deep  channel with steep slopes. 
		(c)   shallow channel with mild slopes. Simplified gGN model in blue,  linear  model of \cite{Chassagne} in red. 
			\label{fig:dispersion}}
	\end{center}
\end{figure}

For the second case,   representative of the experiments by  \cite{Treske},  we can use the dispersion relation to predict the wave length-Froude number dependence.
This is achieved by equating {\color{black} the phase velocity to the  relative bore speed  obtained from  the  Rankine-Hugoniot relations of the non-dispersive equations.
 In particular, one looks at the dependence of the wavelength on the relative Froude number defined as 
\begin{equation}\label{eq:froude}
\textsf{Fr} := \dfrac{|\langle u\rangle_2 - c_b|}{\sqrt{g\overline h_1}}
\end{equation}
where the subscripts ``1'' and ``2'' denote the states ahead of and behind the bore front, whose velocity is denoted by $c_b$.} 
We refer to  \cite{Chassagne,Lemoine} for details. The 
wave lengths  (scaled by  $\overline h_1$)   predicted theoretically 
 using this method are reported in figure  \ref{fig:dispersion1}.  For completeness we also report in the figure the data by \cite{Treske},
which include both those associated to geometrical dispersion (longer waves), and those measured on the channel axis for high Froude, related 
to classical non-hydrostatic  dispersive effects.  
Figure  \ref{fig:dispersion1} shows that the new model provides a good prediction
of the long waves associated to geometrical dispersion. 
The differences  with respect to the linear model   of \cite{Chassagne} for higher Froude numbers is 
in line with the  fact that the simplified gGN  provides  slightly  larger phase celerities for shorter waves,
as seen   in figures \ref{fig:dispersion-b} and  \ref{fig:dispersion-c}. 
Numerical simulation of these experiments using the fully nonlinear equations is discussed in section \ref{sec:treske}.

\begin{figure}
	\begin{center}
\centering\includegraphics[align=c,width=0.4\textwidth]{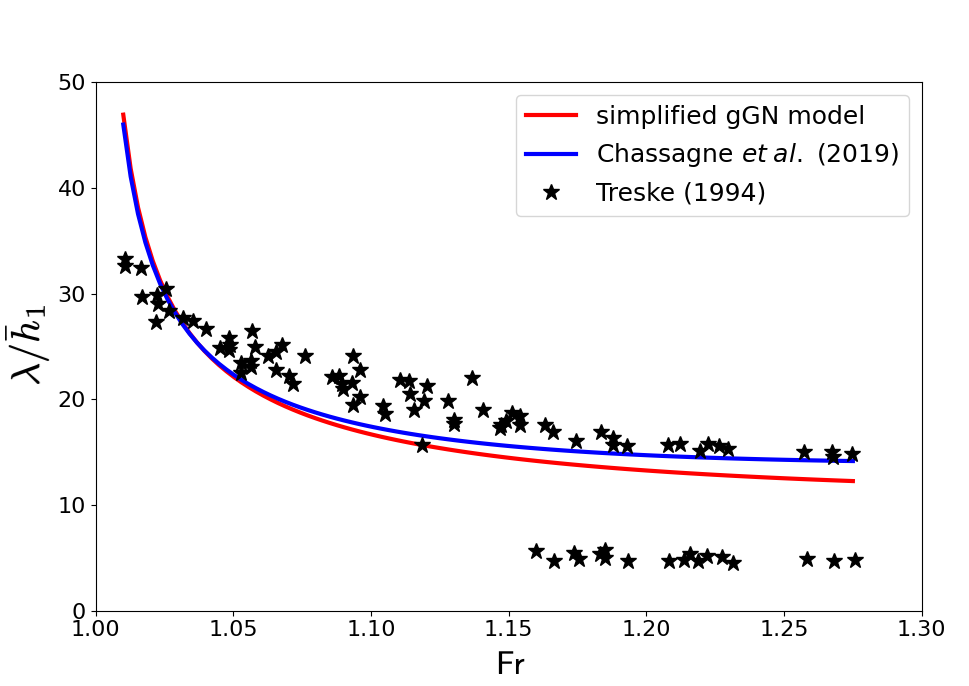} 
		\caption{\color{black} Lemoine analogy for the experiments by \cite{Treske}. Comparison of the theoretical predictions using the dispersion relation
		of the simplified gGN model (blue), with the one of   \cite{Chassagne} (red), and the data by  \cite{Treske} (symbols).
			\label{fig:dispersion1}}
	\end{center}
\end{figure} 
%
%
%
%
%
%
%
\subsection{Travelling  wave solutions:   {\color{black} solitary waves}}
We show in this section that the system derived admits exact travelling wave solutions. We focus here on solitary waves.
We consider  solitary wave of amplitude $a$ and celerity $C$, travelling  on 
on a flat water level  $\overline{h}_{\infty}$ far ahead and far behind.
So, let us  set $\xi = x-Ct$ and assume  that at time $t=0$ the peak is at $x=0$. We now seek solutions of the form
\begin{equation*}
\overline h =  \overline h(\xi), \quad 
\langle u\rangle =  \langle u\rangle (\xi)
\end{equation*}
We can readily write from the mass conservation law that
\begin{equation*}
\overline{h}(\langle u\rangle-C)= m=const,
\end{equation*}
Which, using the conditions at infinity, gives the  relations
\begin{equation}\label{eq:u_sol_h}
\langle u\rangle=  \Big(1- \dfrac{\overline{h}_{\infty} }{\bar h}\Big) C\;,\quad
m=- \overline{h}_{\infty}C
\end{equation}
The momentum equation implies : 
\begin{equation}
 \overline{h}(\langle u\rangle-C)^2+\frac{g\overline{h}^2}{2}-
\chi \ddot\tau=q=const.
\label{generic_equation0}
\end{equation}
Since $\ddot \tau=(\langle u\rangle-C)((\langle u\rangle-C)\tau')'=m^2\tau(\tau \tau')'$ (here `prime' means the derivative with respect to $\xi$), one obtains : 
\begin{equation}
m^2 \tau+\frac{g}{2\tau^2}-\chi m^2\tau(\tau \tau')'=q=const.
\label{generic_equation}
\end{equation}
Let $\tau_\infty=1/\overline{h}_\infty$. Then one can find $q$ and we have : 
\begin{equation}\label{eq:sol_ODE2}
m^2 (\tau-\tau_\infty)+\frac{g}{2}\left(\frac{1}{\tau^2}-\frac{1}{\tau_\infty^2}\right)-\chi m^2\tau(\tau \tau')'=0.
\end{equation}
Multiplying by $\tau'$, one can integrate it again : 
\begin{equation*}
m^2 (\tau-\tau_\infty)^2+g\left(-\frac{1}{\tau}+\frac{1}{\tau_\infty}-\frac{\tau}{\tau_\infty^2}+\frac{1}{\tau_\infty}\right)-\chi m^2(\tau \tau')^2=0.
\end{equation*}
We rewrite this equation as 
\begin{equation}
\chi m^2\tau^2 \tau'^2=F(\tau),\quad F(\tau)= m^2(\tau-\tau_{\infty})^2-\frac{g}{\tau \tau_{\infty}^2}(\tau-\tau_{\infty})^2.
\label{solitary_wave_equation}
\end{equation}
On can remark that that at $\tau=\tau_\infty$, one has $F=0$, $F'=0$. Moreover, $F''>0$ if $m^2>g/\tau_\infty^3$. If the velocity $\langle u\rangle_\infty=0$, the last inequality means that  the Froude number should be large than one (a classical inequality for the existence of solitary waves). Since $F\rightarrow -\infty$ as $\tau \rightarrow 0$, it implies that there exists $\tau_0 <\tau_\infty$ such that $F(\tau_0)=0$. It  implies the existence of supercritical solitary waves having the amplitude $a= \overline h_0-\overline h_\infty, \; \overline h_0=1/\tau_0.   $\\
We can now evaluate the above relation   at the peak, where $\tau'=0$ we have  $\tau_0 =1/\overline{h}_0= 1/(\overline{h}_\infty+a)$.
We thus obtain a  relation celerity/amplitude:
$$
0=m^2 \dfrac{ a^2 }{\overline{h}_\infty^2(\overline{h}_\infty+a)^2} - \dfrac{g a^2}{\overline{h}_\infty+a} 
$$
Using the constancy of $m$ we can also deduce $m^2 = C^2\overline h_{\infty}^2$ which allows to derive the compatibility relations
\begin{equation}\label{eq:celerity}
C^2 = g(\overline{h}_\infty+a) = g\overline{h}_0=\dfrac{g}{\tau_0}\;,\quad
m^2 = g(\overline{h}_\infty+a)\overline{h}_\infty^2=\dfrac{g}{\tau_0\tau_{\infty}^2}
\end{equation}
Combining all of the above results we end with the nonlinear ODE
\begin{equation}\label{eq:ODE}
\tau'^2 =  \dfrac{(\tau-\tau_\infty)^2(\tau-\tau_0)}{\chi\tau^3}. 
\end{equation}
The simplest way to solve it is to construct only a  half-width solitary wave with initial condition $\tau=\tau_0$ at $\xi =0$.

In practice, we have actually solved \eqref{eq:sol_ODE2}  using the basic {\tt odeint} method in {\tt numpy}, 
starting from the peak with given zero derivative, and the known value of $\tau =\tau_0$. Typical shapes of the solitary waves  obtained
for two different values of the non-linearity $\epsilon=a\tau_{\infty}$, and of the dispersion parameter $\chi$ are reported in figure \ref{fig:solitons}.
For sake of comparison we report the SGN solitary wave with the same amplitude. The solitary wave  of the new system 
generally have a more ``peaky''  shape,  and a strong dependence  on the relative values of $\chi$ and $\epsilon$.

\begin{figure}
	\begin{center}
			\begin{subfigure}{0.5\textwidth}
				\includegraphics[width=\textwidth]
				{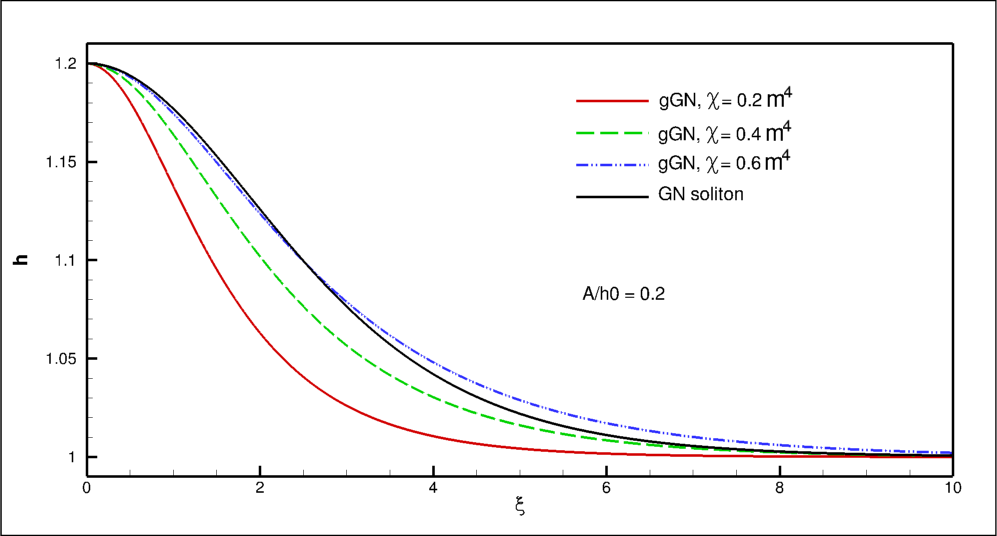}\caption{}
				\label{fig:solitons-a}
			\end{subfigure}\hfill
			\begin{subfigure}{0.5\textwidth}
				\includegraphics[width=\textwidth]
				{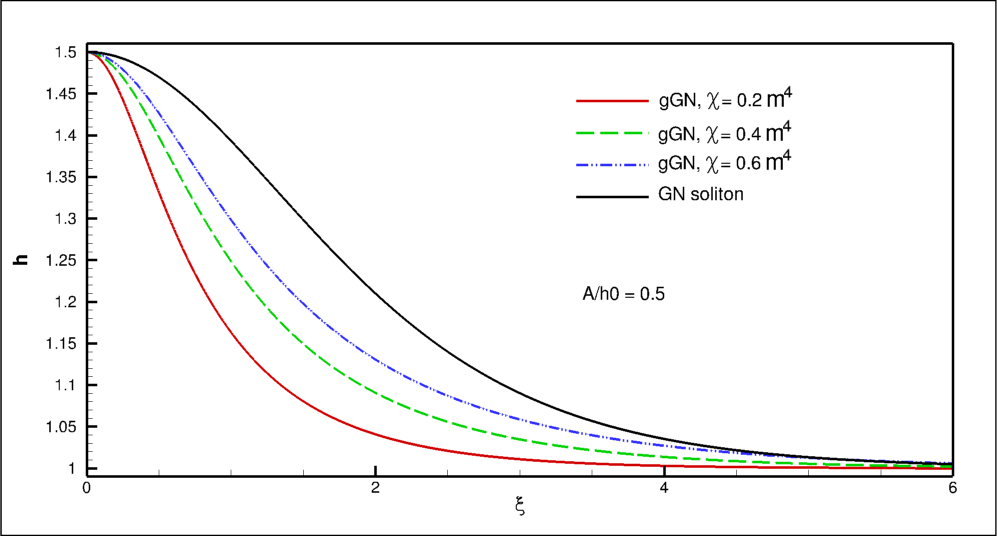}\caption{}
				\label{fig:solitons-b}
			\end{subfigure}\hfill
		\caption{{\color{black} Solitary waves obtained for  values of $\chi\in\{0.2,\,0.4,\,0.6\}\text{m}^4 $, and non-linearity   $ \epsilon=0.2$ (a),   and $ \epsilon=0.5$ (b).
			\label{fig:solitons}}}
	\end{center}
\end{figure}

\subsection{Travelling  waves solutions:  a  composite solution}

{\color{black} Equation \eqref{generic_equation} allows to obtain solutions of other forms than solitary waves}, including e.g. periodic   solutions. 
Following  the approach by \cite{Gavrilyuk_etal}, {\color{black}  we exploit this property to show the existence of exact composite travelling waves 
involving  two  moving profiles  connected   via a generalized Rankine-Hugoniot relation.} 
{\color{black}   
Equation \eqref{generic_equation} multiplied by $\tau'$ can be integrated in a more general form as}
\begin{equation}\label{eq:ODE_periodic}
\chi\tau^3(\tau')^2= (\tau-\tau_1)(\tau-\tau_2)(\tau-\tau_3), \quad 0<\tau_1<\tau_2<\tau_3.
\end{equation}
{\color{black} This equation describes  a periodic wave oscillating between  the depths $\bar h_1=1/\tau_1$ and $\bar h_2=1/\tau_2$. The third root $\tau_3$ controls the wave shape.
If $\tau_3=\tau_2=\tau_\infty$, one obtains  \eqref{eq:ODE} and goes back to a single solitary wave.
On can show that the three roots verify  the constraint (see  \cite{Gavrilyuk_etal} for details),
\begin{equation}\label{eq:ODE_periodic1}
m^2\tau_1\tau_2\tau_3=g.
\end{equation}}
{\color{black} \textcolor{black}{To construct a  stable solution we  follow a  procedure  similar to the one used in \cite{Gavrilyuk_etal} for  the Serre-Green-Naghdi equations: }
\begin{enumerate}
\item Choose   {\color{black} three roots such that} $0<\tau_1 <\tau_2<\tau_3$,  with   $\tau_2$ and $\tau_3$ close to each other.
\item {\color{black} Choose  a sign for the mass flow   obtained  from \eqref{eq:ODE_periodic1}. For $m<0$ we obtain a right travelling wave}.
\item  {\color{black} Give an arbitrary   value  to the velocity} corresponding to the state $\tau_2$. For example, $u_2=0$. 
Compute the  {\color{black} celerity $C$} of the periodic traveling wave from $\bar h_2(u_2- \textcolor{black}{C})=m$, $\bar h_2=1/\tau_2$. {\color{black} Thus,  the periodic traveling wave is perfectly defined. }
\item  From   \eqref{eq:ODE_periodic} we {\color{black} obtain } the value of the second derivative $\tau''\vert_{\tau=\tau_1}$ :
\begin{equation}
2\chi\tau_1^3\tau''\vert_{\tau=\tau_1} =(\tau_1-\tau_2)(\tau_1-\tau_3)>0.
\end{equation}
\item  {\color{black} Compute the values $(h_{\star}, u_{\star})$  which satisfy the generalized Rankine-Hugoniot relations (including dispersive terms) }: 
\begin{equation}
m=(u_{\star}-\textcolor{black}{C})/\tau_{\star},
\end{equation}
\begin{equation}
m^2 \tau_1+\frac{g}{2\tau_1^2}-\chi m^2\tau_1^2  \tau''\vert_{\tau=\tau_1} =m^2 \tau_\star+\frac{g}{2\tau_\star^2}.
\end{equation}
{\color{black} with as usual $\tau_\star=1/h_\star$. As shown in \cite{Gavrilyuk_etal}  if we choose  the root $\tau_\star$ closest to $\tau_2$, then
the solution having the constant  state $\star$ on the left and the corresponding  periodic traveling wave  on the right, is stable.}
\item  Introduce a regularization of the jump using e.g.  a half solitary wave from   $h_\star=1/\tau_\star$   to   $h_1=1/\tau_1$ (in $h$ variables ).
\end{enumerate}

To build the above solution in practice, {\color{black} given as data the tuple $(\tau_1,\tau_2,\tau_3)$,  we first  solve for $\tau_\star$, and then integrate   \eqref{eq:ODE_periodic} 
with initial value  $\tau_1$} to compute the periodic branch.
{\color{black}  The integration is done  in practice recasting \eqref{eq:ODE_periodic} as }
\begin{equation}\label{eq:ODE2_periodic}
\begin{aligned}
2\chi (\tau \tau')' =  &( 1 -\dfrac{\tau_2}{\tau} )( 1-\dfrac{\tau_3}{\tau}) +( 1-\dfrac{\tau_1 }{\tau})( 1-\dfrac{\tau_3}{\tau}) \\ +&  ( 1-\dfrac{\tau_1}{\tau} )( 1-\dfrac{\tau_2}{\tau} ) 
- (1-\dfrac{\tau_1}{\tau} )( 1-\dfrac{\tau_2}{\tau} )( 1-\dfrac{\tau_3}{\tau}) 
\end{aligned}
\end{equation}
{\color{black} which is  integrated  with initial conditions $(\tau, \tau') = (\tau_1, 0)$. The results presented have been obtained  using } the 8th order Runge-Kutta method in the {\tt SciPy} class {\tt solve\_ivp}, with relative and absolute tolerances {\tt rtol=$10^{-8}$} and 
 {\tt atol=$10^{-12}$}. Concerning the  regularization, we have computed the half solitary wave as discussed in the previous section.
 
As an example, let us take {\color{black}   $(\tau_1, \tau_2, \tau_3)=(1 \mathrm{m}^{-1} , 1.3\mathrm{m}^{-1}, 1.301\mathrm{m}^{-1}$), and   $\chi=0.4 ${\rm m$^4$}.
For this choice }  we obtain    $\tau_\star=1.30050031985595${\rm m$^{-1}$} $<\tau_1$.
Integrating the ODEs we obtain the solution plotted in figure  \ref{fig:disco}. {\color{black} In the figure we also compare three treatments of the transition: } the one obtained using a gGN solitary wave, a Gaussian $e^{-(x/L)^2}$ with $L=8${\rm m}, and  no regularization. We will compare the evolution of the three later in the results section.

\begin{figure}
	\begin{center}
			\begin{subfigure}{0.66\textwidth}
				\includegraphics[width=\textwidth]
				{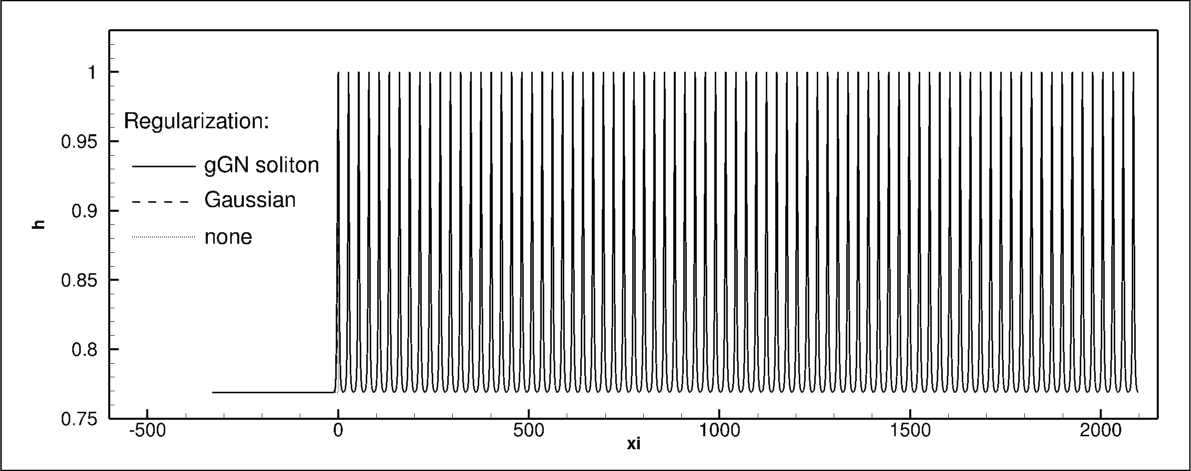}\caption{}
				\label{fig:disco-a}
			\end{subfigure}\hfill
			\begin{subfigure}{0.3\textwidth}
				\includegraphics[width=\textwidth]
				{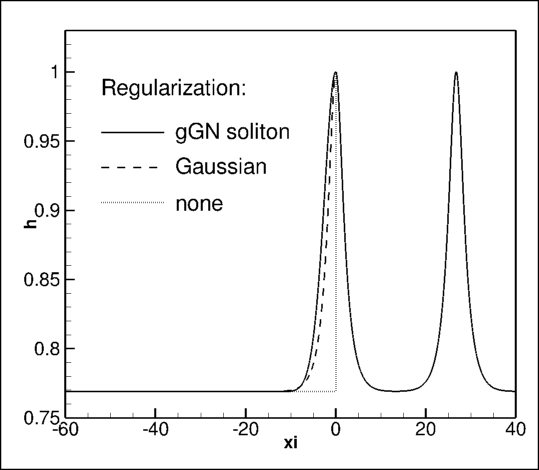}\caption{}
				\label{fig:disco-b}
			\end{subfigure}
		\caption{{\color{black}Composite travelling wave solution obtained for $\chi=0.4 ${\rm m$^4$}, setting  $\tau_1=1${\rm m$^{-1}$}, $\tau_2=1.3${\rm m$^{-1}$},  and $\tau_3=1.301${\rm $m^{-1}$}. 
		(a)    solution   on the whole domain [-350, 2100]{\rm m}.  (b)  zoom of the regularized jump and first peak.
			\label{fig:disco}}}
	\end{center}
\end{figure}

%
%
%
%
%

 \section{System reformulations and  numerical approximation}
 \label{4}

{\color{black} Using   well-established numerical methods, numerical tests are performed for two different reformulations of the dispersive model.
This section describes in some detail the form of the  equations used  for the discretization, recalling their main properties.
The  description of the numerics is left for completeness in    appendix \ref{app:schemes}.\\

For simplicity,  and unless otherwise specified, in the remainder of the paper  we will  use the notation $h$ and $u$ for    the average depth and Favre averaged
velocity.}

\subsection{Elliptic-Hyperbolic approximation}

{\color{black} We first consider   a splitting of the original system in an elliptic and  an hyperbolic step. This a rather classical
idea used in   many works    \cite{wei1995fully,Metayer10,bonneton2011splitting,cpr2,Gavrilyuk_Shyue_2024}.}

{\color{black} More specifically, here we follow the  reformulation introduced in   \cite{Filippini} and \cite{Kazolea} for the Serre-Green-Naghdi equations,
 and applied  to  other  Boussinesq models in \cite{Cauquis,Torlo,Jouy}.   The idea is to write   the system as a perturbation to the shallow water equations 
 via an algebraic source.}
The compatibility with the original equations leads to an elliptic problem for the latter.
To obtain this form, we start by using the relation
\begin{equation}
\label{dispersion4}
\begin{split}
 h^2\ddot\tau =   h ( u_x)^2 +   h(    u_{xt}   +   u    u_{xx}) = h ( u_x)^2 +   h\dot u_x 
=  h(    u_t +   u     u_x )_x = h ( \dot u)_x
\end{split}
 \end{equation}
 where   
$$
  \dot u_x = (    u_{xt}   +   u    u_{xx})  \ne (\dot u)_x=(    u_t +   u     u_x )_x .
$$
Using the above,   equation \eqref{eq:4b} can  be recast as 
\begin{equation}
\label{GN-explicit-pressure}
(hu)_t +(hu^2+ g\frac{h^2}{2}  )_x
= [ \dfrac{\chi}{ h}   (\dot u)_x ]_x
 \end{equation}
Now following \cite{Filippini},  we { \color{black}   introduce the variable $\phi$ satisfying}
\begin{equation}\label{phi-source}
  h \phi:= (hu)_t +   (hu^2 + g h ^2/2)_x \Rightarrow \phi  - gh_x= \dot u
\end{equation}
We can now combine the above definition with the compatibility requirement
$$
h\phi = [ \dfrac{\chi}{ h}   (\dot u)_x ]_x
$$
and obtain the equation for $\phi$:
\begin{equation*}\begin{split}
h\phi=[ \dfrac{\chi}{ h}   ( \phi  - gh_x)_x ]_x.
\end{split}\end{equation*}
The dispersive system \eqref{eq:4a}, \eqref{eq:4b}, \eqref{dispersion3} can thus be recast as   
\begin{equation}\label{eq:HES}\begin{split}
 h \phi - \chi [\dfrac{1}{ h}( \phi - g h_x       )_x]_x = &0\\
 h_t + (  h u)_x =&0\\
(hu)_t + (hu^2 + g  h^2/2)_x =& h \phi
%
\end{split}\end{equation}
which gives two operators. The first  an elliptic problem for the auxiliary variable $\phi$,
whose    variational form  reads  
\begin{equation}\label{eq:EHS-variational}
\int v   h  \phi +\chi \int v_x \tau \phi_x= \chi \int v_x\tau   \delta_x 
\end{equation}
where $v$ is a  test function with compact support $v \in H^1_0$, and having   set $\delta := gh_x$.
We can have the following simple characterization.
\begin{proposition}[Coercivity of the elliptic operator\label{prop:EHS-coercive}]  
Provided there exist  solution independent 
uniformly bounded   values $h_{\min}>0$, and $h_{\max}<\infty$ such that $h\in  (h_{\min},h_{\max})$, then the variational form \eqref{eq:EHS-variational}
is uniformly   coercive. 
\end{proposition}
{\it Proof.} Under the hypotheses made, we can easily show the $H^1$ norm equivalence   
$$
 \dfrac{\chi C_{\min}}{h_{\max}}  \|\phi\|_{H^1}^2  \le  
  \int    h  \phi^2 +\chi \int   \tau (\phi_x)^2
\le \dfrac{\chi C_{\max}}{h_{\min}}    \|\phi\|_{H^1}^2 
$$
for any $C_{\max} \ge  \max( 1 ,\dfrac{h_{\min}h_{\max}}{\chi} )$  and $C_{\max} \le  \min( 1 ,\dfrac{h_{\min}h_{\max}}{\chi} )\;\; \qedsymbol$.\\

The last two equations in \eqref{eq:HES} define  an hyperbolic evolution operator for the 
shallow water variables, with an extra algebraic forcing  term linear in  $\phi$. 

The system can be effectively  integrated in time and space. Here   the elliptic step is discretized using a standard finite element method.
The hyperbolic part is evolved with an explicit two-stages residual distribution scheme. 
Details are given in  appendix \ref{app:schemes}.

%

\subsection{Hyperbolic relaxation}

{\color{black} To avoid the  inversion of an elliptic operator, we replace the original system by an  approximate hyperbolic system which 
can be numerically treated with classical techniques  also used for the Saint-Venant equations \cite{Alireza,Favrie,Busto}.
Because of the   variational  formulation discussed in section 3.2, 
we will use   the method of augmented  Lagrangian proposed in \cite{Favrie,Firas_2019}. 
Compared to other approaches,  this technique preserves the variational structure of the  equations, thus guaranteeing the existence of   an exact 
energy conservation law also for the approximate system. 
A rigorous asymptotic analysis and justification of this approach is given in   \cite{Duchene_2019}. }
 

{\color{black}  In this method we  introduce  an auxiliary  variable $\eta(t,x)$ satisfying the 
second order    equation  
\begin{equation}\label{eq.wave-eta}
\ddot\eta +\frac{\mu}{\chi} \left(\eta-\tau\right)=0.
\end{equation}
This equation can be written as a first order quasilinear system 
\begin{equation*}
 \eta_t + u \eta_x  =   w,
\end{equation*}
\begin{equation*}
w_t + uw_x  =
-\frac{\mu }{\chi}  \left( \eta  -  \tau\right).
\end{equation*}
To couple the  evolution of $\eta$ to the other quantities, we consider the augmented system}
\begin{equation}\label{eq:hyp-form}
\begin{split}
h_t +&(hu)_x =0, \\
 (hu)_t +&(hu^2+ p_{tot} )_x=0,\\
 \left( h \eta \right)_t +&\left(hu \eta \right)_x  =h  w,\\
\left(hw \right)_t +&\left(hu w  \right)_x  =
-\frac{\mu }{\chi}  \left(h \eta  -  1\right).
\end{split}
\end{equation}
with the closure relation 
\begin{equation}\label{eq:hyp-pressure1} 
p_{tot}= g\frac{h^2}{2}  +  \mu \tau \left( h \eta   -1\right)= g\frac{h^2}{2}  +  \mu \left(  \eta   -\tau\right).
\end{equation}
As shown  in   \cite{Duchene_2019},   $\eta$ is  an order $\mathcal{O}(\mu)$ asymptotic approximation of  $\displaystyle \tau(t,x)=1/ h(t,x)$.
So  $\mu$ is a large positive parameter, which allows to relax $\eta$ to $\displaystyle \tau(t,x)$.}
In the limit $\mu\rightarrow \infty$  we recover the original system, and, in particular, the total pressure $p_{tot}$ becomes 
\begin{equation}
p_{tot}=\frac{gh^2}{2}-\chi\ddot{\tau}.
\end{equation}

 The system is hyperbolic with  eigenvalues 
\begin{equation}
\lambda_1=u-c, \quad \lambda_{2, 3 }=u, \quad\lambda_4=u+c,
\end{equation}
{\color{black} having set}
\begin{equation}
c^2=gh+ \mu\tau^2 .
\end{equation}
Following \cite{Firas_2019} and \cite{Sergey}, one can   show that  \eqref{eq:hyp-form} are  the Euler-Lagrange equations for the Lagrangian :
 $$
L=\int_{-\infty}^{+\infty}\left(h\frac{u^2}{2}-W(h, \eta, \dot{\eta})\right)dx, 
$$
with 
 $$
W(h, \eta, \dot{\eta})=\frac{gh^2}{2}-\frac{\chi \, h\, \dot{\eta}^2}{2}+\frac{\mu \, h}{2} \left(\eta-\frac{1}{h}\right)^2.
$$
The pressure $p_{tot}$ is defined as the partial Legendre transform with respect to $h$ 
\begin{equation}
p_{tot}=h\frac{\partial W}{\partial h}-W=\frac{gh^2}{2}+\mu \left(\eta-\tau\right),
\end{equation}
and  equation \eqref{eq.wave-eta}  is  equivalent to the Euler-Lagrange equation for $\eta $ 
\begin{equation}
\frac{\partial W}{\partial \eta}-\frac{\partial}{\partial t}\left(\frac{\partial W}{\partial \dot \eta}\right)-\frac{\partial }{\partial x}\left(\frac{\partial W}{\partial \dot \eta}u\right)=0.
\end{equation}

To integrate these hyperbolic equations numerically we use  an explicit two-stages residual distribution scheme,
coupled to a Strang splitting in time to treat the stiff source term.
Details are given in  appendix \ref{app:schemes}.

\section{Numerical tests}
\label{5}

\subsection{Verification using analytical solutions}

\subsubsection{Solitary waves}

We have performed a verification of  the one-dimensional solver using the solitary wave solution of section~\ref{3}. 
We discuss here the case of a solitary wave of non-linearity $\epsilon=a\tau_{\infty}=0.2$, 
and we consider the case $\chi=0.4$ (cf.   figure \ref{fig:solitons-a}), and $h_{\infty}=1${\rm m}. {\color{black}The behaviour with other values of the parameters is very similar.}

{\color{black} We have considered a solitary wave whose  peak is initially positioned   at $x_0 = -100${\rm m} in the the spatial   domain  $[-200, 200]${\rm m}. 
The soliton propagation  over a  distance   $L=200${\rm m}  has been computed  on several mesh resolutions  using  both the elliptic-hyperbolic formulation,
and the hyperbolic relaxation form.    We report in figure \ref{fig:soliton-conv-dx}  the resulting grid  convergence of the errors,
defined as the difference between the computed and exact depth, and velocity.  
Figures \ref{fig:soliton-conv-dx-a} and \ref{fig:soliton-conv-dx-c} show that with the elliptic-hyperbolic formulation 
we can achieve  the expected  second order of convergence in all the norms, for both depth and velocity.

For the hyperbolic reformulation,  figures  \ref{fig:soliton-conv-dx-b} and \ref{fig:soliton-conv-dx-d} show a dependence of the convergence rates 
on the value of the relaxation constant $\mu$. }
Already for $\mu =1000$ the error convergence {\color{black} rates reduce on finer meshes.  
 This is due essentially  to the modelling error for finite $\mu$}.
Values of $\mu$  of  order $10^4$ are necessary to correct this. However, such values affect negatively 
 the computational time, due to the time step restriction (cf. equation \eqref{eq:cfl-hyp} in appendix \ref{app:schemes}).

\begin{figure}
\begin{subfigure}{0.5\textwidth}
\centering\includegraphics[width=0.8\textwidth]{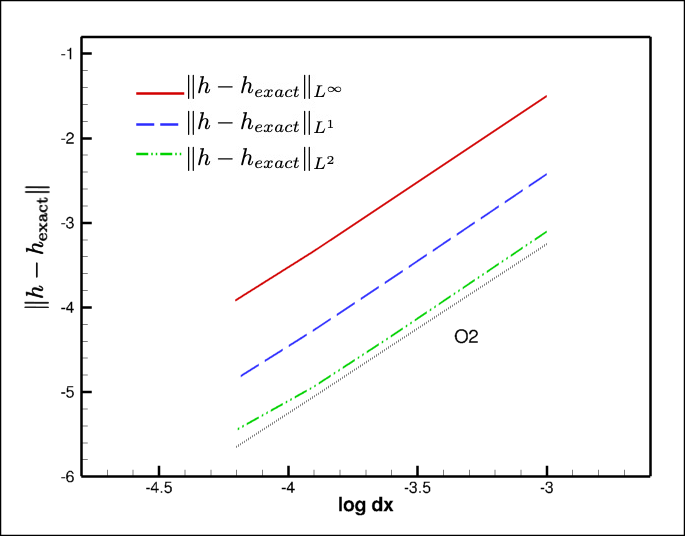}
\caption{}
\label{fig:soliton-conv-dx-a}
\end{subfigure}\hfill
\begin{subfigure}{0.5\textwidth}%
\centering\includegraphics[width=0.8\textwidth]{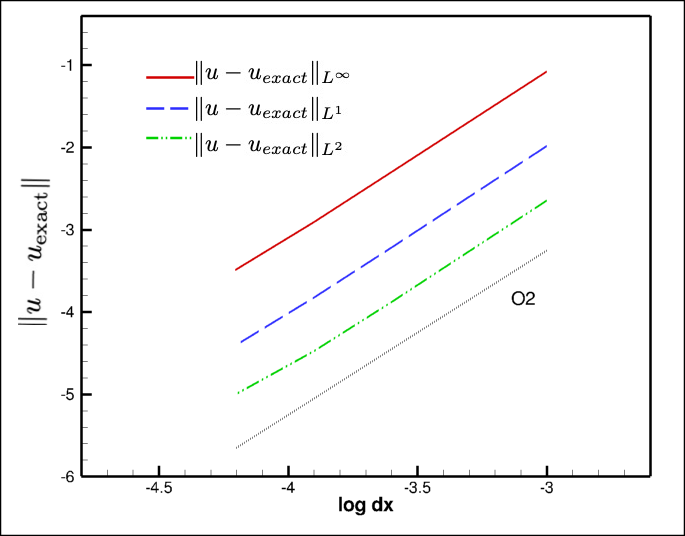}
\caption{}
\label{fig:soliton-conv-dx-b}
\end{subfigure} \\ 
\begin{subfigure}{0.5\textwidth}
\centering\includegraphics[width=0.8\textwidth]{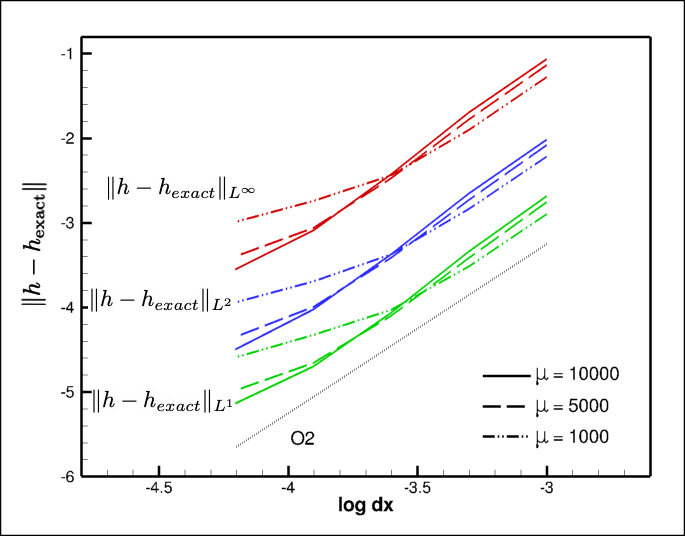}
\caption{}
\label{fig:soliton-conv-dx-c}
\end{subfigure}\hfill
\begin{subfigure}{0.5\textwidth}%
\centering\includegraphics[width=0.8\textwidth]{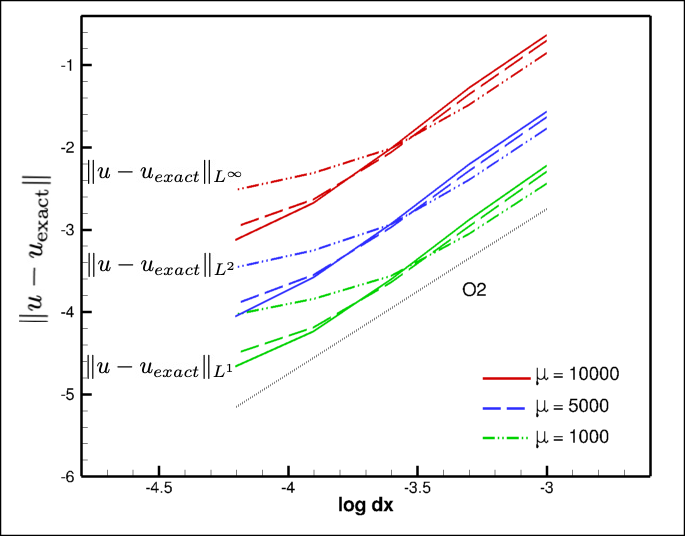}
\caption{}\label{fig:soliton-conv-dx-d}
\end{subfigure}
\caption{{\color{black} Grid convergence to a  solitary wave  with  non-linearity $\epsilon=0.2$, and $\chi=0.4$m$^4$. (a) depth error for  the elliptic-hyperbolic formulation. 
(b) depth error for  the hyperbolic formulation for different  $\mu$.  (c) velocity error for  the elliptic-hyperbolic formulation. 
(d) velocity error for  the hyperbolic formulation for different  $\mu$. \label{fig:soliton-conv-dx}}}
\end{figure}

To clarify the impact of this on the qualitative behaviour of the model, we have looked at the convergence {\color{black} of the solution with  $\mu$.} 
  The results are summarized in figure \ref{fig:soliton-conv-mu}. In particular, figure \ref{fig:soliton-conv-mu-a}
shows the wave profile  of the solitary wave obtained numerically   at final time. Despite what the error convergence suggests, 
clearly there is no visual difference in the solutions when using values of $\mu$  above $~ 10^3$.  {\color{black} This is reassuring concerning
the use of the hyperbolic approximation in physical applications, although the  specific 
optimal value of $\mu$ may depend on the  problem parameters.}
Figure \ref{fig:soliton-conv-mu-b} depicts the convergence of the depth with respect to $\mu$, showing  {\color{black} the expected} first order rate.

\begin{figure}
\begin{subfigure}{0.6\textwidth}
\includegraphics[width=\textwidth]{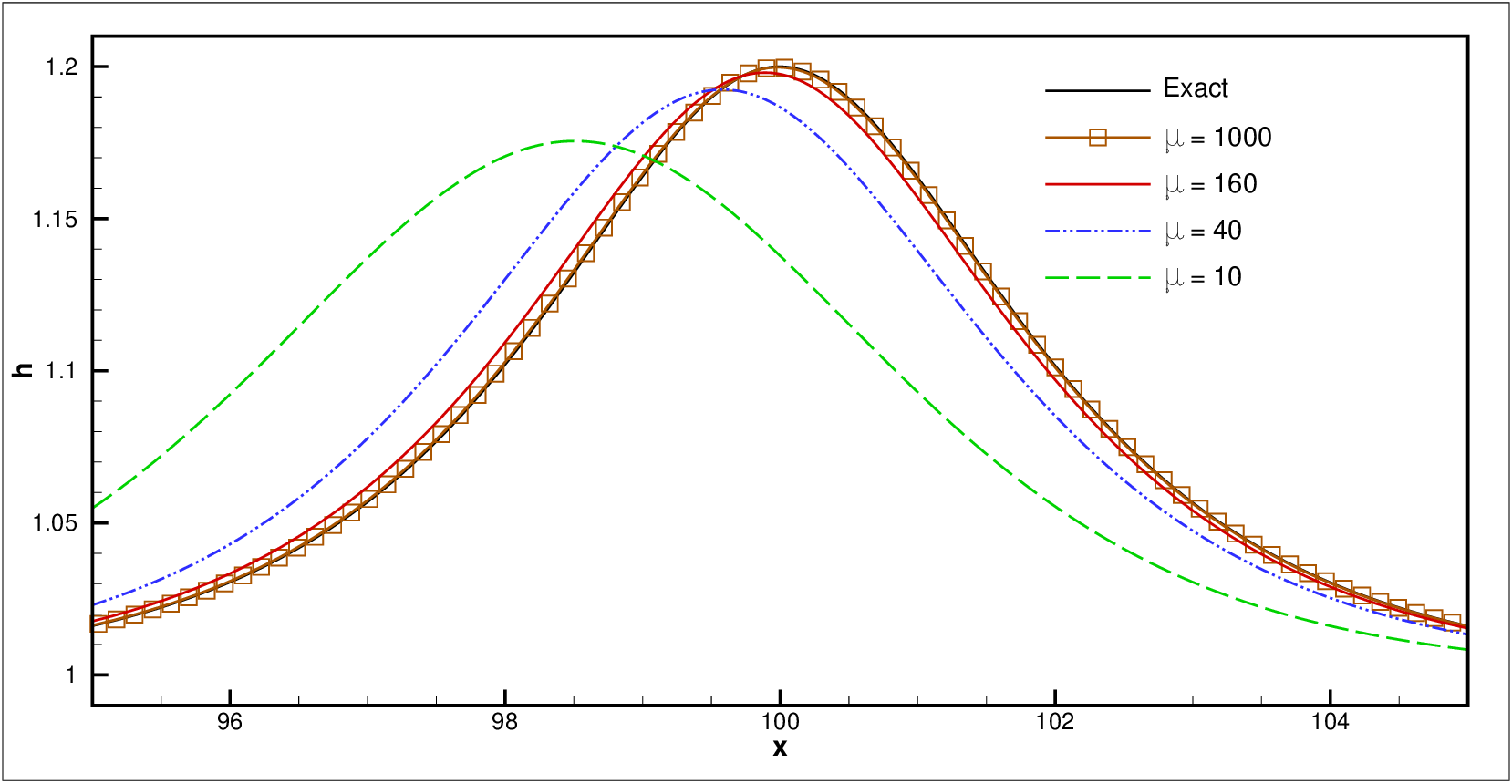}
\caption{}\label{fig:soliton-conv-mu-a}
\end{subfigure}\hfill
\begin{subfigure}{0.4\textwidth}
\includegraphics[width=\textwidth]{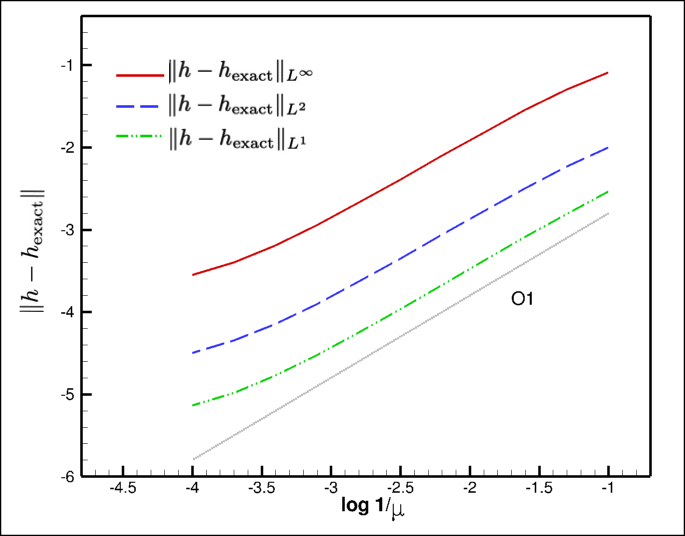}
\caption{}\label{fig:soliton-conv-mu-b}
\end{subfigure}
\caption{{\color{black}Convergence to a solitary wave  with  non-linearity $\epsilon=0.2$, and $\chi=0.4$m$^4$. (a)  depth distribution at final time obtained with the hyperbolic 
reformulation and different $\mu$. (b) depth error  convergence with $\mu$.\label{fig:soliton-conv-mu}}}
\end{figure}

\subsubsection{{\color{black}Composite solution}}

{\color{black}We now consider    computations of  the composite travelling wave solution of section 3.4. 
Besides verifying further the implementation, the numerical study tries to clarify the role of the regularization.
As a first exercise, we perform a grid convergence study. We start from  the initial solution 
with values provided at the end of \S3.4 (cf figure  \ref{fig:disco}), using the regularization with half gGN solitary wave.
Simulations are run on the computational domain $[-100, 2300]${\rm m}, with the initial jump at the origin. Free  boundary conditions are imposed on the right,
while the constant solution is imposed on the left.
We let the model evolve the flow for $50${\rm s}, and then compare with the initial solution translated by 
the exact celerity  computed analytically.  The results are reported in figure \ref{fig:disco1}. }

\begin{figure}
\begin{subfigure}{0.5\textwidth}
\includegraphics[width=\textwidth]{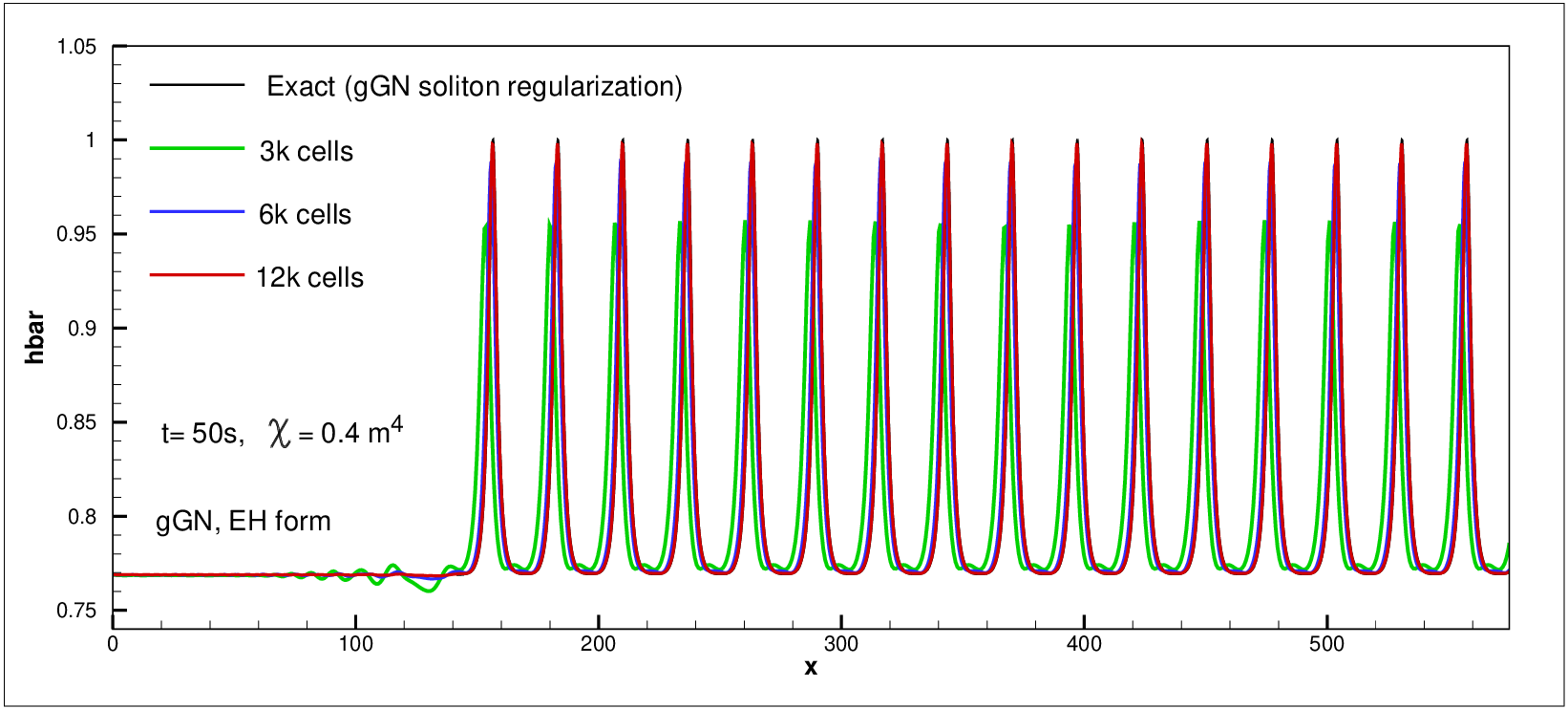} 
\caption{}\label{fig:disco1-a}
\end{subfigure}\hfill
\begin{subfigure}{0.5\textwidth}
\includegraphics[width=\textwidth]{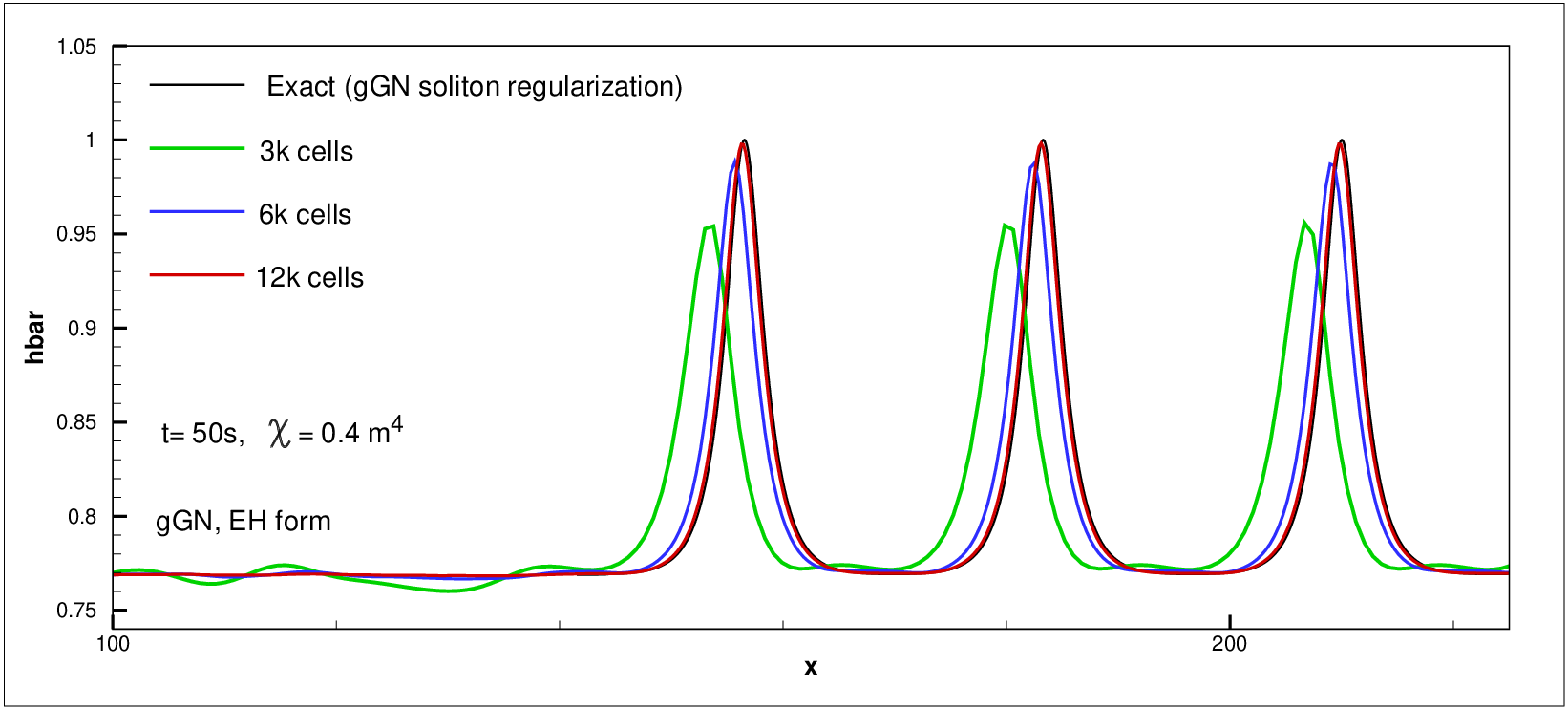} 
\caption{}\label{fig:disco1-b}
\end{subfigure}\\
\begin{subfigure}{0.5\textwidth}
\includegraphics[width=\textwidth]{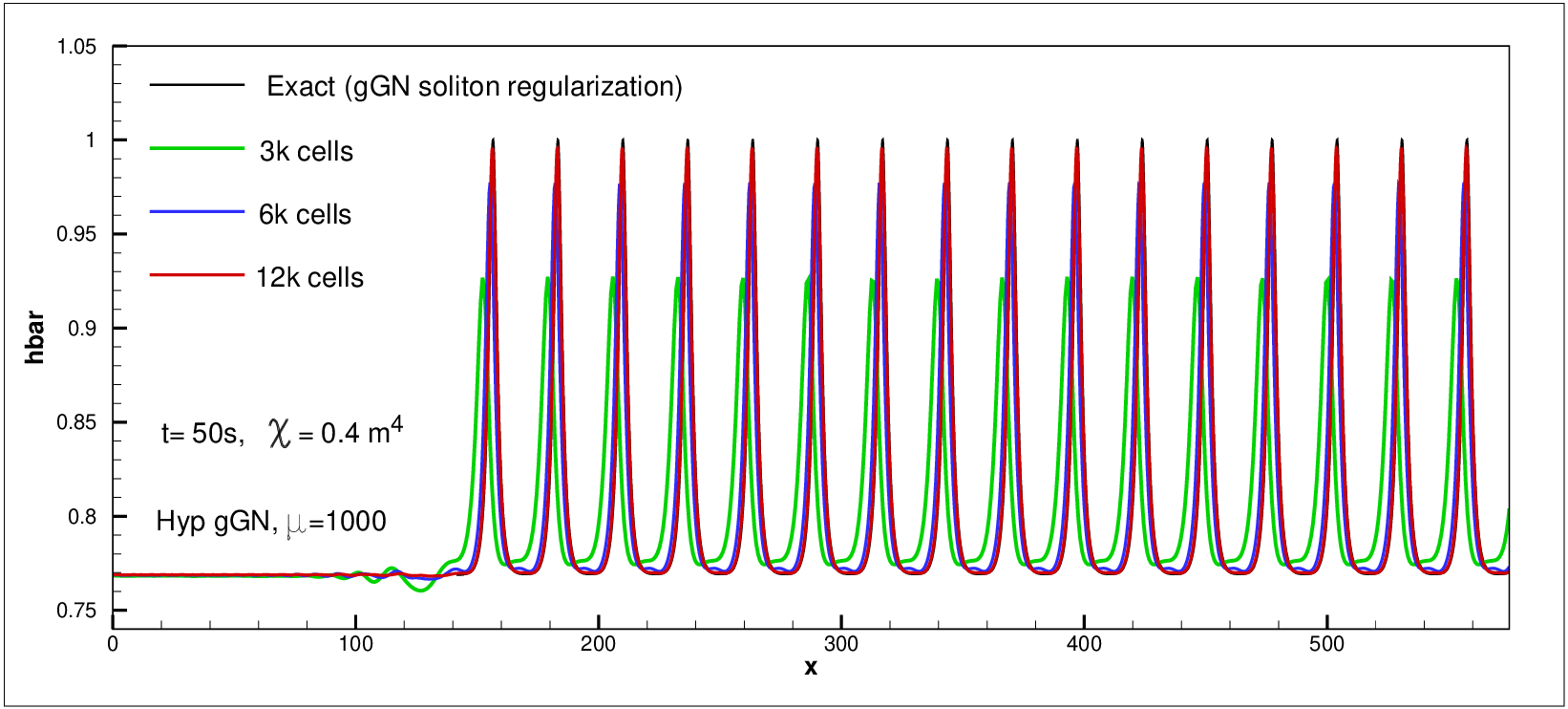} 
\caption{}\label{fig:disco1-c}
\end{subfigure}\hfill
\begin{subfigure}{0.5\textwidth}
\includegraphics[width=\textwidth]{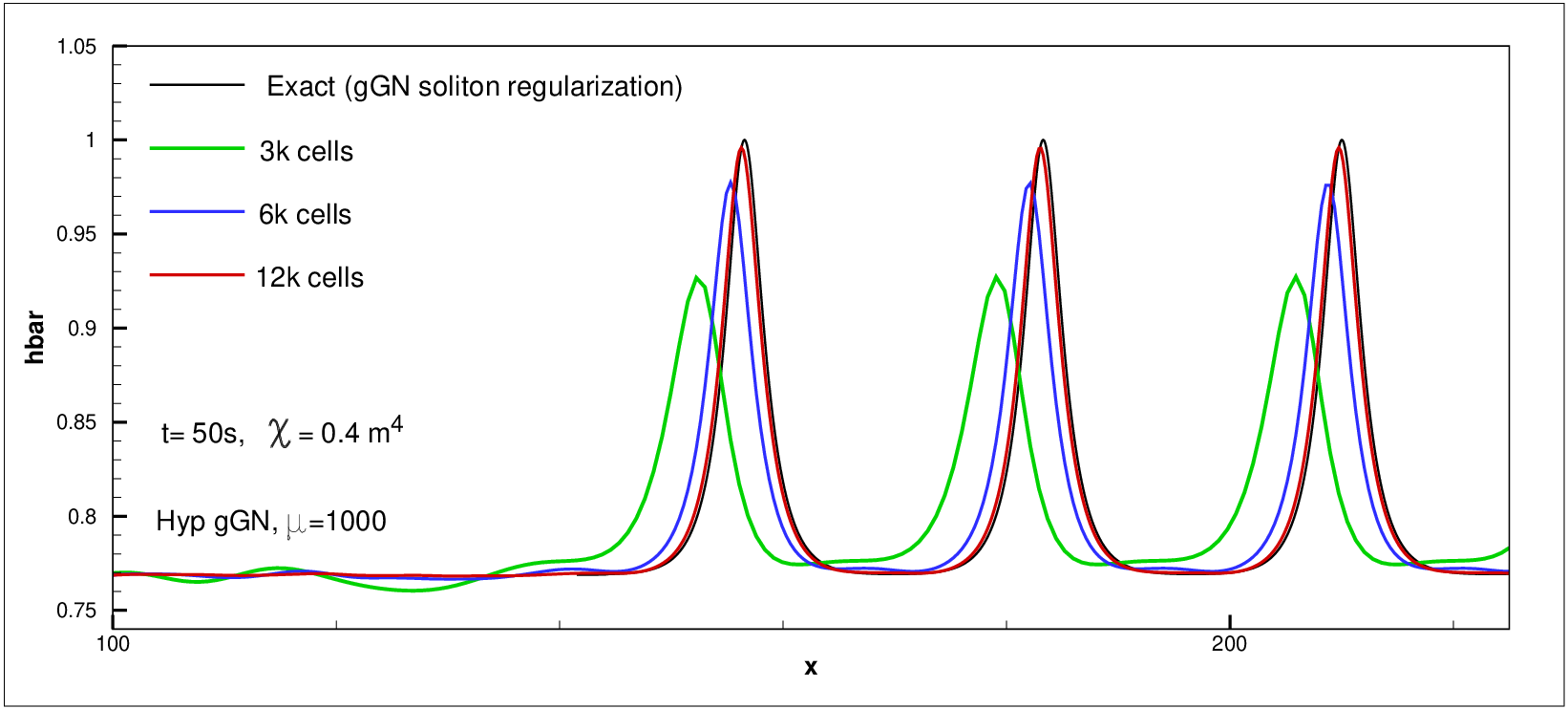} 
\caption{}\label{fig:disco1-d}
\end{subfigure}
\caption{ {\color{black} Grid convergence toward the regularized  composite solution  of section 3.5. (a)  
depth   at $t=50$s obtained with the    elliptic-hyperbolic approximation. (b) close up view of the first peaks
obtained with the    elliptic-hyperbolic approximation. 
(b) depth   at $t=50$s obtained with the  hyperbolic formulation  with $\mu=1000$.
(d)  close up view of the first peaks obtained with the  hyperbolic formulation.  \label{fig:disco1}  }}
\end{figure}

To give a more general view, we report in the figure  both the solutions in the subdomain [0,575]{\rm m}, and zoom of the first peaks, including the regularized one.
As one can see, both numerical approximations of the model converge nicely to the  travelling solution.   
 Quite interestingly, the results converge to the translated
initial solution, including the regularization. This suggests that the actual solution admitted is composed from a constant state, obtained from the Rankine-Hugoniot relations,
and an infinite series of oscillations between the values $\tau_1$ and $\tau_2$, much resembling an  infinite series of solitary waves {\color{black} propagating at the level $\tau_2$. }

\begin{figure}
\begin{subfigure}{0.5\textwidth}
\includegraphics[width=\textwidth]{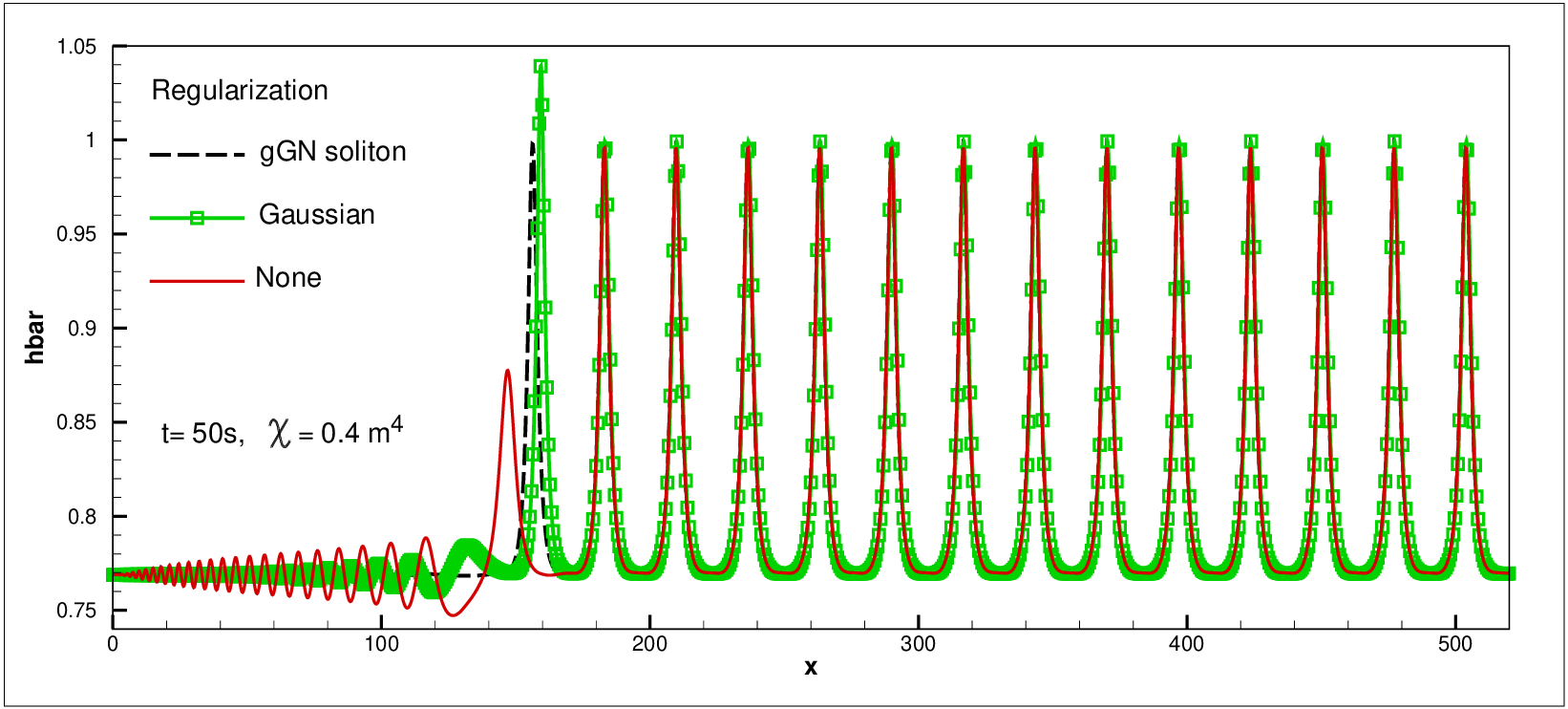}
\caption{}\label{fig:disco2-a}
\end{subfigure}\hfill
\begin{subfigure}{0.5\textwidth}
\includegraphics[width=\textwidth]{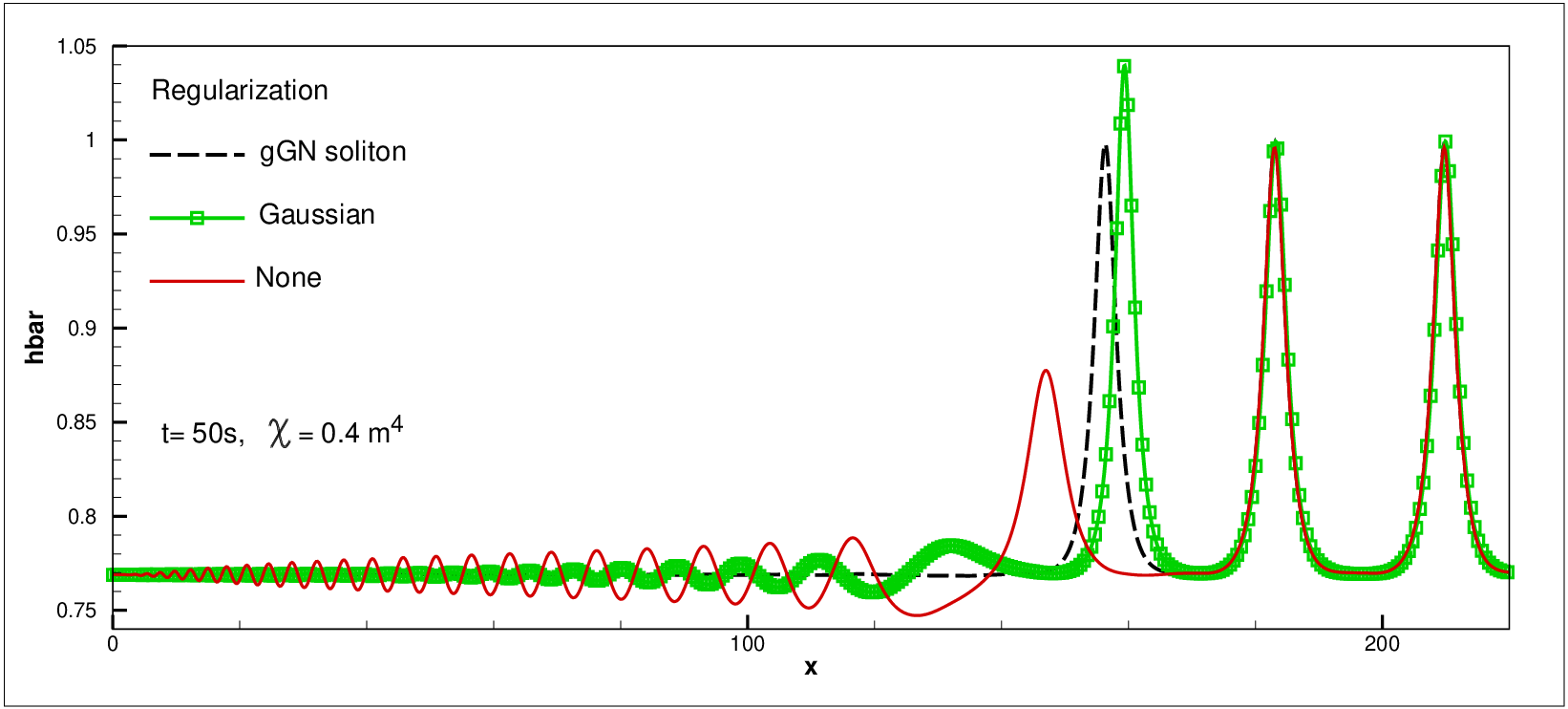}
\caption{}\label{fig:disco2-b}
\end{subfigure}
\caption{Composite solution: influence of the regularization of the initial solution.  \label{fig:disco2} }
\end{figure}

To get more insight into this we have performed the same simulation with the different regularization strategies   discussed in section 3.4.
Grid converged results are reported in figure \ref{fig:disco2}.  Clearly, past the regularized peak,  the three cases provide an  identical
infinite train of waves. Secondly, the  initialization with a half solitary wave  is the only case which does not introduce any secondary waves, to which the numerical approximation
converges as the mesh is refined.   This indicates that the exact solution is not a discontinuous one  in a classical sense, but rather a composite solution composed of a constant state $\star$, and a periodic wave train of waves of large length. This traveling wave solution verifies the generalized Rankine-Hugoniot conditions coming from the mass and momentum equations. In this sense, it is a ``discontinuous" solution of the gGN equations. 
Such   solutions are related to   shock solutions  of  the corresponding  Whitham system    (see e.g. \cite{Gavrilyuk_Shyue_2021} for the study of the  Benjamin-Bona-Mahony  
equation). 

\subsection{Breakdown of a Gaussian water column}

We pass now to the study of the behaviour of the model. To begin with, we consider a simple case consisting in the breakdown of a Gaussian water column
for which the initial state is given by
$$
h(x,t=0) = h_{\infty}( 1 + \epsilon e^{- (x-x_0)^2/L^2})\;,\quad  u(x,t=0) =0.
$$
We start by  investigating the propagation properties of the {\color{black} simplified} geometrical Green-Naghdi model   with respect to 
the {\color{black} dispersion and non-linearity} parameters $\chi$ and $\epsilon$.  To this end we consider   six cases corresponding to two values  $\chi=0.1${\rm m$^4$}
and $\chi=0.6${\rm m$^4$}, and three non-linearity values: $\epsilon=0.125$,  $\epsilon=0.25$, and $\epsilon=0.5$.
We consider a computational domain $[-150,150]${\rm m}, on which we set the initial Gaussian centered in $x_0=0$, and we take $L=2${\rm m}, 
and $h_{\infty}=2${\rm m}  as well. We simulate with both solution methods the evolution of the initial column up to time $16${\rm s} using $N=10^4$ cells, 
and taking a value $\mu=10^{3}$ in the hyperbolic relaxation formulation.

\begin{figure}
\begin{subfigure}{0.5\textwidth}
\includegraphics[width=\textwidth]{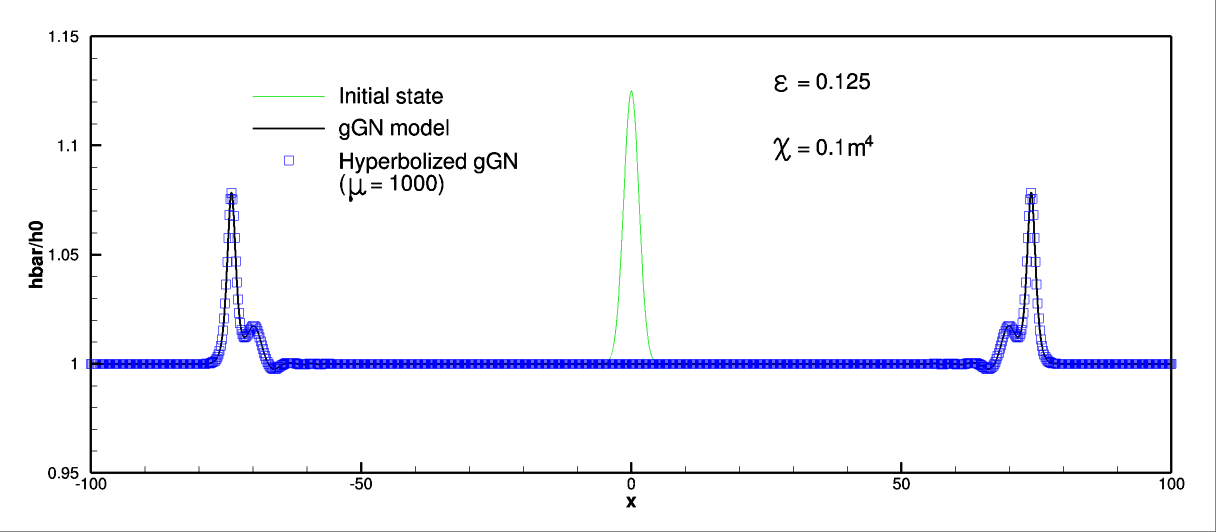}
\caption{}
\label{fig:bell1d-a}
\end{subfigure}\hfill
\begin{subfigure}{0.5\textwidth}
\includegraphics[width=\textwidth]{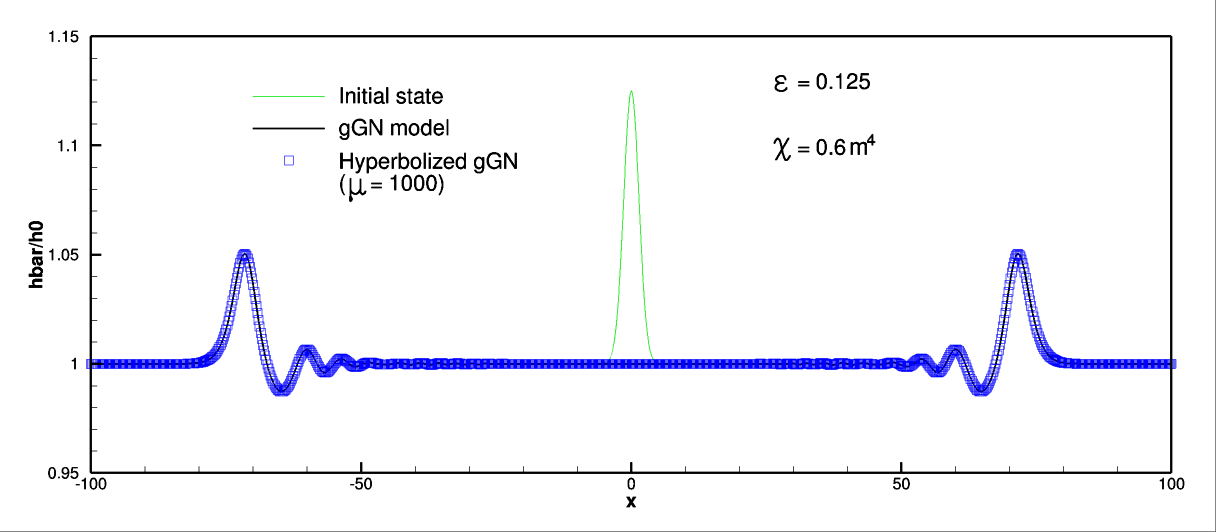}
\caption{}
\label{fig:bell1d-b}
\end{subfigure} \\
\begin{subfigure}{0.5\textwidth}
\includegraphics[width=\textwidth]{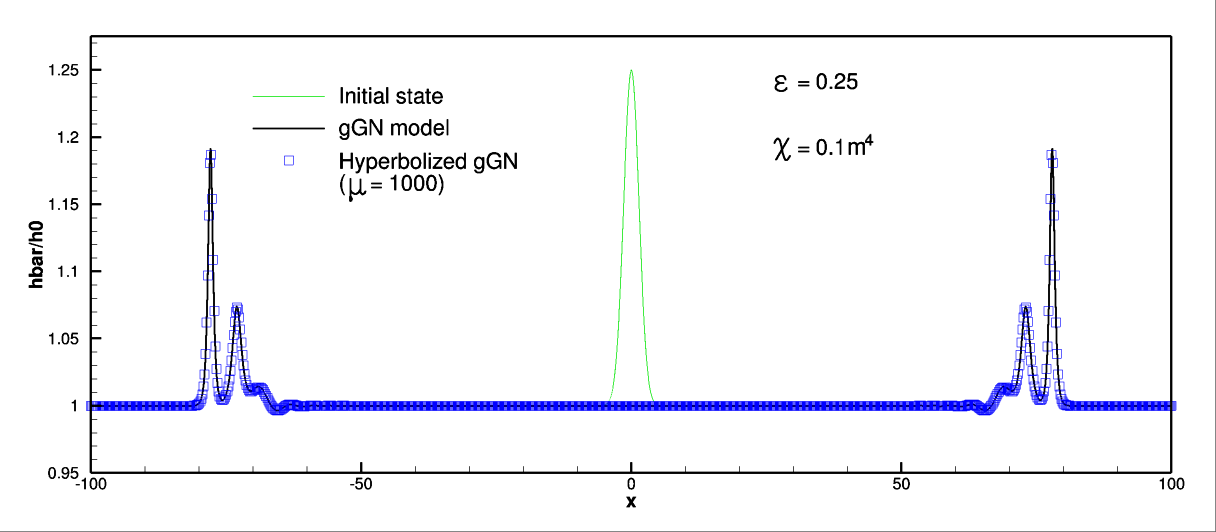}
\caption{}
\label{fig:bell1d-c}
\end{subfigure}\hfill
\begin{subfigure}{0.5\textwidth}
\includegraphics[width=\textwidth]{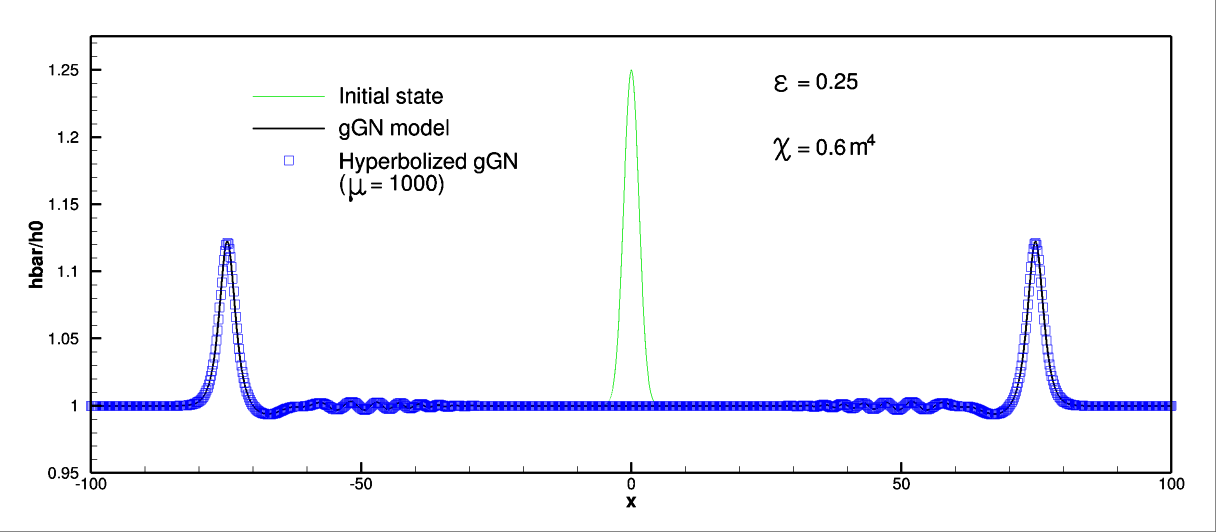}
\caption{}
\label{fig:bell1d-d}
\end{subfigure} \\
\begin{subfigure}{0.5\textwidth}
\includegraphics[width=\textwidth]{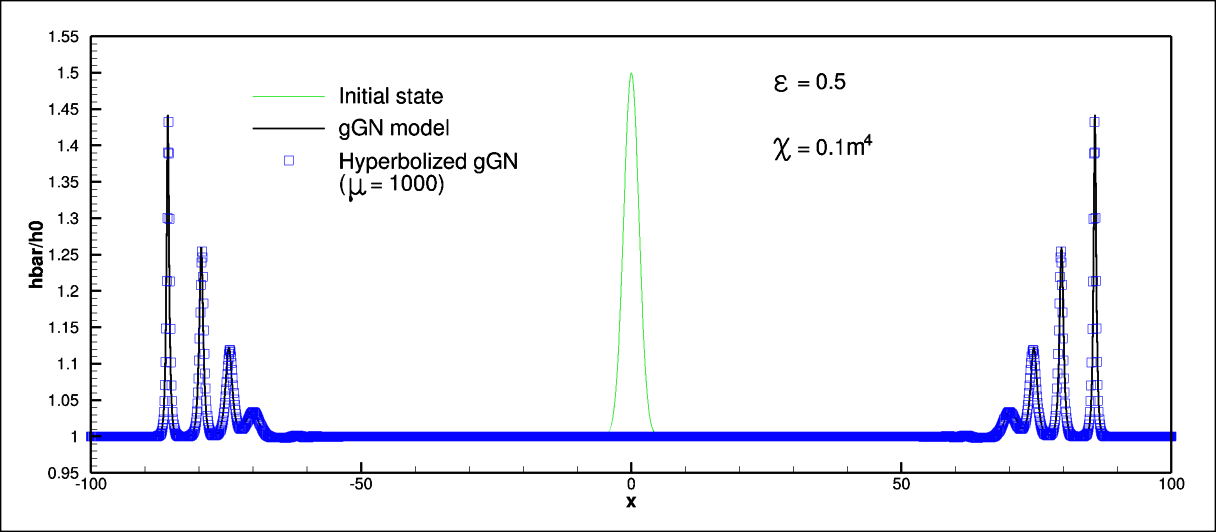}
\caption{}
\label{fig:bell1d-e}
\end{subfigure}\hfill
\begin{subfigure}{0.5\textwidth}
\includegraphics[width=\textwidth]{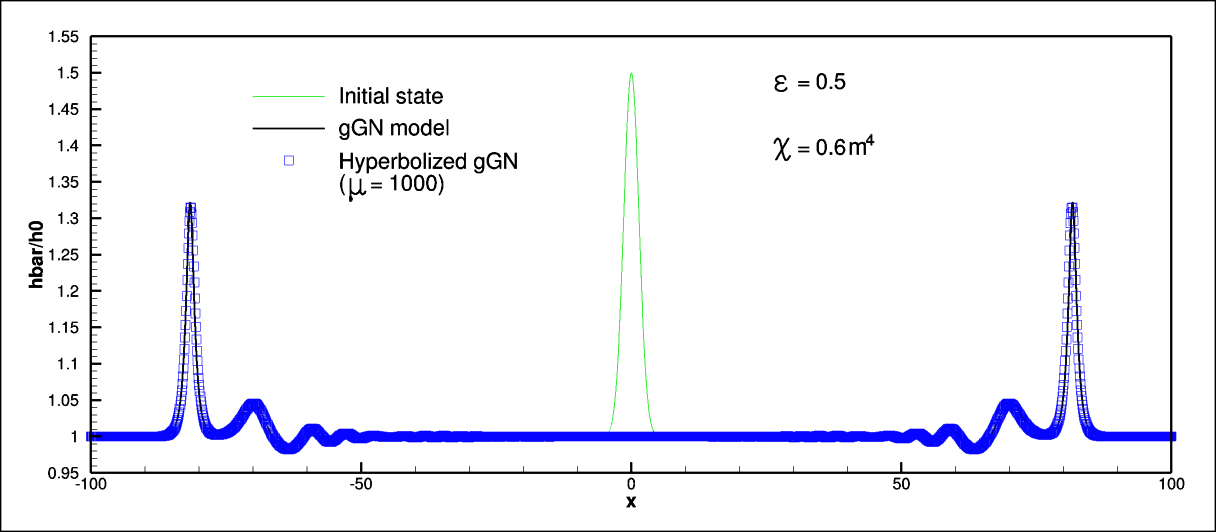}
\caption{}
\label{fig:bell1d-f}
\end{subfigure}  
\caption{{\color{black} Breakdown of a Gaussian water column: solutions at time $t=16${\rm s}.  (a)   $\epsilon= 0.125$, $\chi =0.1\text{m}^4$.
(b)   $\epsilon= 0.125$, $\chi =0.6\text{m}^4$. (c)   $\epsilon= 0.25$, $\chi =0.1\text{m}^4$.
(d)   $\epsilon= 0.25$, $\chi =0.6\text{m}^4$. (e)   $\epsilon= 0.5$, $\chi =0.1\text{m}^4$.
(f)   $\epsilon= 0.5$, $\chi =0.6\text{m}^4$. Solid line: elliptic-hyperbolic system.  {\color{blue} \small $\square$}  hyperbolized system with $\mu=1000$.\label{fig:bell1d}}}
\end{figure}

The results are visualized in figure \ref{fig:bell1d}. Several remarks can be made. First of all, the two different {\color{black} formulation} provide almost identical results,
which is a further verification (albeit indirect) of our implementation.  Furthermore, comparing the left and right {\color{black} column}, we can see
that higher values of $\chi$ result in longer waves, as one may expect since in this case the dispersive regularization has more weight. 
The lower value of $\chi$ results in shorter and taller waves, with stronger and more localized  oscillations. 
 These differences are enhanced when considering increasing values of non-linearity.
 Note that as the discussion of   section 2.4 shows,  for a given section shape, higher values of $\chi$  may correspond either  to deeper  channels,
or to  larger  ones. 

\subsubsection{Validation against section averaged 2D shallow water simulations}

Since for this case there is no exact solution, we  consider a validation against section averaged full 2D shallow water computations.
The latter have been run with the solver developed in the series of works \cite{r15,ar18,ar20}, and thoroughly validated on 
a wide range of cases against analytical and experimental data. 

The setup of the 2D simulations is the following. We consider three sections:   symmetric trapezoidal; symmetric  triangular; non-symmetric trapezium. 
{\color{black} Note that the last is out of the hypotheses of the model. It is included in the study for sake of generality, and to test the limit of our derivation.}
Following the notation of {\color{black} figures \ref{fig:periodic structure-b} and \ref{fig:periodic structure-c} in section~\S2, 
we fix the value of $b_0$, and then choose appropriate values of   the remaining lengths for each section type to obtain the correct value of $\chi$.
For the case $\chi=0.1${\rm m}$^4$, we set   $b_0 =2.5${\rm m}. 
We then choose  $({\it l}_1, {\it l}_2, {\it l}_3) = (1.07, 0.36,1.07)${\rm m} in the symmetric trapezoidal case,}
and  $({\it l}_1, {\it l}_2, {\it l}_3) = (1.38575, 0.0,1.38575)${\rm m}  in the triangular one. For the     non-symmetric trapezium we choose 
$({\it l}_1, {\it l}_2, {\it l}_3) = (1.284, 0.36,0.7249)${\rm m}. 
For $\chi =0.6${\rm m}$^4$ we have  set $b_0 =2.75${\rm m}, and chosen
  $({\it l}_1, {\it l}_2, {\it l}_3) = (2.165, 1.1175,2.165)${\rm m} in the trapezoidal case,
and  $({\it l}_1, {\it l}_2, {\it l}_3) = (3.0855, 0.0,3.0855)${\rm m}  in the triangular one. 
Also in this case we have computed solutions  with a non symmetric trapezoidal section obtained with  $({\it l}_1, {\it l}_2, {\it l}_3) = (2.598, 1.1175,1.14668)${\rm m}.

In all cases, the initial solution is then set using  equation \eqref{h_expression}: 
$$
h(x,y,t=0) =   \overline{h}(x,t=0)  - b(y) + \overline{b}  = h_{\infty}( 1 + \epsilon e^{- (x-x_0)^2/L^2}) - b(y) + \overline{b}.
$$
with $b(y)$  as in {\color{black} figures \ref{fig:periodic structure-b} and \ref{fig:periodic structure-c}}, and $ h_{\infty}=2${\rm m}.   Only the case $\epsilon=0.25$ is commented for brevity.  
The computational domain is the rectangle $[-150,150]\textrm{m}\times[-{\it l}_1-{\it l}_2/2,{\it l}_2/2+{\it l}_3 ]\textrm{m}$. 
Both velocity components are set to zero at $t=0$, and reflective boundary conditions   are set on all boundaries.  
With the above initialization, and the chosen values for $h_{\infty}$ and $b_0$, 
we satisfy all the hypotheses of the model.
 Fine triangulations with  mesh sizes $\Delta x\approx 6.25 ${\rm cm} and 
 $\Delta y\approx 2${\rm cm}  are used. As a first step,  we run a case with $b(y)=0$. 
 The result is reported for completeness in figure \ref{fig:3dbell_no_b} showing the expected behaviour  from the shallow water system with shock formation
 in finite time. {\color{black} Figures \ref{fig:3dbell_no_b-c} and \ref{fig:3dbell_no_b-d}  } show $y-$averaged data confirming this behaviour.

We now consider the simulations  including bathymetric variations. {\color{black} A close up three dimensional view 
of the  right-going waves is  reported in figures \ref{fig:3dbell_chi01} 
for the case $\chi=0.1${\rm m}$^4$,  and \ref{fig:3dbell_chi06} for $\chi=0.6${\rm m}$^4$.}
%
The figures clearly show the  formation and propagation of smooth waves, with roughly a  first wave resembling  a solitary wave with variable amplitude along $y$,
followed by  secondary smooth waves. The first wave amplitudes are higher on the shallower parts than in the center.
These results are similar to those  presented in \cite{Chassagne} for Favre waves: no discontinuities are produced despite {\color{black} of the 
the nonlinear hyperbolic character of the system, and of the non-linearity of the problem}.
Comparing the two values of $\chi$ we  see that for  $\chi=0.1${\rm m}$^4$ (thinner or shallower channels) we obtain shorter waves, with secondary ones
clustered close to the leading wave. For higher $\chi$ we   see more dispersion,  longer waves, and secondary waves with  lower celerity.
This behaviour is identical to the one of the gGN system in  one dimension. \\

To go further, we now compute numerically $y-$ averages of the 2D data, and compare them to the solutions obtained with the geometrical Green-Naghdi equations
developed in this paper, which for the value of $\epsilon=0.25$ corresponds to the second row of figure~\ref{fig:bell1d}. The comparisons are reported in figure~\ref{fig:bell_1d_2d}.
The plots show that the averaged data for the different sections are very close and present an excellent match   with the solutions of the  1D gGN model. 
{\color{black} Not surprisingly,  the non-symmetric case is more poorly reproduced by the one-dimensional simulations.
This is expected as symmetry  plays an important role in the derivation of the model.}
These results 
provide further evidence of the existence of these dispersive-like waves due to transverse refraction. These are indeed dispersive waves, 
which can be modelled with the section averaged geometrical Green-Naghdi equations proposed in this paper.

\begin{figure}
\begin{center}
\begin{subfigure}{0.5\textwidth}
\includegraphics[width=6.7cm, height = 2.78cm,trim={2.5cm 2.5cm 3cm 0},clip]{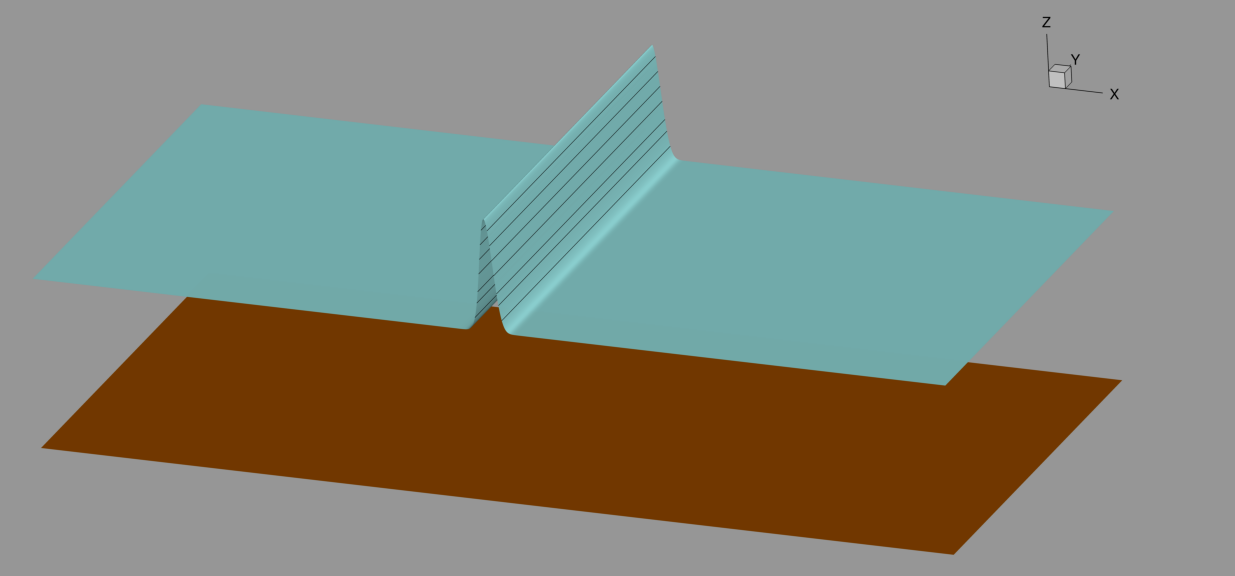}
\caption{}
\label{fig:3dbell_no_b-a}
\end{subfigure}\hfill
\begin{subfigure}{0.5\textwidth}
\includegraphics[width=6.7cm, height = 2.78cm,trim={2.5cm 2.5cm 3cm 0},clip]{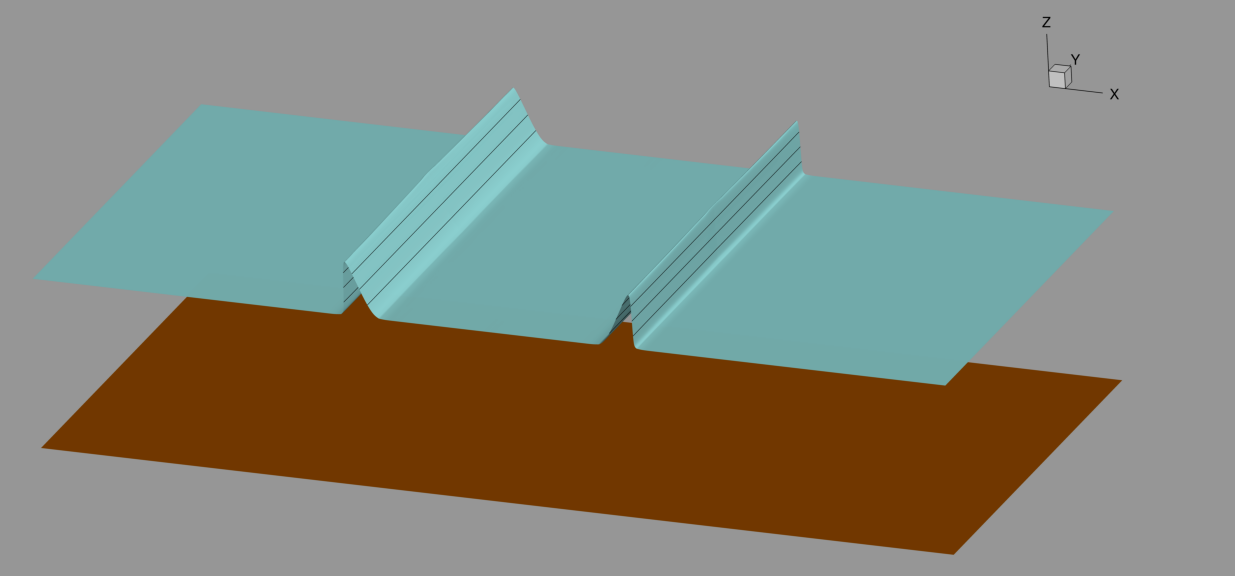}
\caption{}
\label{fig:3dbell_no_b-b}
\end{subfigure}\\
\begin{subfigure}{0.5\textwidth}
\includegraphics[width=\textwidth]{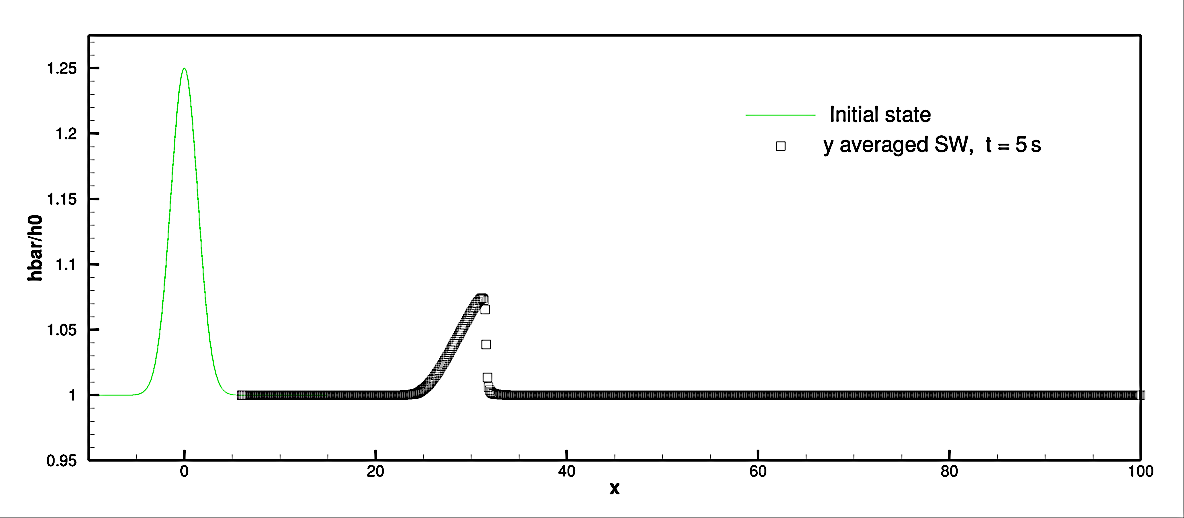}
\caption{}
\label{fig:3dbell_no_b-c}
\end{subfigure}\hfill
\begin{subfigure}{0.5\textwidth}
\includegraphics[width=\textwidth]{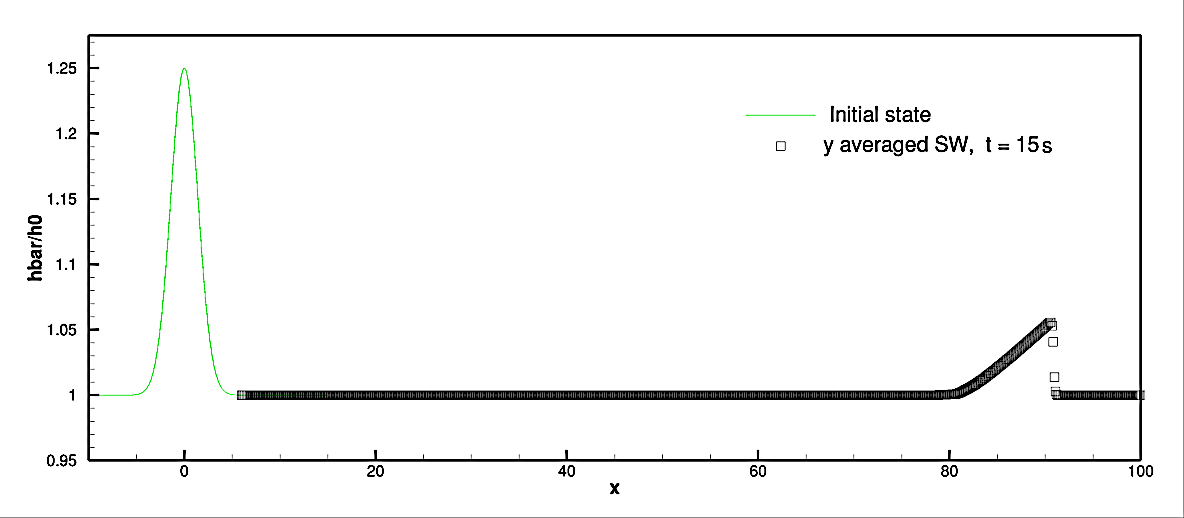}
\caption{}
\label{fig:3dbell_no_b-d}
\end{subfigure}
\end{center}
\caption{{\color{black} Breakdown of a Gaussian water column:     shallow water simulations with flat bathymetry  for   $\epsilon=0.25$. 
(a)  initial free surface. (b) free surface    at time $t=5${\rm s}.
(c)  $y-$averaged data at times $t=5${\rm s}. (d)  $y-$averaged data at times $t=15${\rm s}.
 \label{fig:3dbell_no_b}}}
\end{figure}

\begin{figure}
\begin{subfigure}{0.33\textwidth}
\includegraphics[width=\textwidth,trim={17cm 0 0 0},clip]{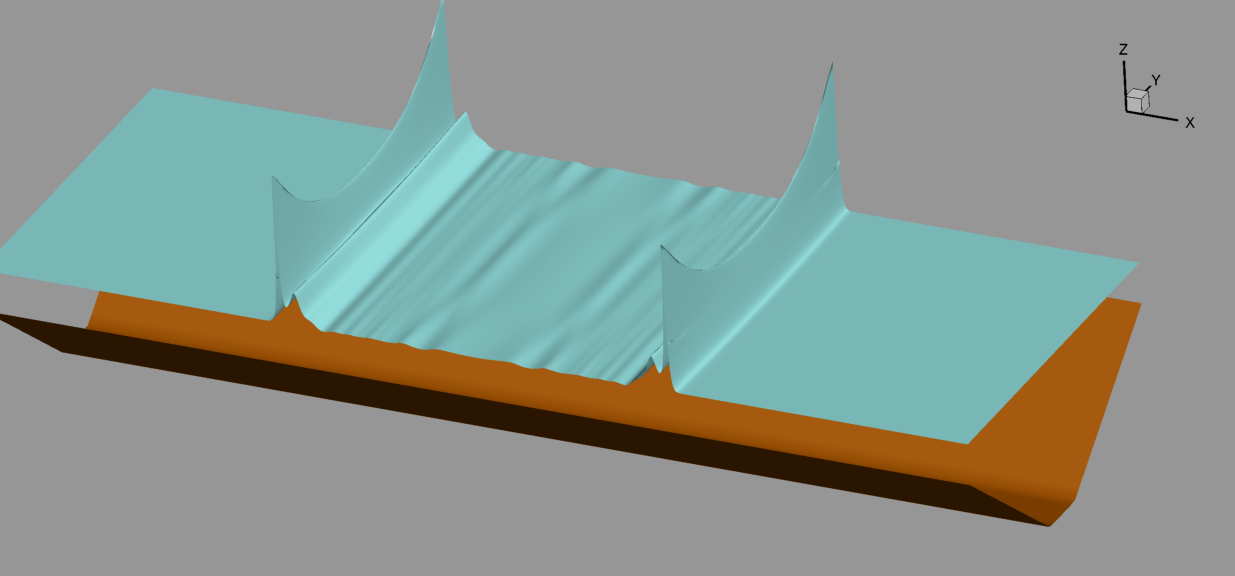}
\caption{}
\label{fig:3dbell_chi01-a}
\end{subfigure}\hfill
\begin{subfigure}{0.33\textwidth}
\includegraphics[width=\textwidth,trim={17cm 0 0 0},clip]{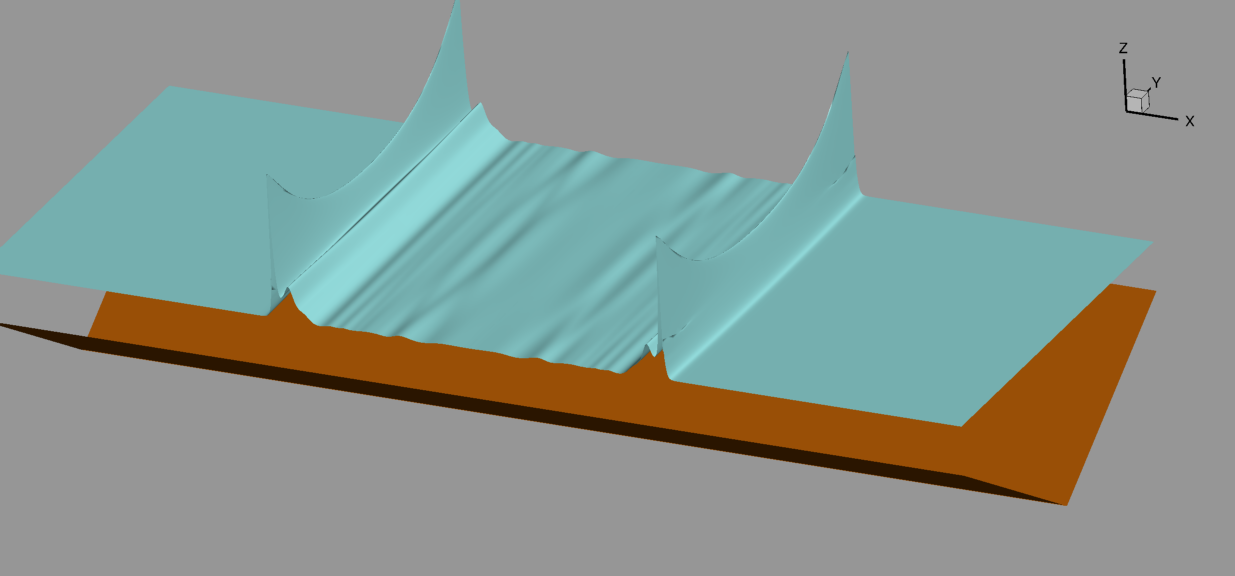}
\caption{}
\label{fig:3dbell_chi01-b}
\end{subfigure}\hfill
\begin{subfigure}{0.33\textwidth}
\includegraphics[width=\textwidth,trim={17cm 0 0 0},clip]{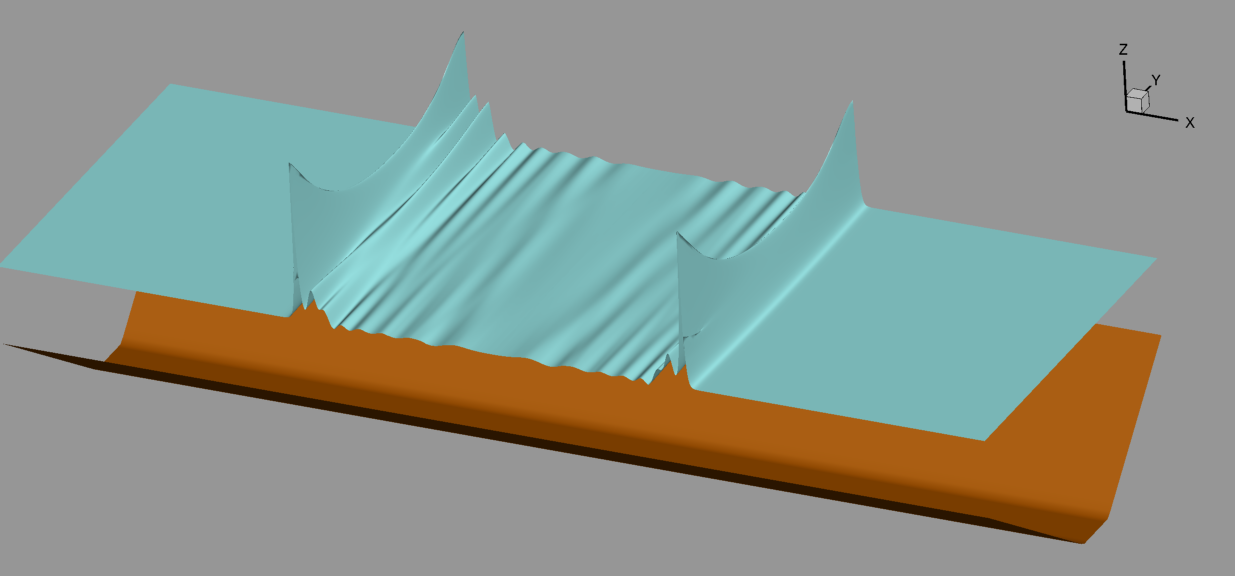}
\caption{}
\label{fig:3dbell_chi01-c}
\end{subfigure}
\caption{{\color{black}Breakdown of a Gaussian water column with $\chi=0.1${\rm m}$^4$ and $\epsilon=0.25$. Enhanced view of the right-going wave at time $t=8${\rm s}. Shallow water simulations for a
 symmetric trapezoidal section (a),  symmetric triangular section (b), and   
  asymmetric trapezoidal  section (c). \label{fig:3dbell_chi01}}}
\end{figure}

\begin{figure}
\begin{subfigure}{0.33\textwidth}
\includegraphics[width=\textwidth,trim={17cm 0 0 0},clip]{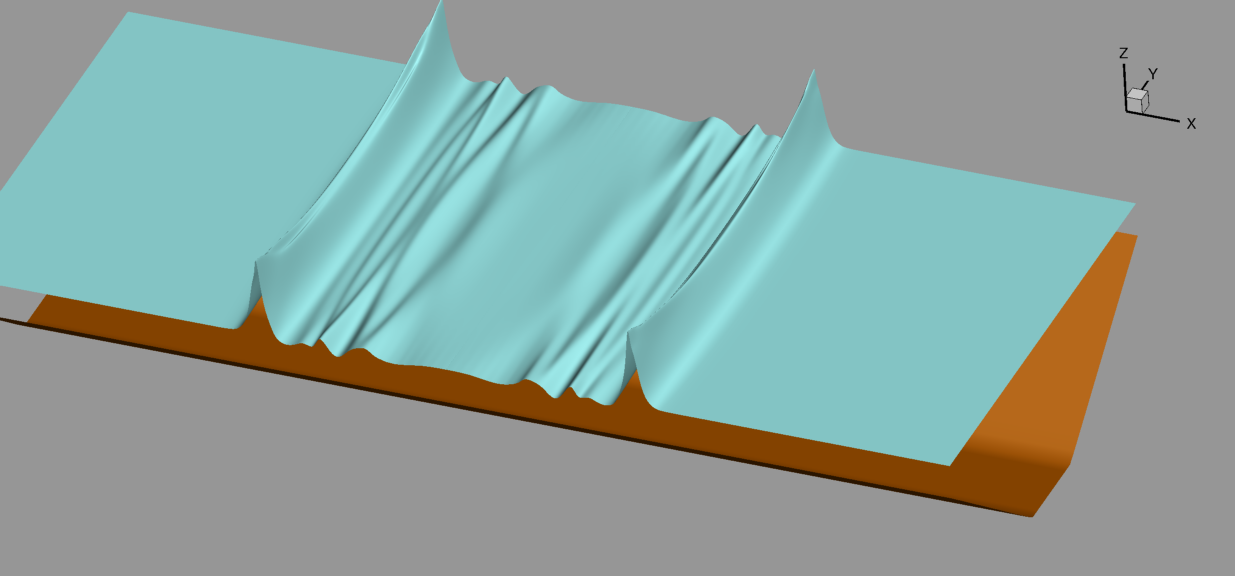}
\caption{}
\label{fig:3dbell_chi06-a}
\end{subfigure}\hfill
\begin{subfigure}{0.33\textwidth}
\includegraphics[width=\textwidth,trim={17cm 0 0 0},clip]{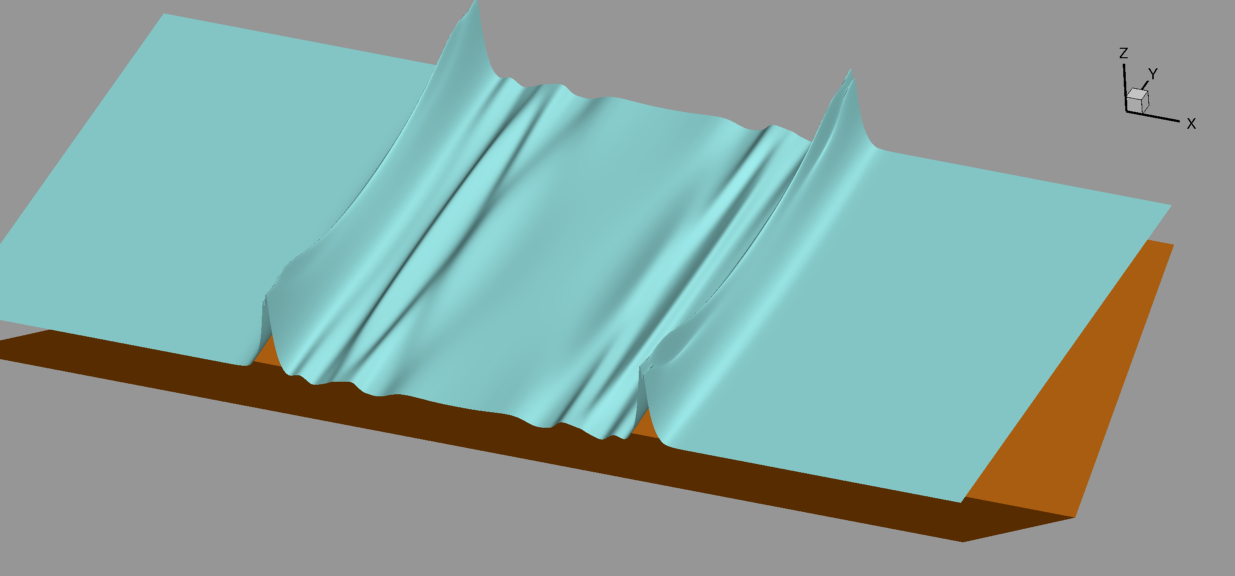}
\caption{}
\label{fig:3dbell_chi06-b}
\end{subfigure}\hfill
\begin{subfigure}{0.33\textwidth}
\includegraphics[width=\textwidth,trim={17cm 0 0 0},clip]{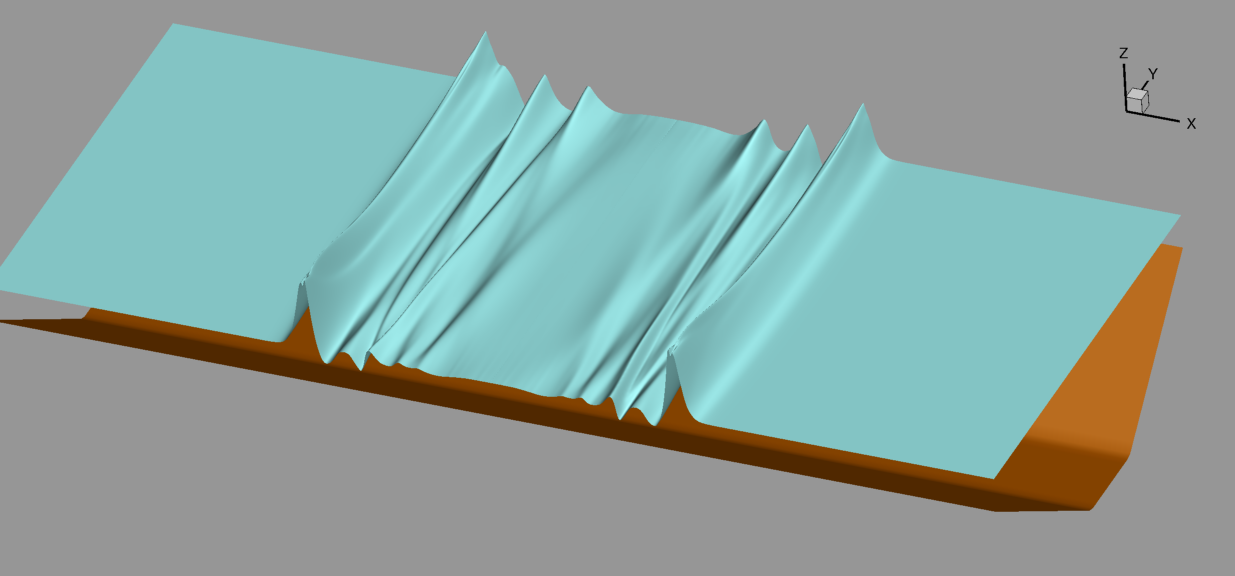}
\caption{}
\label{fig:3dbell_chi06-c}
\end{subfigure}
\caption{{\color{black}Breakdown of a Gaussian water column with $\chi=0.6${\rm m}$^4$ and $\epsilon=0.25$. Enhanced view of the right-going wave at time $t=8${\rm s}. Shallow water simulations for a
 symmetric trapezoidal section (a),  symmetric triangular section (b), and   
  asymmetric trapezoidal  section (c). \label{fig:3dbell_chi06}}}  
\end{figure}

\begin{figure}
\begin{subfigure}{0.5\textwidth}
\includegraphics[width=\textwidth]{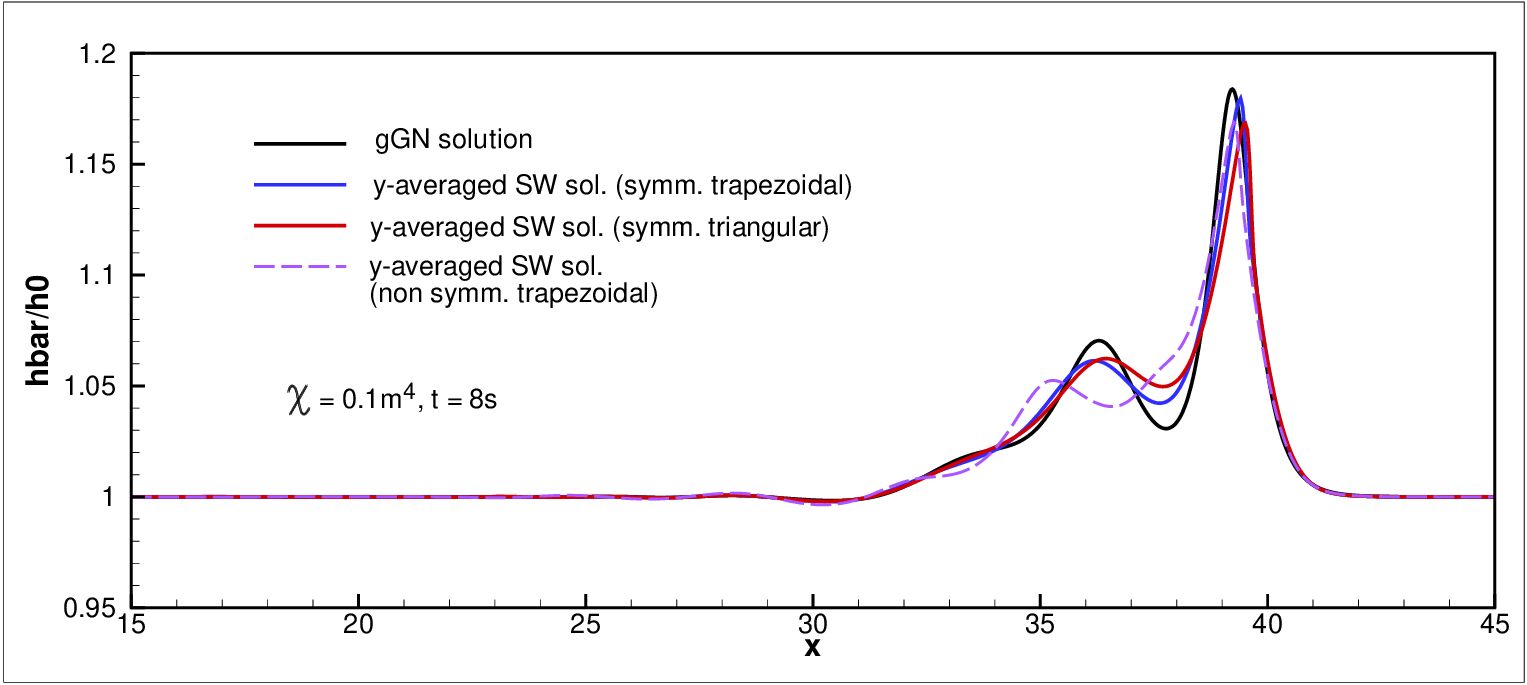}
\caption{}
 \label{fig:bell_1d_2d-a}
\end{subfigure}\hfill
\begin{subfigure}{0.5\textwidth}
\includegraphics[width=\textwidth]{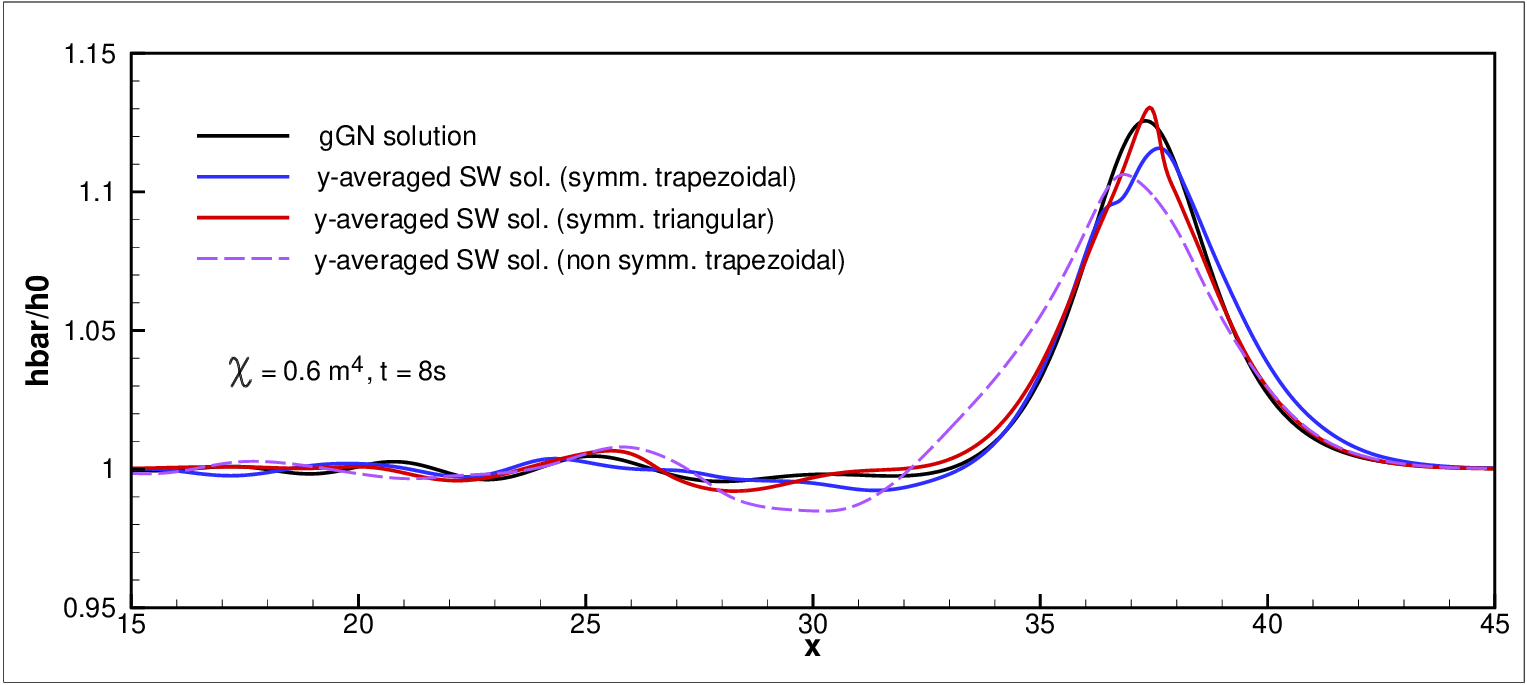}
\caption{}
 \label{fig:bell_1d_2d-b}
\end{subfigure}
\caption{{\color{black}Breakdown of a Gaussian water column, time  $t=8${\rm s}. Comparison of width averaged multi-dimensional shallow water 
results with solutions of the gGN model.  (a)  $\chi=0.1${\rm m}$^4$.  (b)  $\chi=0.6${\rm m}$^4$.  \label{fig:bell_1d_2d}}}
\end{figure}

\subsection{Riemann problem}

We consider now the computation of Riemann problems. As in \cite{Sergey} and \cite{rr24} we consider {\color{black} initially a flow at rest, with
average depth defined by a  smoothed  discontinuity defined as }: 
\begin{equation}\label{eq.rp0}
h_0^{\text{1D}}(x) := h_0 \big( 1 + \dfrac{\epsilon}{2}( 1- \text{tanh}\,x )\big)\,.
\end{equation}
As done for the column fall, we will start by comparing the two reformulations of the model for different values of the parameters.
We will consider here three values of the non-linearity $\epsilon = 0.125$, $\epsilon = 0.25$,  and $\epsilon = 0.5$, and
again the  values of the dispersion coefficient $\chi=0.1${\rm m}$^4$,   and $\chi=0.6${\rm m}$^4$.
We compute mesh converged numerical solutions for both approaches discussed in section \ref{4}, using a value $\mu=10^3$ for the hyperbolized formulation.
The results are summarized in figure \ref{fig:RP-1d}. The plots show that the two formulations give almost identical results,
except for the highest values of the non-linearity, for which the hyperbolized model provide slightly lower peaks and a  small  phase shift.
{\color{black} To better see this effect, a zoom of the leading waves in figures \ref{fig:RP-1d-c} and \ref{fig:RP-1d-f}
is reported on   figure \ref{fig:RP-1db}. This difference   is most certainly  due to the finite  value of $\mu=10^3$ used in the hyperbolized model.}

\begin{figure}
\begin{subfigure}{0.32\textwidth} 
\includegraphics[width=\textwidth]{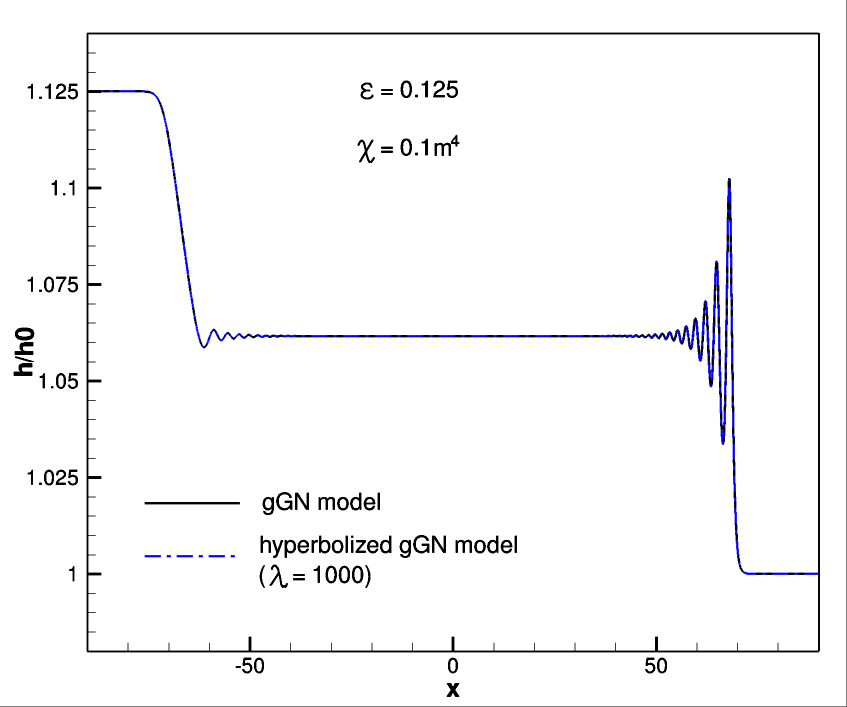}
\caption{}
 \label{fig:RP-1d-a}
\end{subfigure}\hfill
\begin{subfigure}{0.32\textwidth} 
\includegraphics[width=\textwidth]{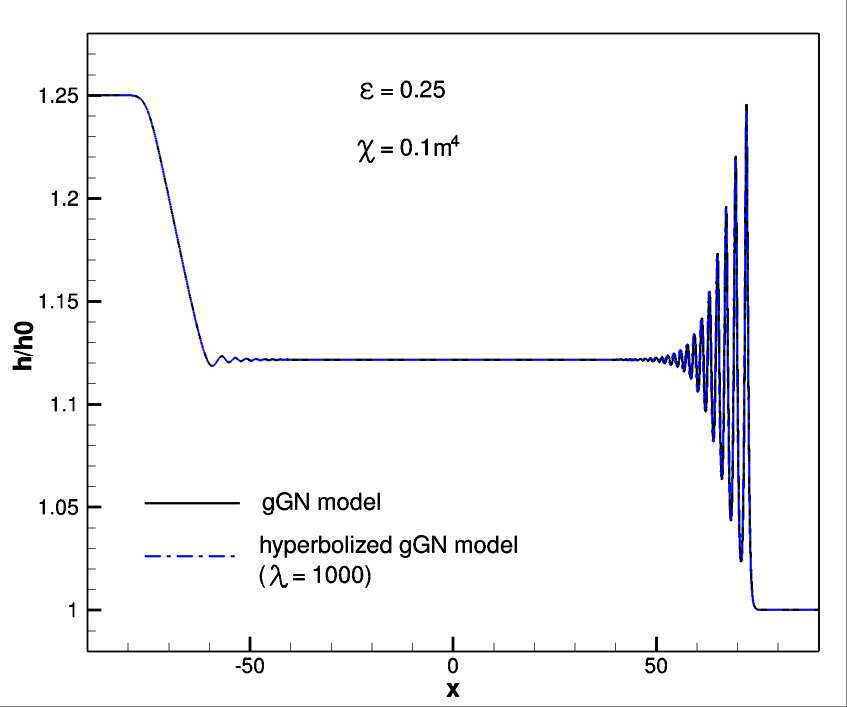}
\caption{}
 \label{fig:RP-1d-b}
\end{subfigure}\hfill
\begin{subfigure}{0.32\textwidth} 
\includegraphics[width=\textwidth]{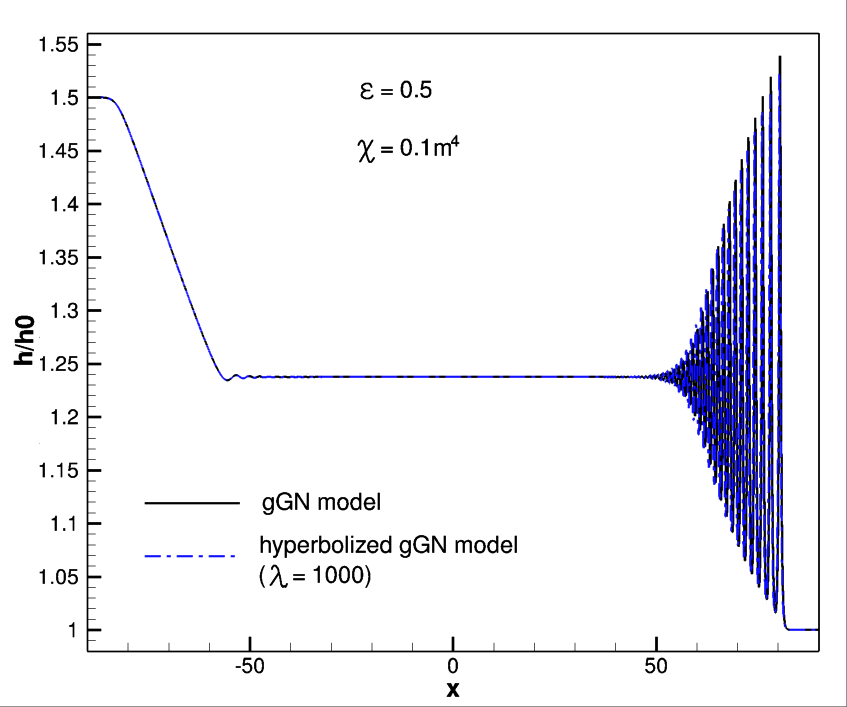}
\caption{}
 \label{fig:RP-1d-c}
\end{subfigure}\\
\begin{subfigure}{0.32\textwidth} 
\includegraphics[width=\textwidth]{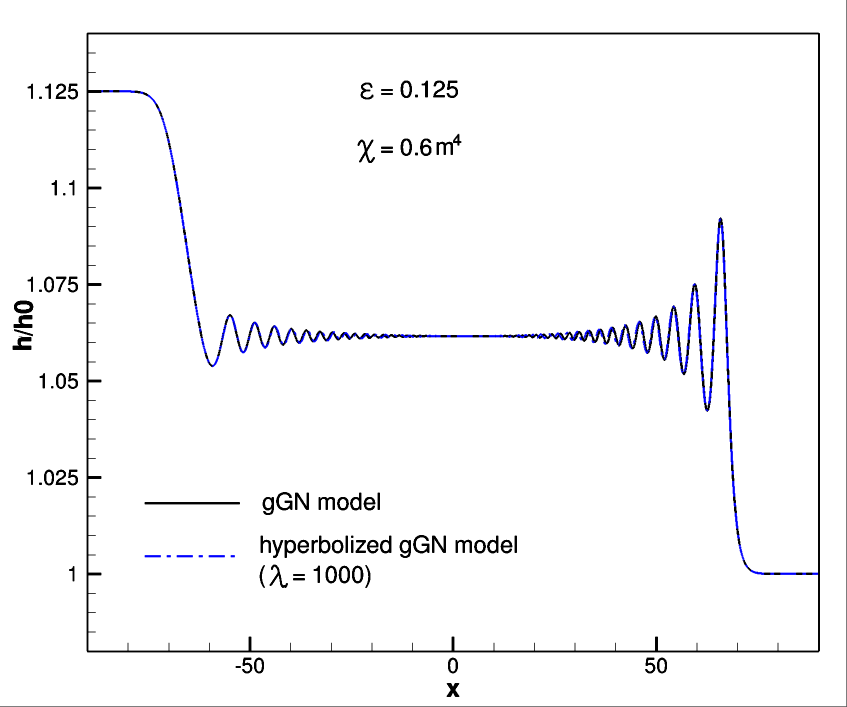}
\caption{}
 \label{fig:RP-1d-d}
\end{subfigure}\hfill
\begin{subfigure}{0.32\textwidth} 
\includegraphics[width=\textwidth]{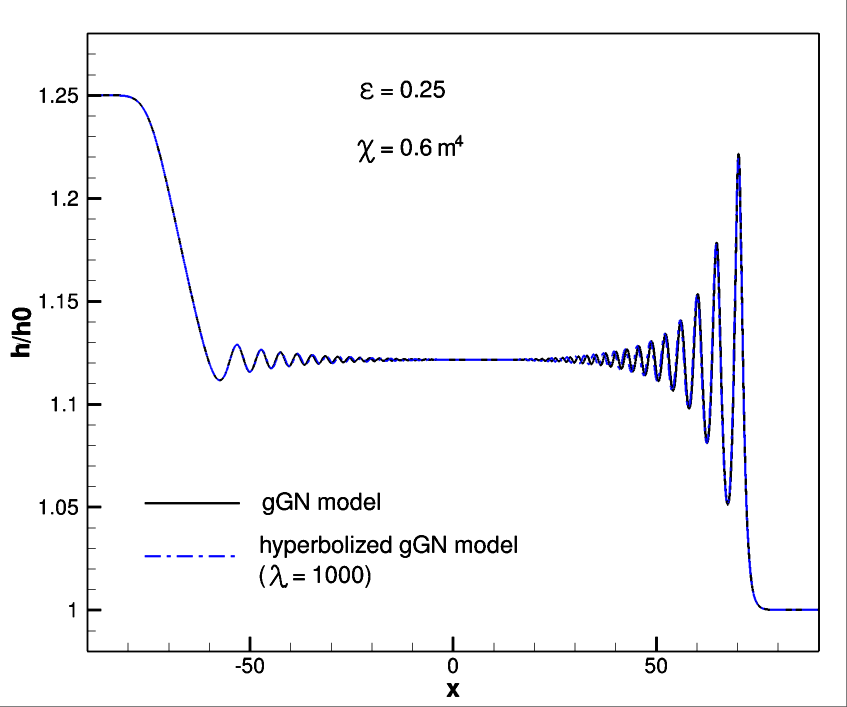}
\caption{}
 \label{fig:RP-1d-e}
\end{subfigure}\hfill
\begin{subfigure}{0.32\textwidth} 
\includegraphics[width=\textwidth]{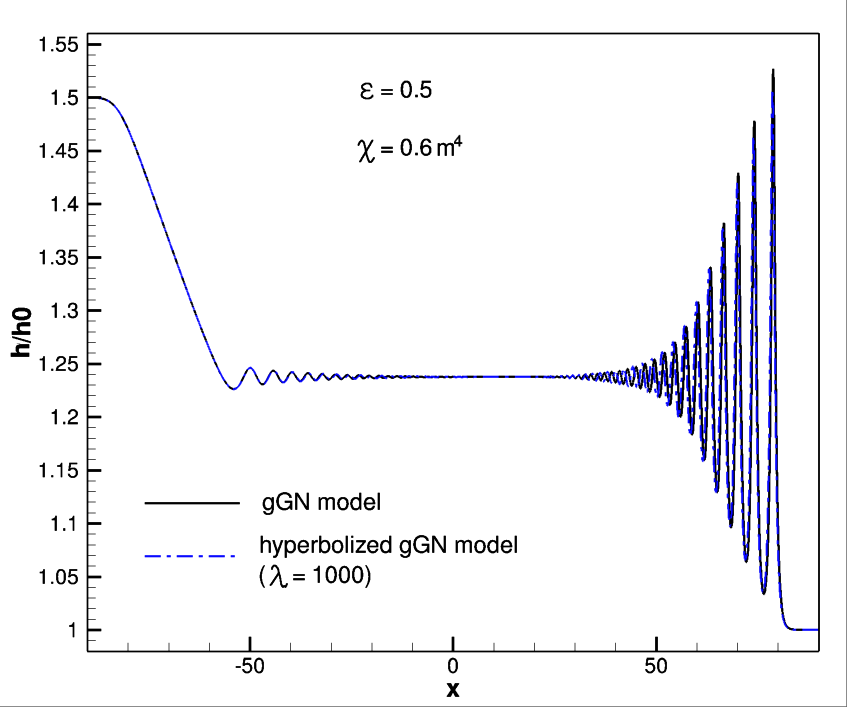}
\caption{}
 \label{fig:RP-1d-f}
\end{subfigure}
\caption{{\color{black} Riemann problem. Solutions at $t=15${\rm s} obtained with   the elliptic-hyperbolic  (black solid line) and hyperbolic relaxation  (blue dash dotted line) of the gGN model.
(a)     $\epsilon=0.125$ and $\chi=0.1${\rm m}$^4$. (b) $\epsilon=0.25$ and $\chi=0.1${\rm m}$^4$. (c) $\epsilon=0.5$ and $\chi=0.1${\rm m}$^4$.
(d)     $\epsilon=0.125$ and $\chi=0.6${\rm m}$^4$. (e) $\epsilon=0.25$ and $\chi=0.6${\rm m}$^4$. (f) $\epsilon=0.5$ and $\chi=0.6${\rm m}$^4$.   \label{fig:RP-1d}}}
\end{figure}

\begin{figure}
\begin{center}
\begin{subfigure}{0.33\textwidth} 
\includegraphics[width=\textwidth]{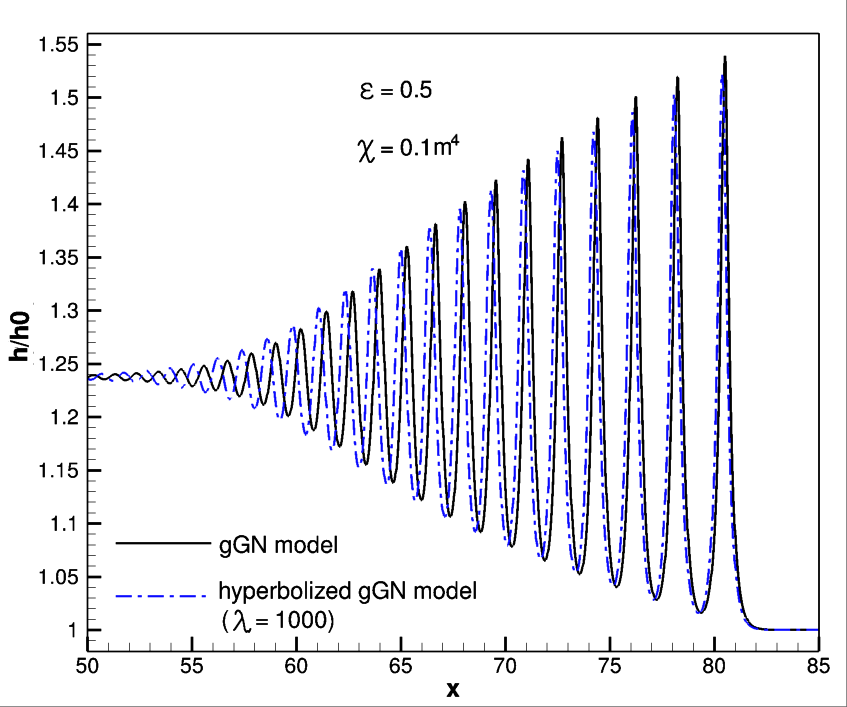}
\caption{}
\label{fig:RP-1db-a}
\end{subfigure}
\begin{subfigure}{0.33\textwidth} 
\includegraphics[width=\textwidth]{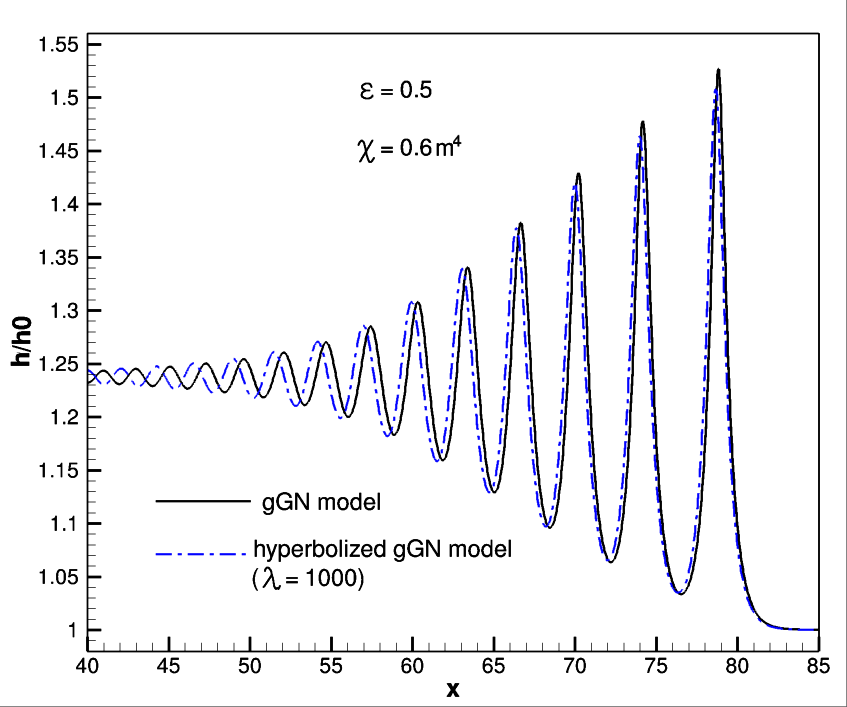}
\caption{}
\label{fig:RP-1db-b}
\end{subfigure}\hfill
\caption{{\color{black}Riemann problem: zoom of the leading waves  for  $\epsilon=0.5$. (a) zoom of  figure \ref{fig:RP-1d-c}. (b) zoom of figure \ref{fig:RP-1d-f}.    \label{fig:RP-1db}}}
\end{center}
\end{figure}

Concerning the qualitative behaviour of the model, as expected, smaller values of $\chi$ 
provide higher  frequency dispersive shocks. {\color{black} Conversely}, higher non-linearity lead to stronger secondary waves, all travelling faster,
and ultimately also dispersive shocks with shorter wavelengths at this final time.

\subsubsection{Validation against section averaged 2D shallow water simulations}

As done for the previous case, as a means of validation we have run full 2D shallow water computations. 
We proceed as before. We consider equivalent sections providing values of the geometrical dispersion coefficient      $\chi=0.1${\rm m}$^4$,  and $\chi=0.6${\rm m}$^4$.
The geometrical  definitions are the same {\color{black} used for the Gaussian bell breakdown of  section 5.2.1.} 
Then we initialize the solution based on relation \eqref{h_expression}:
$$
h(x,y,t=0) =   h_0^{\text{1D}}(x)  - b(y) + \overline{b},
$$
with $h_0^{\text{1D}}(x)$ defined {\color{black} as in \eqref{eq.rp0}}. As before, with these values we can initialize the flow on a constant upstream average depth of $2${\rm m},
without incurring in dry areas. We will consider here   the case with lowest non-linearity $\epsilon=0.125$.

To begin with, we look at the free surface obtained at $t=15${\rm s}, which are reported in figure \ref{fig:RP-3d}. 
The first thing we can see  is that none of the computations with bathymetry exhibits the formation of classical shock waves,
{\color{black} as in the flat bathymetry results, reported in  figure \ref{fig:RP-3d-g}  for completeness}.
All the results show a free surface level which has an almost
one dimensional phase,  with finite wavelength oscillations. These wavelengths are several orders of magnitude larger than the mesh size, and have nothing to do
with numerical dispersion, or lack of monotonicity, but are related to dispersion induced by physical refraction in the $y$ direction.
We {\color{black} also note  that  for a given $\chi$   different sections} provide very close results to one another. {\color{black} Perhaps the main
difference  is the one between the symmetric trapezoidal shapes of figures \ref{fig:RP-3d-b} to  \ref{fig:RP-3d-d}, and the non-symmetric ones of figures \ref{fig:RP-3d-e} and \ref{fig:RP-3d-f}}. 
The latter ones give a more three dimensional field, richer in secondary waves.
As for the water column case, {\color{black} the wave heights obtained are more important in shallower regions}.
{\color{black} As in the one dimensional results,  longer waves are observed  for larger $\chi$.} 
 
To validate  the gGN model, we  compare its results with  transverse averages of the 2D  solutions.  The resulting comparison is reported in figure \ref{fig:RP-1d-2d}. 
{\color{black} The match between  one dimensional and averaged multi-dimensional results is excellent for small values of $\chi$ and symmetric sections, as visible in figures  
 \ref{fig:RP-1d-2d-a} and  \ref{fig:RP-1d-2d-b}. These values are relevant for shallower channels for example. 
The comparison is still satisfactory for  the higher value of $\chi$, especially for symmetric sections, as visible in figures  \ref{fig:RP-1d-2d-c} and  \ref{fig:RP-1d-2d-d}.}
 For both values of $\chi$ the symmetric 2D results provide a solution with 4 or 5 peaks, while the 1D model predicts
in general more  secondary oscillations.  The amplitude and phases of  the leading front are excellent, and those of the first   two secondary waves
are also captured nicely by the gGN model.
{\color{black} As before the non-symmetric case provides a worse match. This is expected, since symmetry plays a key role
in the derivation of our model.} \\

Overall these results, as those of the previous section, validate our model, showing a very satisfactory  agreement.
We also confirm the findings of    \cite{Chassagne} on more simulations, and highlight  the existence
of these  little known  dispersive-like waves due to transverse refraction, which
can be modelled with our geometrical Green-Naghdi equations.

\begin{figure}
\begin{subfigure}{0.5\textwidth}
\includegraphics[width=6.5cm, height = 2.8cm,trim={3.5cm 1.cm 3cm 0},clip]{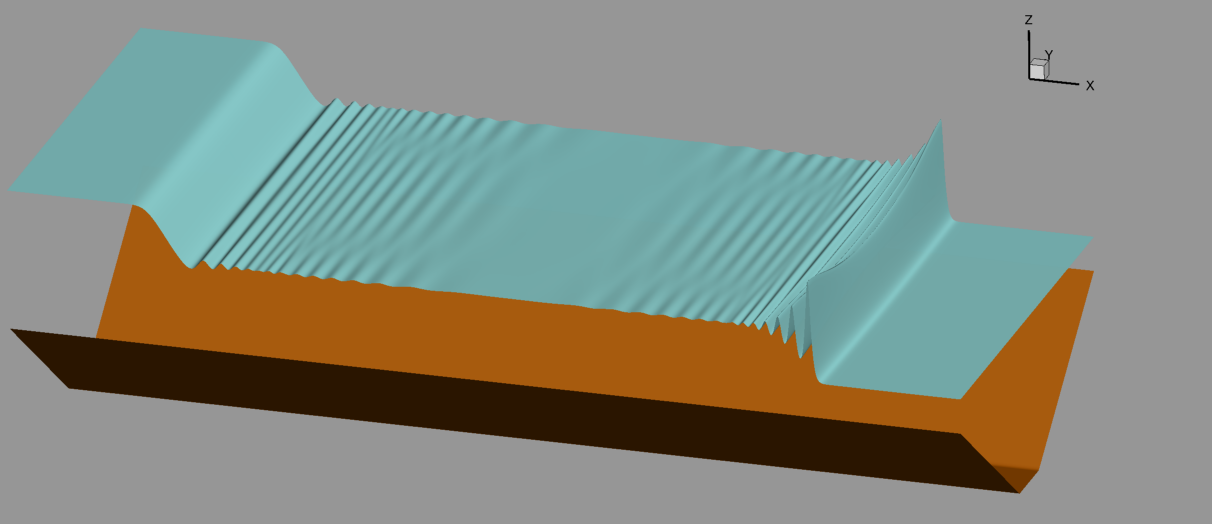}
\caption{}
 \label{fig:RP-3d-a}
 \end{subfigure}\hfill
\begin{subfigure}{0.5\textwidth}
\includegraphics[width=6.5cm, height = 2.8cm,trim={3.5cm 1.cm 3cm 0},clip]{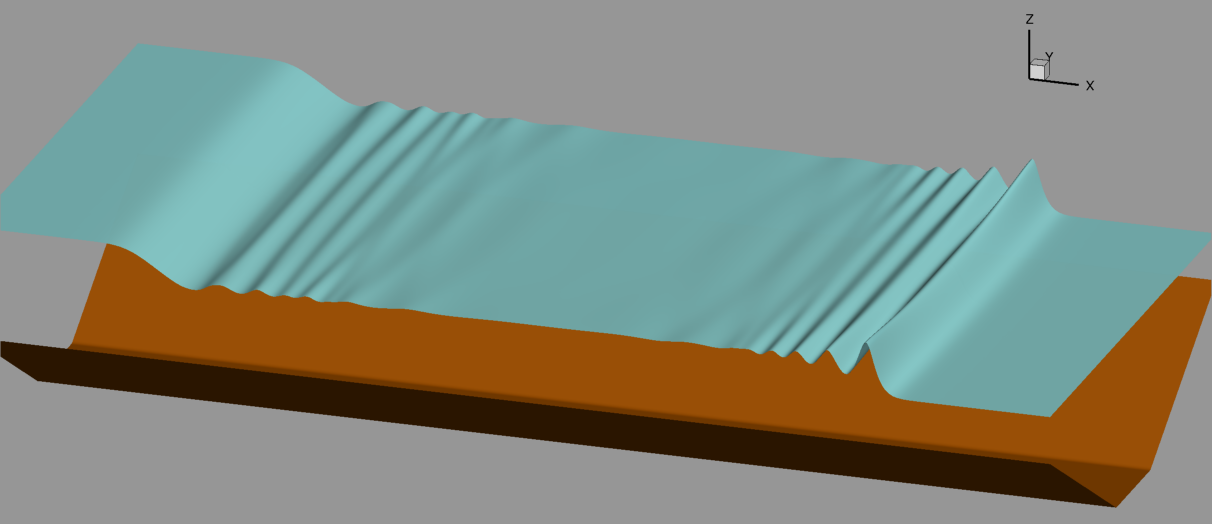}
\caption{}
 \label{fig:RP-3d-b}
 \end{subfigure}\\
 \begin{subfigure}{0.5\textwidth}
 \includegraphics[width=6.5cm, height = 2.8cm,trim={3.5cm 1.cm 3cm 0},clip]{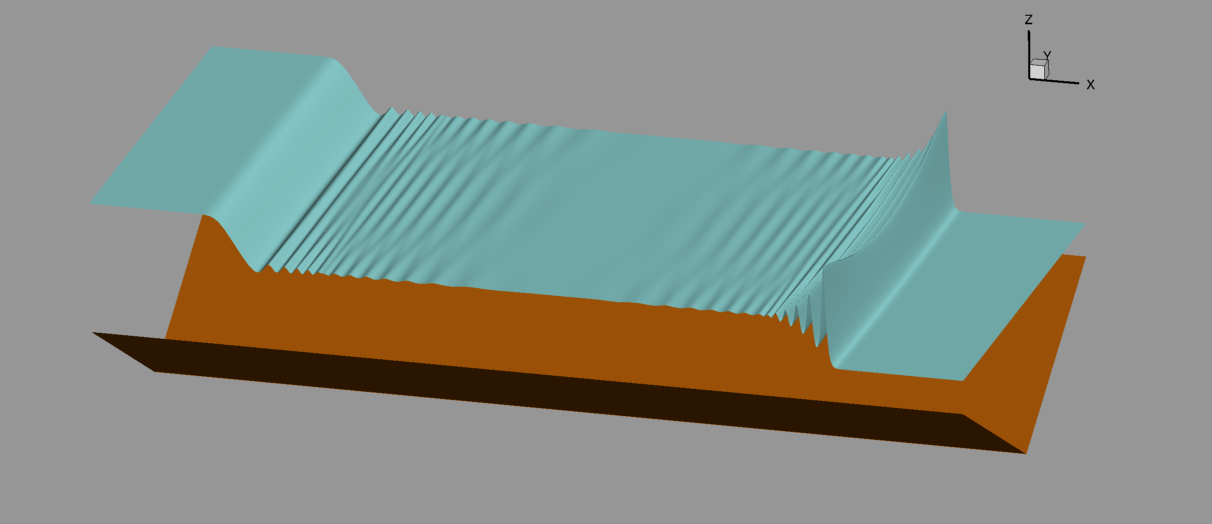}
\caption{}
 \label{fig:RP-3d-c}
 \end{subfigure}\hfill
\begin{subfigure}{0.5\textwidth}
\includegraphics[width=6.5cm, height = 2.8cm,trim={3.5cm 1.cm 3cm 0},clip]{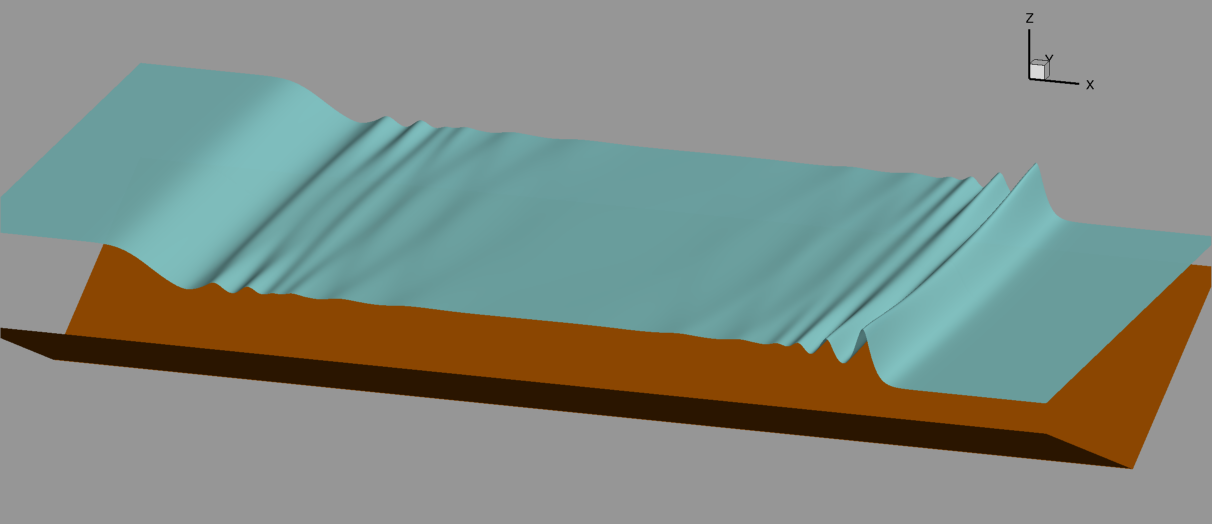}
\caption{}
 \label{fig:RP-3d-d}
 \end{subfigure}\\
 \begin{subfigure}{0.5\textwidth}
 \includegraphics[width=6.5cm, height = 2.8cm,trim={3.5cm 1.cm 3cm 0},clip]{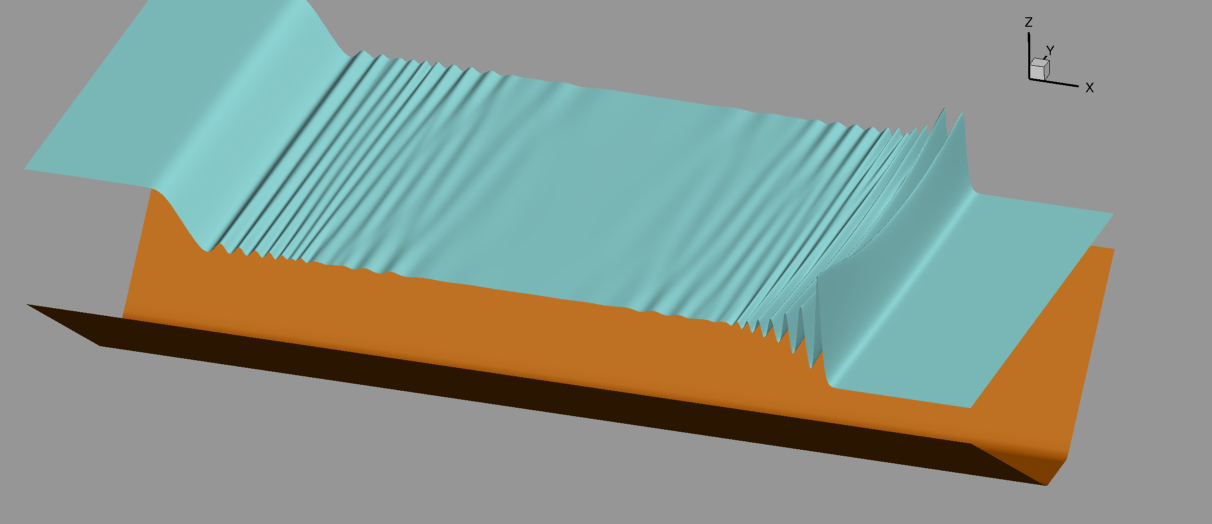}
\caption{}
 \label{fig:RP-3d-e}
 \end{subfigure}\hfill
\begin{subfigure}{0.5\textwidth}
\includegraphics[width=6.5cm, height = 2.8cm,trim={3.5cm 1.cm 3cm 0},clip]{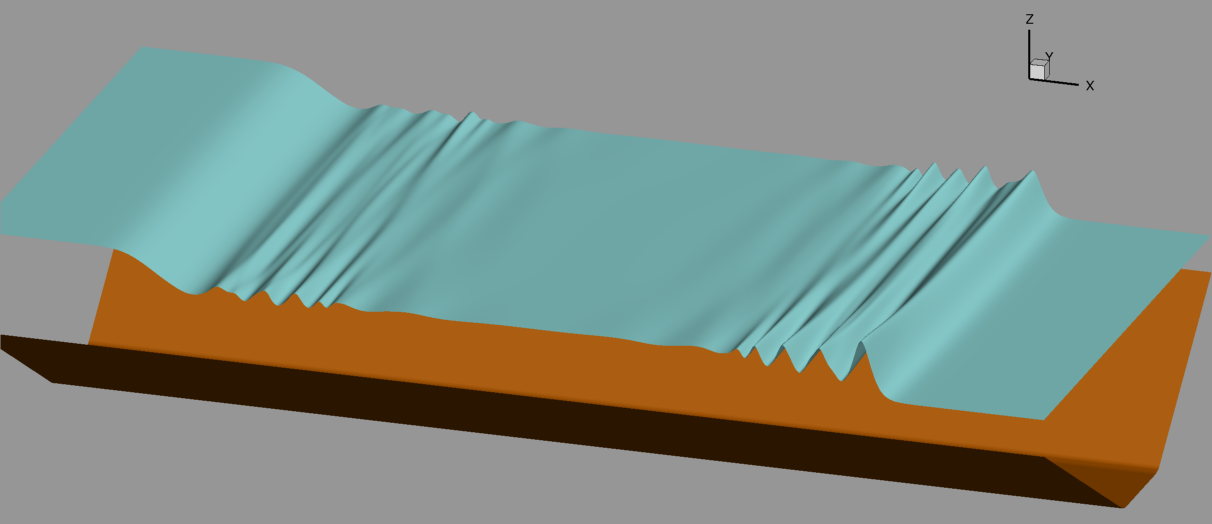}
\caption{}
 \label{fig:RP-3d-f}
 \end{subfigure}\\
 \begin{subfigure}{\textwidth}
\centering\includegraphics[width=6.5cm, height = 2.8cm]{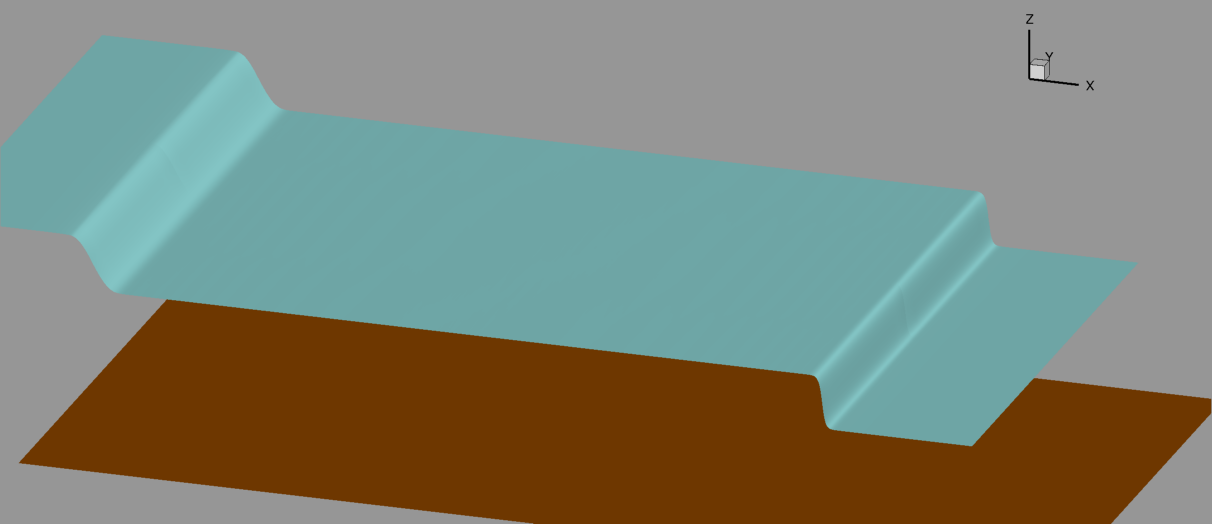}
\caption{}
 \label{fig:RP-3d-g}
 \end{subfigure}
\caption{{\color{black}Riemann problem with $\epsilon=0.125$ . Free surface  at $t=15${\rm s}  obtained from   shallow water solutions.
(a)    symmetric   trapezoidal   section with  $\chi=0.1${\rm m}$^4$. (b)     symmetric  trapezoidal   section with  $\chi=0.6${\rm m}$^4$.
(c)  symmetric     triangular   section with  $\chi=0.1${\rm m}$^4$. (d)   symmetric    triangular   section with  $\chi=0.6${\rm m}$^4$.
(e)   non-symmetric   trapezoidal   section with  $\chi=0.1${\rm m}$^4$. (f)     non-symmetric  trapezoidal   section with  $\chi=0.6${\rm m}$^4$.
(g) shallow water simulation with  flat bathymetry.  \label{fig:RP-3d}}}
\end{figure}

\begin{figure}
\begin{subfigure}{0.65\textwidth}
\centering\includegraphics[width=\textwidth]{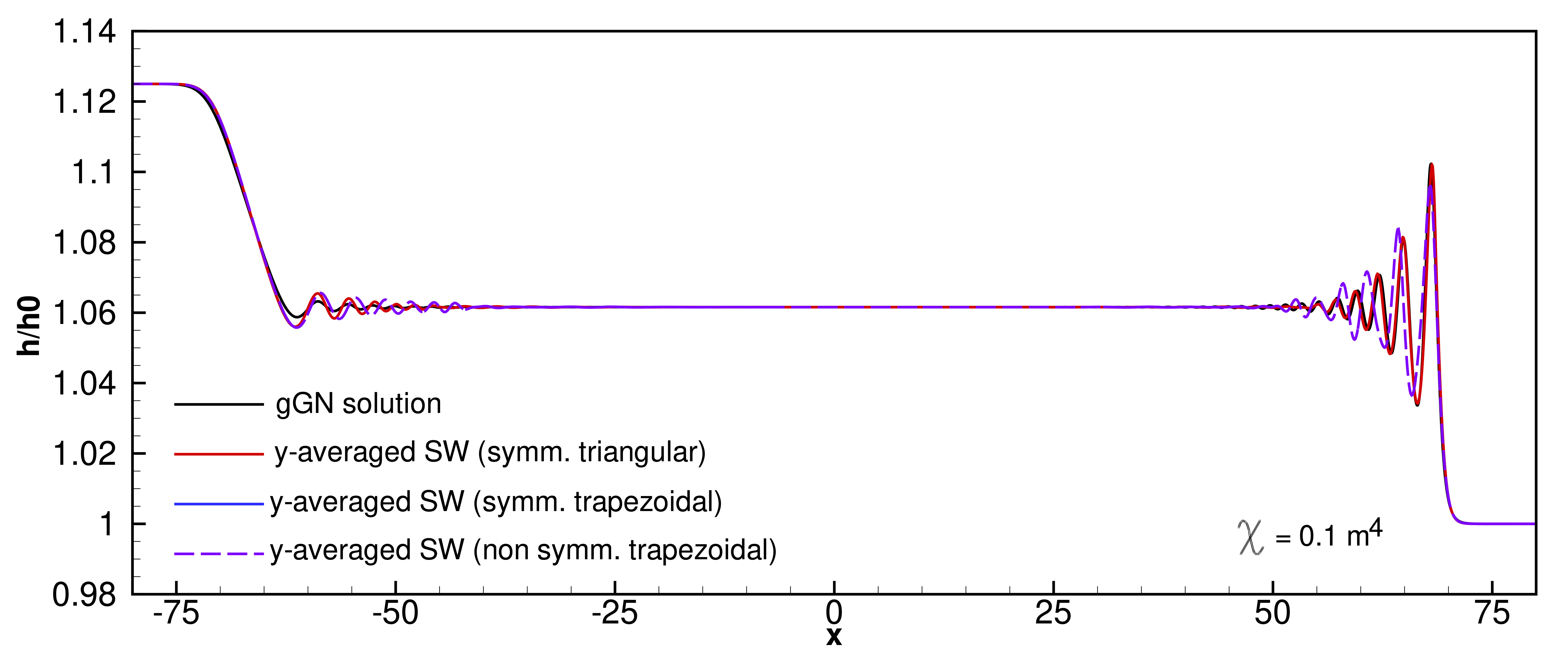}
\caption{}
 \label{fig:RP-1d-2d-a}
 \end{subfigure}\hfill
\begin{subfigure}{0.35\textwidth}
\centering\includegraphics[width=0.85\textwidth]{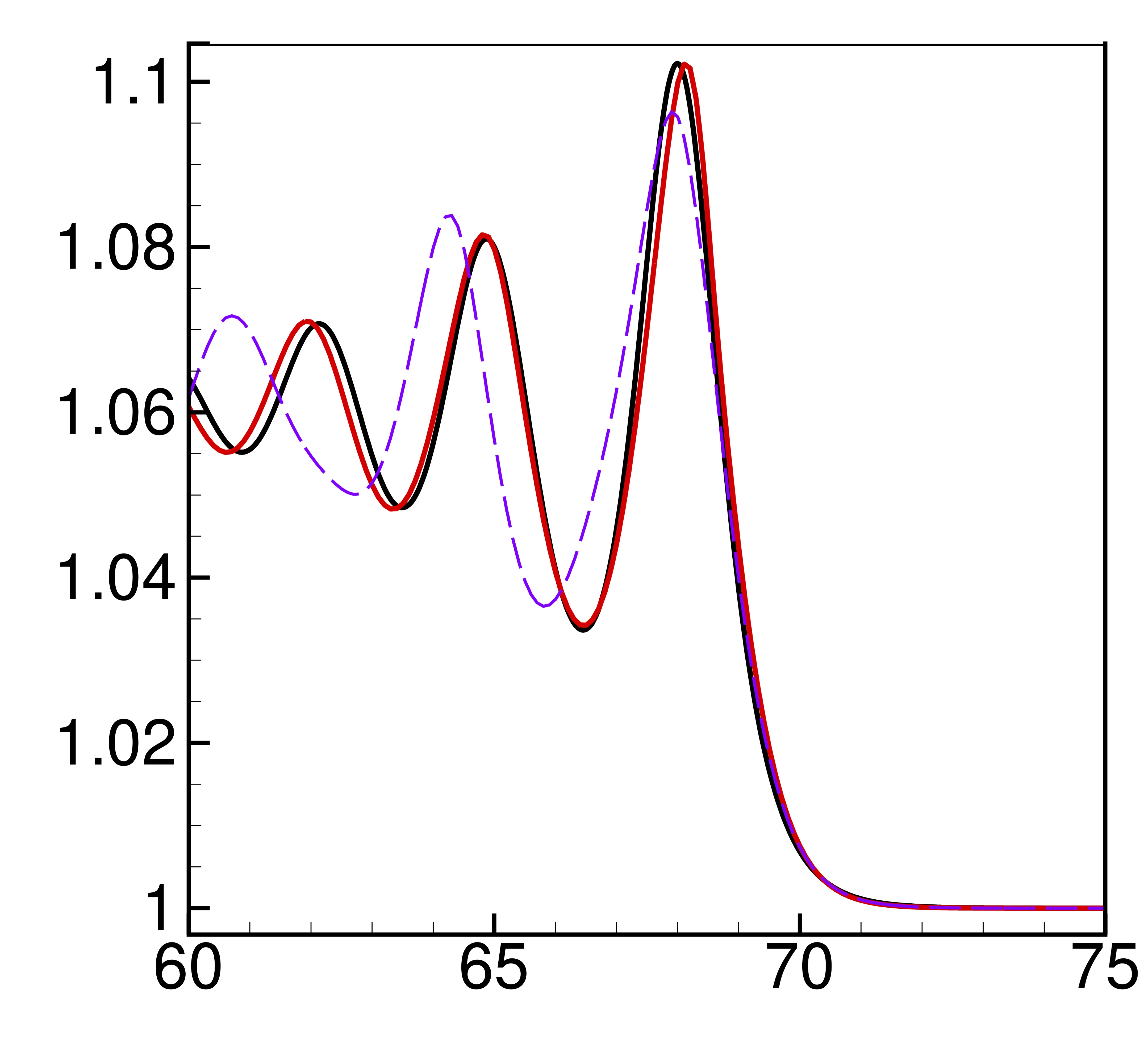} 
\caption{}
 \label{fig:RP-1d-2d-b}
 \end{subfigure}\\
\begin{subfigure}{0.65\textwidth}
\centering\includegraphics[width=\textwidth]{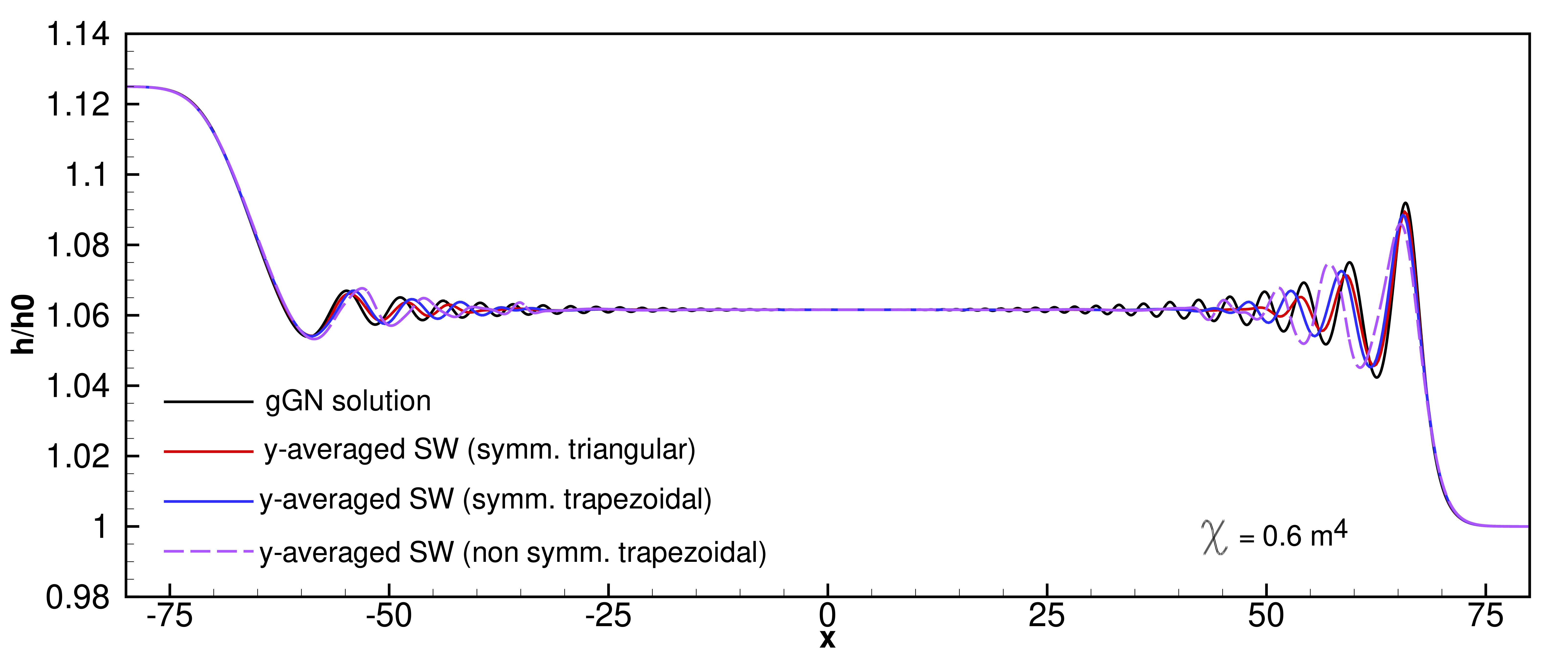}
\caption{}
 \label{fig:RP-1d-2d-c}
 \end{subfigure}\hfill
\begin{subfigure}{0.35\textwidth}
\centering\includegraphics[width=0.85\textwidth]{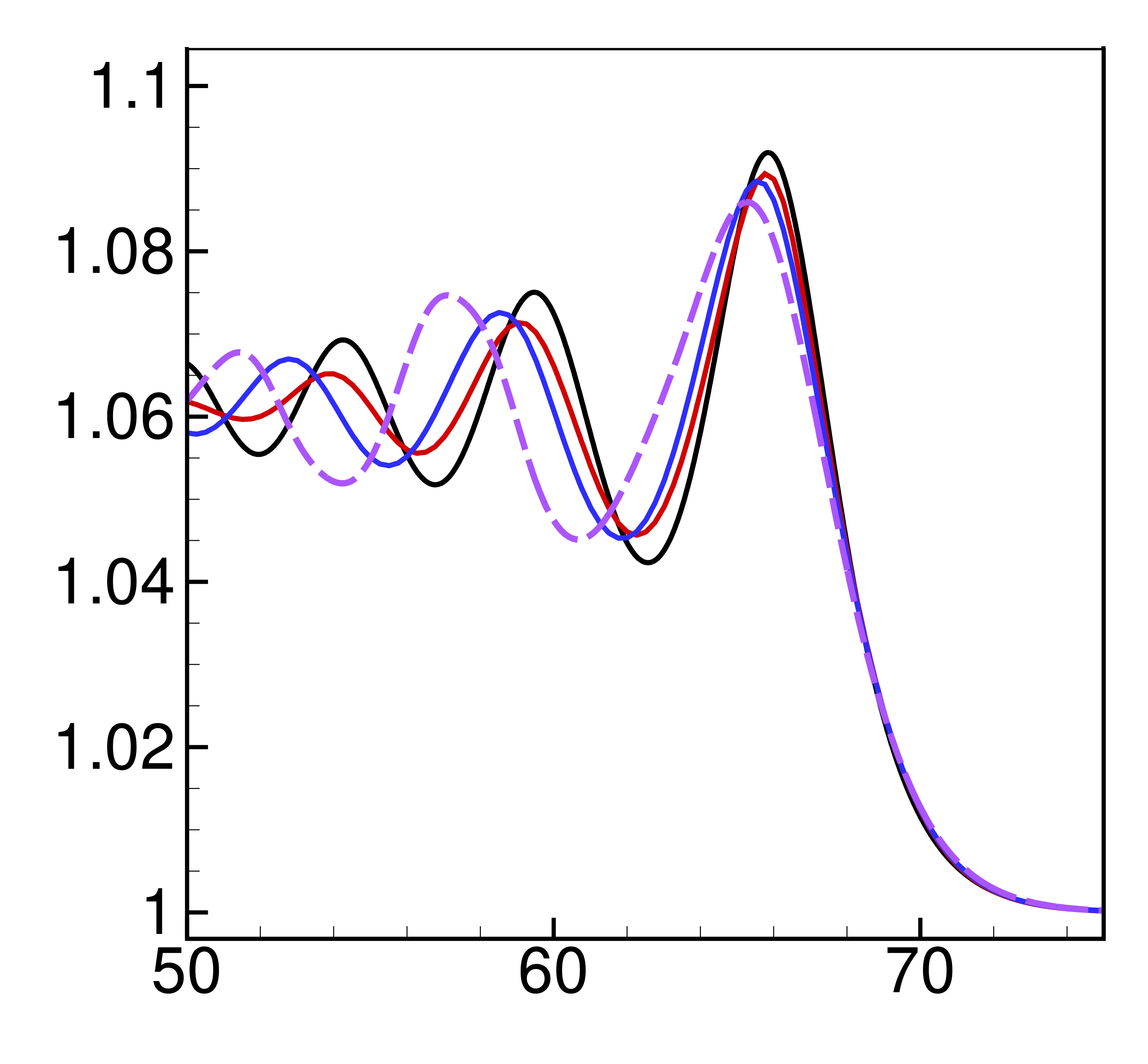}
\caption{}
 \label{fig:RP-1d-2d-d}
 \end{subfigure}
\caption{{\color{black}Riemann problem   with $\epsilon=0.125$, time  $t=15${\rm s}. Comparison 
of width averaged free surface profiles from multi-dimensional shallow water simulations with one dimensional profiles obtained with the gGN model.
(a) $\chi=0.1${\rm m}$^4$, full domain. (b) $\chi=0.1${\rm m}$^4$, zoom of the leading waves.
(c) $\chi=0.6${\rm m}$^4$, full domain. (d) $\chi=0.6${\rm m}$^4$, zoom of the leading waves. \label{fig:RP-1d-2d}}}
\end{figure}

\subsubsection{Solitary wave  fission}

To conclude the study of the Riemann problem, we have  run   {\color{black} long time 1D simulations for  the
largest values   $\chi=0.6${\rm m}$^4$ and $\epsilon=0.5$. The final time is now set to $600${\rm s},}
in order to verify the occurrence of solitary wave  fission, and compare the resulting solitary wave  with the ones studied in   sections~\S3.3 and \S5.1.
We report in figure \ref{fig:soliton-fission} a visualization of the front of the wave superposing several instances of the leading waves. 
In the figure  we {\color{black} align the}  second peak of the dispersive  front, so that we may visualize the separation of the first solitary wave.
For the last instance, we  have measured an amplitude $a= 0.56 h_0$ which
we used  to generate a corresponding exact gGN solitary wave solving
\eqref{eq:sol_ODE2} with  matching amplitude.  
The exact solitary wave corresponding to the measured amplitude is reported in green in the picture, such that it maximum is superposed it to the first solitary wave resulting 
from the Riemann problem. The agreement is excellent.

\begin{figure}
\centering\includegraphics[width=0.7\textwidth]{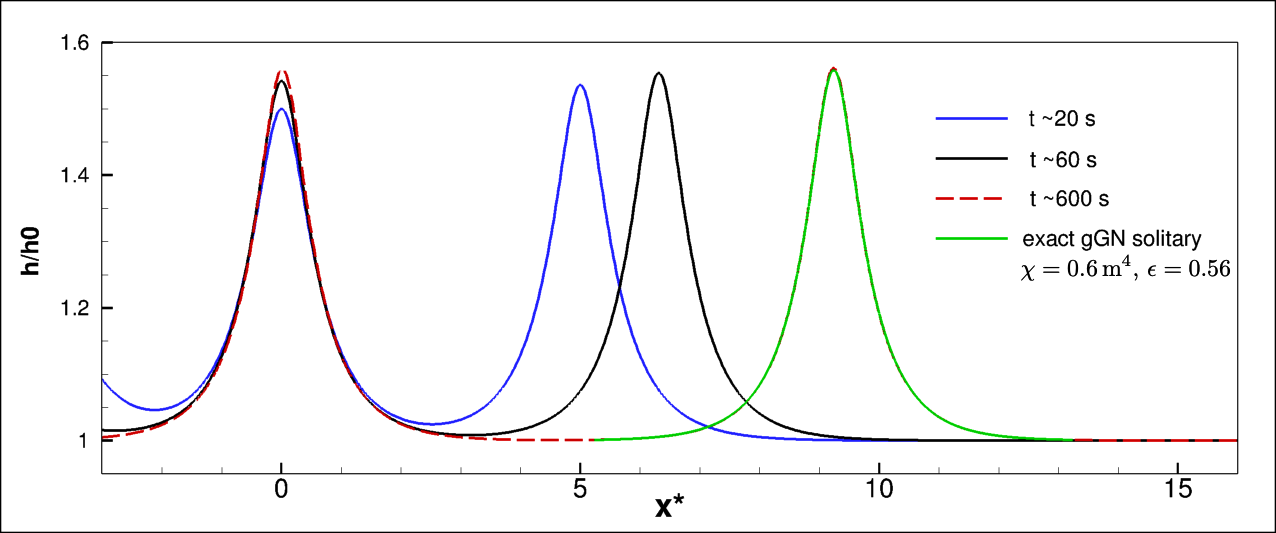} 
\caption{Riemann problem: solitary wave  fission for the case    $\epsilon=0.5$, and   $\chi=0.6${\rm m}$^4$.
Solutions at times $t  =20${\rm s} (blue), $t = 60${\rm s} (black), and  $t = 600${\rm s} (red).
All secondary peaks are  aligned at the origin. The last solution is used to generate a corresponding exact gGN solitary wave (in green) obtained solving
\eqref{eq:sol_ODE2} with  matching amplitude.   \label{fig:soliton-fission}}
\end{figure}

\subsection{\color{black}Treske's experiments in a trapezoidal channel\label{sec:treske}} 

{\color{black}As a final test, we consider the simulation of the experiments in channels with trapezoidal sections by \cite{Treske}.
We are interested in the issue of generation of long waves associated with
geometrical dispersion. 
%
The setup of the problem is very similar to the one described in \cite{Chassagne,Jouy}. 
The channel considered has a symmetric trapezoidal shape. Using the notation of figure \ref{fig:dispersion-a} in section \ref{sec:dispersion},  
we consider as in the experiments a width $w=1.24$m, and a slope with $\beta=1/3$.  
 The  initial state consists of a smoothed depth  and speed discontinuity. {\color{black} For 
 clarity, we introduce here again the notation $\overline h$ for the transverse average depth. The initial condition is  defined as }  
$$
\begin{aligned}
\bar h_{\text{ini}}(x) = &  \bar h_1 + \dfrac{\bar h_2 - \bar h_1}{2}( 1 -\text{tanh}(x/\alpha))\\ 
u_{\text{ini}}(x) = &   \dfrac{u_2}{2}( 1 -\text{tanh}(x/\alpha))\\ 
\end{aligned}
$$
{\color{black} The values $(\bar h_2, u_2)$ are  obtained from  the section averaged Rankine-Hugoniot conditions 
of the shallow water model obtained for $\chi=0$.}
 The value $\bar h_1$ is computed using the experimental height in the channel center $h_1=0.16$m, 
using the formula valid for a trapezium
$$
\bar h = \dfrac{\beta w + h}{\beta w + 2 h}h.
$$ 
To evaluate the dispersion coefficient, we use \eqref{eq:chi_trap} with $b_0= h_2$, {\color{black} obtained
by inverting the above formula for a given $\bar h_2$. The remaining geometrical quantities are
defined as   ${\it l}_2 = w$ and ${\it l}_1={\it l}_3=b_0/\beta$.  }

The simulations are run on a domain of length $L=100$ m using  the elliptic-hyperbolic reformation. The results of the figures are obtained on a 
largely converged resolution of 20.000 mesh points.  As in previous works,  the simulations are stopped after  
 the first peak  has travelled a distance of  72m.   Finally, to choose the parameter $\alpha$ in the initialization, we use the steepness of the waves.
In particular, we have  set  $\alpha = L (h_2-h_1)/\lambda$ with the wavelength $\lambda$ obtained from the Lemoine theory (cf. section \ref{sec:dispersion}). 
We simulate the waves for values of Froude  {\color{black} $Fr\in[1.0125, 1.20]$  with
$$
Fr:=\dfrac{|u_2-c_b|}{\sqrt{g\overline h_1}}
$$
where as usual $c_b$ denotes the speed of the bore.}

\begin{figure}
\begin{subfigure}{0.6\textwidth}
\centering\includegraphics[align=c,width=0.8\textwidth]{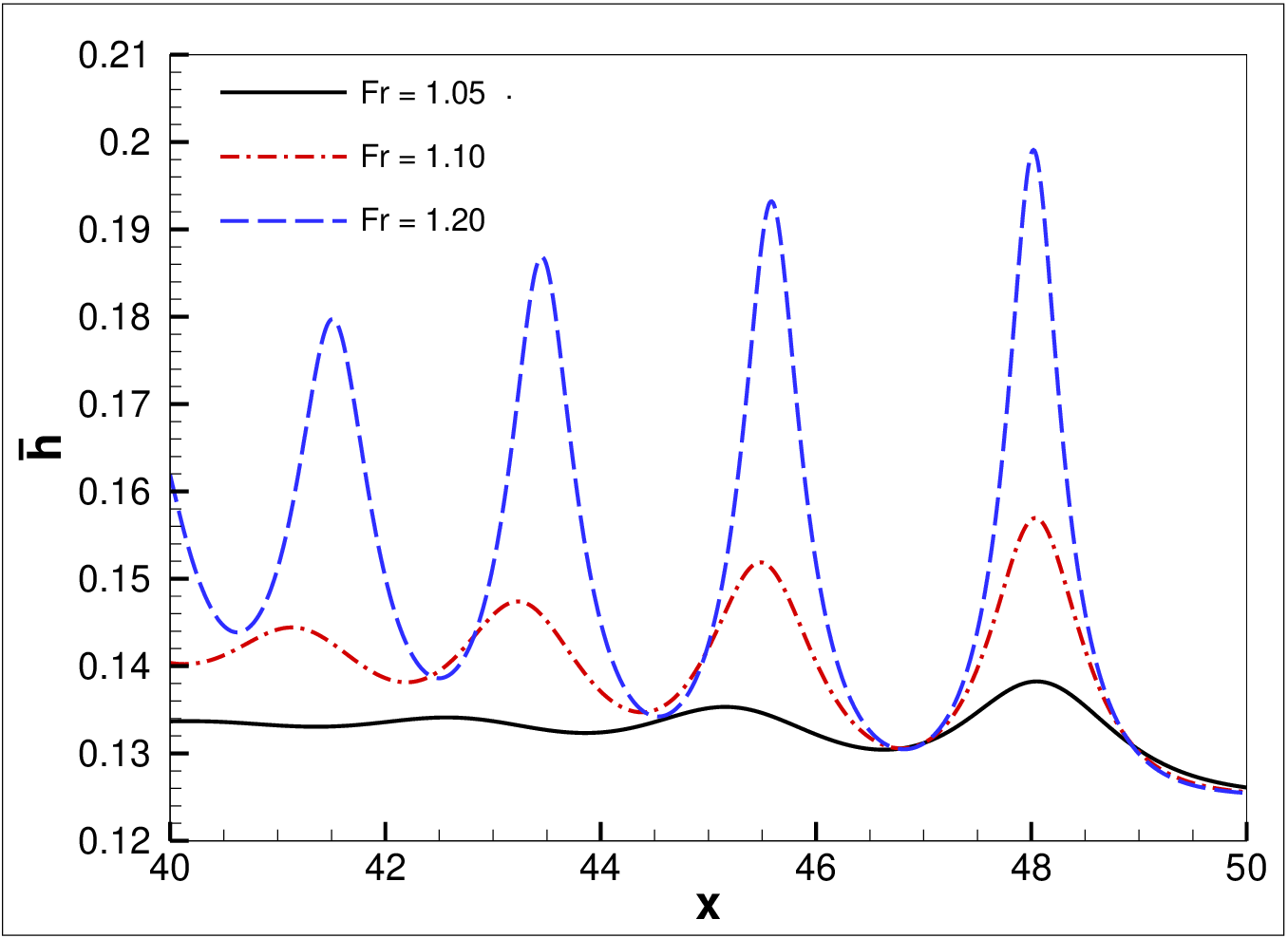}
\caption{}
\label{fig:treske0-a}
\end{subfigure}\hfill
\begin{subfigure}{0.4\textwidth}
\centering\includegraphics[align=c,width=0.8\textwidth]{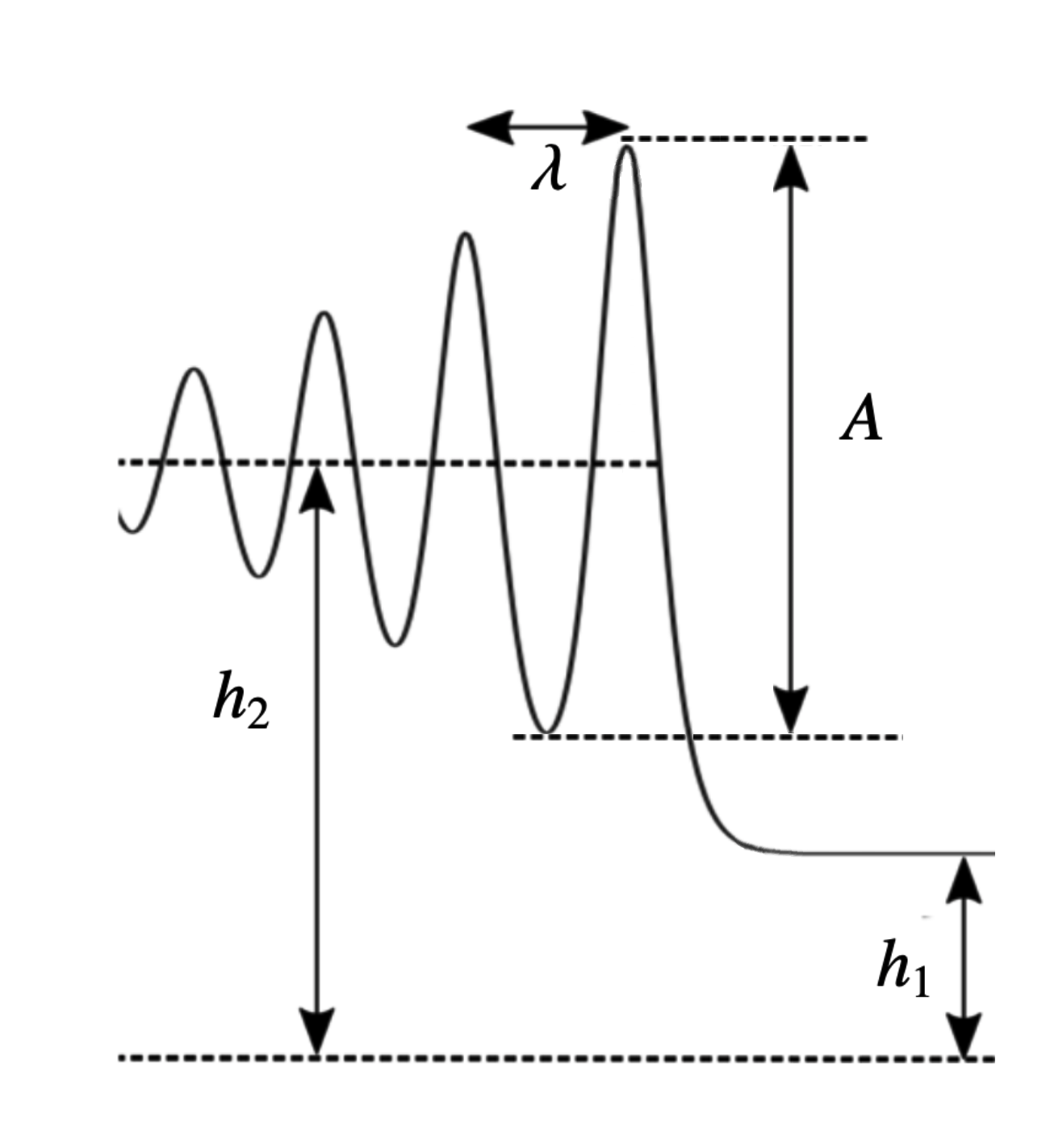}
\caption{}
\label{fig:treske0-b}
\end{subfigure}
\caption{{\color{black}Treske experiments in a trapezoidal channel. (a)  section averaged heights obtained with  the simplified gGN model for Froude numbers 1.05 (solid black), 1.1 (blue dash-dot), and 1.2 (red dashed).
(b) definition of measured wavelength and amplitude. \label{fig:treske0}}}
\end{figure}

{\color{black} Figure \ref{fig:treske0-a} shows } the typical wave profiles obtained for a low, intermediate and high value of the Froude number.
We can clearly see the waves becoming shorter and steeper. As it is customary for this experiment we measure
the wavelengths and amplitudes defined respectively as   the distance between the first two peaks, and
the height difference between the first crest and the first trough {\color{black}(cf.  figure \ref{fig:treske0-b})}. These measures are made non-dimensional by dividing with $\bar h_1$,
and compared to the data by \cite{Treske} in figure \ref{fig:treske}. 
The results show a very reasonable agreement with the data, and fully validate our derivation. 
{\color{black} In figure   \ref{fig:treske-a} we can observe} an over-prediction of the  wavelength is observed   for higher values of Froude. This  
may be related to the lack of higher order non-linear effects (and possibly dispersion) in the model. 
 Concerning the amplitudes, as proposed in \cite{Jouy} we have averaged the experimental data on the channel  banks  and axis to obtain a fair comparison.
 The average is obtained as $A_{\text{banks}}/3+2A_{\text{axis}}/3$ corresponding to integrating a   symmetric parabola.
{\color{black} Figure   \ref{fig:treske-b}  shows that our model provides a correct prediction of the amplitude growth with the Froude number.}

\begin{figure}
\begin{subfigure}{0.5\textwidth}
\centering\includegraphics[align=c,width=0.9\textwidth]{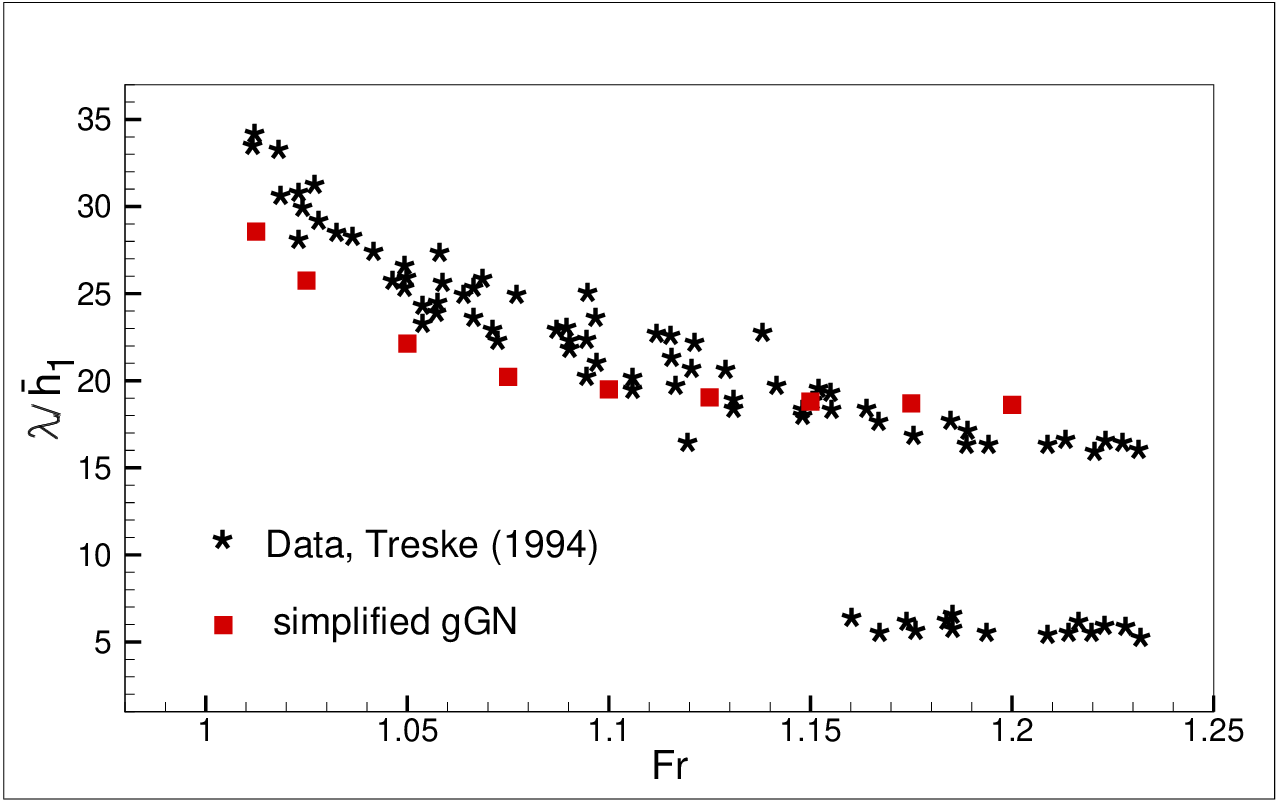}
\caption{}
\label{fig:treske-a}
\end{subfigure}\hfill
\begin{subfigure}{0.5\textwidth}
\centering\includegraphics[align=c,width=0.9\textwidth]{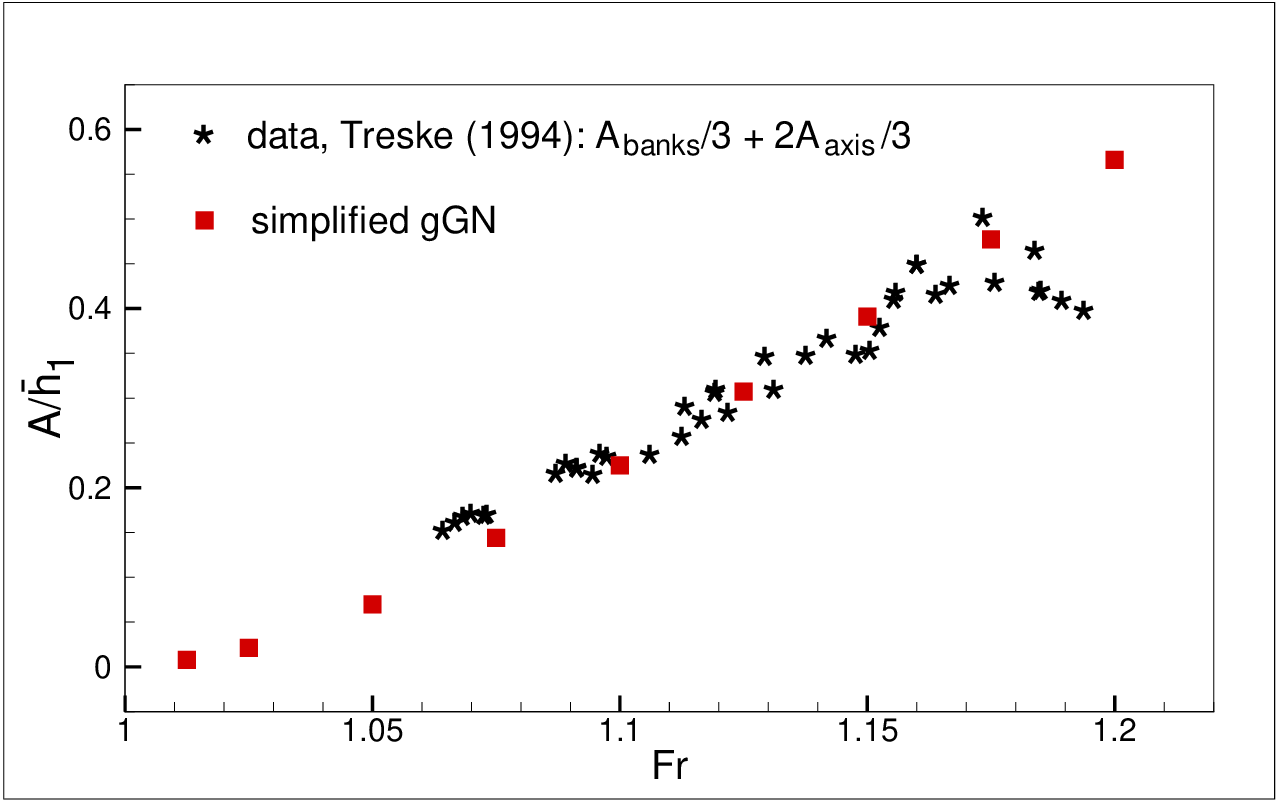}
\caption{}
\label{fig:treske-b}
\end{subfigure}
\caption{{\color{black}Treske experiments in a trapezoidal channel. (a)  wave length as function of the Froude number.
(b) amplitude as as function of the Froude number. Predictions of the simplified gGN model (squares) compared with the data
by  \cite{Treske} (asterisks).   \label{fig:treske}}}
\end{figure}

\section{Conclusion}

From the  Saint-Venant equations over topography we have derived  one-dimensional  geometrical Green-Naghdi (gGN) type equations describing the evolution of section averaged depth and velocity
The dispersive effects arise  from   transverse refraction {\color{black} solely} due to bathymetric variations.
The model proposed has several appealing properties: it is Galilean invariant;  it admits the  energy conservation law, and  exact travelling waves solutions.
Its dispersion relation allows to obtain quantitative predictions of the wavelengths of undular bores for low Froude numbers. 

We have {\color{black} proposed } two approaches to solve the model numerically, {\color{black} and discussed 
numerical  tests going from comparisons with analytical travelling {\color{black} wave} solutions,  to validation with respect to 
 multidimensional shallow water simulations, and  experimental data}.
In particular,  {\color{black} the multi--dimensional} shallow water simulations {\color{black} confirm} the occurrence of these dispersive-like waves {\color{black} in channels of 
finite width} adding evidence to the one already presented  e.g. in \cite{Chassagne}. {\color{black} The  section averages of the multi-dimensional computations
have been used to validate the geometrical gGN model proposed. Finally, the new model predicts with satisfactory accuracy   the data from       \cite{Treske},
fully confirming its ability to describe the non-linear geometrical waves observed experimentally}.\\

Many questions remain open concerning  {\color{black} the interaction of the different  processes at play, and in particular
the emergence of geometrical dispersive effects.} 
There is certainly some  the interaction between non-linearity and dispersion which controls this process, and 
 probably {\color{black}some} equivalent of  a wave breaking range in which the shallow water equations do predict shock formation.
 {\color{black} A major issue, focus of ongoing work,   is  also the interaction of different dispersive processes and 
 the formulation
 of a model including  with clear hypotheses} both horizontal dispersion (dispersive-like waves), and {\color{black} usual non-hydrostatic
 effects as those modelled by}  the classical Serre-Green-Naghdi model.

\section*{Acknowledgements}
SG would like to thank the Isaac Newton Institute for Mathematical Sciences, Cambridge, for support and hospitality during the program "Emergent phenomena in nonlinear dispersive waves", where work on this paper was undertaken. This work was supported by EPSRC grant EP/R014604/1.
MR is a member of the CARDAMOM research team of the Inria center at University of Bordeaux. The authors are grateful to the anonymous reviewers  for  
important remarks and suggestions  which helped in improving both   the quality of the manuscript,  and the  derivation of the model.

\appendix 
{\color{black}
\section{Geometrical coefficient properties for symmetric channels}\label{app:geo1}
We prove  that for  $b(y)=b(l-y)$ for any $0<y<l$, i.e. {\color{black} for} $b(y)$  symmetric with respect to the center  the channel, we have  
	\begin{equation*}
	\overline{S}=0.
	\end{equation*}
	Indeed, 
	\begin{equation*}
	l\, \overline{S}=\int_{0}^{l}\int_{0}^{y}\left( \overline{b}-b(s) \right) ds dy=\frac{\overline{b}\,  l^2 }{2}-\int_{0}^{l}\int_{0}^{y}b(s) ds dy
	\end{equation*}
	The last term is just the  integration over the isosceles right triangle 
	$A=\left\{ (s,y) \vert\; 0<s<y<l \right\} $ in the square 
	$B=\left\{ (s,y)\vert\; 0<s<1, 0<y<l \right\}$. If the function $b(y)$ is symmetric, one can replace the integration over $A$ by the integration over the whole square  
	$B=\left\{ (s,y)\vert\; 0<s<1, 0<y<l  \right\} $ divided by $2$  : 
	\begin{equation*}
	l\, \overline{S}=\frac{\overline{b}\,  l^2 }{2}-\frac{1}{2}\left(\int_{0}^{l}\int_{0}^{y}b(s) ds dy+\int_{0}^{l}\int_{y}^{l}b(s) ds dy\right)=\frac{\overline{b}\,  l^2 }{2}-\frac{1}{2}\int_{0}^{l}\int_{0}^{l}b(s) ds dy
	\end{equation*}
	\begin{equation*}
	=\frac{\overline{b}\,  l^2 }{2}-\frac{\overline{b}l}{2}\int_{0}^{l} dy=0.
	\end{equation*}

\section{Potential for symmetric channels}\label{app.prop1}
We prove here proposition  \ref{prop1}. To this end we consider the direct computation of 
	\begin{equation*}
	-\left(\frac{\partial N}{\partial \tau }-\frac{D}{Dt}\left(\frac{\partial N}{\partial \dot \tau}\right)\right)\frac{d\sigma}{dy}=\frac{d\sigma}{ds}\int_0^y\left(\frac{\sigma(s) \ddot{\tau}}{\displaystyle 1-\tau \frac{d \sigma(s)}{ds}}+\frac{\sigma(s)\displaystyle\frac{d\sigma(s)}{ds}\dot{\tau}^2}{2\left(\displaystyle 1-\tau \frac{d \sigma}{ds}\right)^2}\right)ds,
	\end{equation*}
	\begin{equation*}
	\frac{\partial }{\partial y}\left(\sigma(y)\frac{\partial N}{\partial \tau}\right)=\frac{d\sigma}{dy}\frac{\partial N}{\partial \tau}+\sigma(y)\frac{\partial }{\partial \tau}\left(\frac{\partial N}{\partial y}\right)
	\end{equation*}
	\begin{equation*}
	=\frac{d\sigma}{dy}\frac{\dot\tau^2}{2}\int_0^y\frac{\sigma(s)\displaystyle\frac{d\sigma(s)}{ds}}{\left(\displaystyle 1-\tau \frac{d \sigma}{ds}\right)^2}ds+\frac{d\sigma}{dy}\frac{\dot\tau^2}{2}\frac{\sigma^2(y)}{\left(\displaystyle 1-\tau \frac{d \sigma}{dy}\right)^2}.
	\end{equation*} 
	Summing the last expressions, we obtain
		\begin{equation}
	M\frac{d\sigma}{dy}=-\left(\frac{\partial N}{\partial \tau }-\frac{D}{Dt}\left(\frac{\partial N}{\partial \dot \tau}\right)\right)\frac{d\sigma}{dy}+\frac{\partial }{\partial y}\left(\sigma(y)\frac{\partial N}{\partial\tau}\right).
	\label{important_relation}
	\end{equation}
  Now, consider the  case  $\overline S=\mathcal{O}(\varepsilon^\beta)$.  This property reduces to $\bar S =0$   for symmetric bathymetries $b(y)$. In this case
  by virtue of \eqref{eq.sigma} we have 
  $$
  \sigma(0)=\sigma(l)=\mathcal{O}(\varepsilon^\beta).
  $$
   Using this fact we can write  
	\begin{equation}
	\overline{M\frac{d\sigma}{dy}}=-\overline{\left(\frac{\partial N}{\partial \tau }-\frac{D}{Dt}\left(\frac{\partial N}{\partial \dot \tau}\right)\right)\frac{d\sigma}{dy}}+\left.\sigma(y)\frac{\partial N}{\partial\tau}\right\vert_{0}^{l}
	\end{equation}
	\begin{equation}
	=-\overline{\left(\frac{\partial N}{\partial \tau }-\frac{D}{Dt}\left(\frac{\partial N}{\partial \dot \tau}\right)\right)\frac{d\sigma}{dy}}+\mathcal{O}(\varepsilon^\beta)=-\overline{\left(\frac{\displaystyle\partial N\frac{d\sigma}{dy}}{\partial \tau }-\frac{D}{Dt}\left(\frac{\displaystyle\partial N\frac{d\sigma}{dy}}{\partial \dot \tau}\right)\right)}+\mathcal{O}(\varepsilon^\beta).
	\end{equation}
	Denoting 
	\begin{equation*}
	{\cal{L}}(\tau, \dot\tau)=\overline{N\frac{d\sigma}{dy}}
	\end{equation*}
	we finally get in the case of quasi-symmetric channels, and within $\mathcal{O}(\varepsilon^\beta)$  : 
	\begin{equation}
	\overline{M\frac{d\sigma}{dy}}=-\left(\frac{\partial {\cal{L}}}{\partial \tau }-\frac{D}{Dt}\left(\frac{\partial {\cal{L}}}{\partial \dot \tau}\right)\right) =-\dfrac{\delta  {\cal{L}} }{\delta \tau}
	\label{variational_formulation_1_0} 
	\end{equation}
 }

 \section{Numerical methods}\label{app:schemes}

 This appendix provides some more  details concerning   the schemes used to obtain
 the discrete  equations used to obtain the results of section \ref{5}.  
First we discuss the  time integration method used with the different reformulations of the system,
and then the approximation of the spatial differential operators.

\subsection{Time stepping}

\paragraph{\normalfont \emph{Elliptic-Hyperbolic approximation.}} For   system \eqref{eq:HES}  
 we start from initial conditions simply set on physical quantities $(h,u)$. The system is then
 evolved in time by means of an explicit multistage method.
Let us set
\begin{equation}\label{eq:UEH}
U_{\textsf{EH}} = (h, hu)^T,
\end{equation}  
and 
\begin{equation}\label{eq:UEH1}
\Psi_{\textsf{SW}} =   ( (hu)_x, (hu^2 + gh^2/2)_x - \phi )^T,
\end{equation}  
given $(h,u)$     at time $n$ and at intermediate stages $(0,\, k)$, 
as well as $\phi$  at intermediate stages $(0,\, k-1)$,  the generic stage $k$ reads: 
\begin{equation}\label{eq:EH-steps}
\begin{split}
h^{k}\phi^k  - & \chi [\dfrac{1}{ h^k}( \phi^k - g h^k_x       )_x]_x = 0\longrightarrow \phi^k\\[5pt]
U_{\textsf{EH}}^{k+1}=   & \sum_{\ell\ge1}\alpha_{\ell} U_{\textsf{EH}}^{k+1-\ell} - \Delta t \sum_{\ell\ge1}\beta_{\ell} \Psi_{\textsf{SW}}(U_{\textsf{EH}}^{k+1-\ell}, \phi^{k+1-\ell}),
\end{split}
\end{equation}
In this approach we need to assemble and invert the elliptic operator   at each intermediate stage.
In practice here we use  \eqref{eq:EH-steps} with two stages.  

\paragraph{\normalfont \emph{Hyperbolic system.}} In this case initial conditions are set on physical quantities  $(h,u)$
as well as the auxiliary variables $w$ and $\eta$. {\color{black} For the latter}    initial conditions are deduced from the limit $\mu\rightarrow\infty$:
$$
\eta(0,x) = \tau(0,x)\;,\quad  w(0,x) =(u(0,x))_x/h(0,x).
$$
For solitary waves, relations \eqref{eq:u_sol_h}  and \eqref{eq:ODE} have been used to evaluate these  quantities.

System  \eqref{eq:hyp-form}  is  then     integrated in time  using the  method proposed in \cite{Favrie}.
In particular, setting  
\begin{equation}\label{eq:Uhyp}
U_{\textsf{Hyp}} = ( h, hu, h\eta, hw )^T,
\end{equation} 
and 
\begin{equation}\label{eq:Uhyp1}
\Psi_{\textsf{H}}  =   ( (hu)_x, (hu^2 + p_{tot})_x,  (hu\eta)_x, (huw)_x  )^T,
\end{equation}  
given the solution value $U_{\textsf{Hyp}}^n$ at time $ t^n$,
we use a classical second order Strang splitting method:
 $$
U_{\textsf{Hyp}}^{n+1} =O_1(\Delta t/2)O_{\textsf{Hyp}}(\Delta t)O_1(\Delta t/2)
 $$
 where the operator $O_1$ is  the system of ODEs  
\begin{equation}\label{eq:ODE_hyp}
\begin{split}
h_t  = &0 \\
 (h u)_t  =&0\\
 \left(h \eta \right)_t  = & {h}\,  w,\\ \left(h\, w \right)_t   =& 
-\frac{\mu }{\chi}  \left(h \eta  -  1\right).
\end{split}
\end{equation}
Given   initial data $(\tau_0,  u_0, \eta_0, w_0)$ the above is  integrated exactly  as 
\begin{equation}\label{eq:ODEsol}
\begin{split}
\eta= & \eta_0 \cos( \sqrt{\alpha} t) + \tau_0 \left(1  - \cos( \sqrt{\alpha} t) \right) + \dfrac{w_0}{\sqrt{\alpha}}\sin( \sqrt{\alpha} t) \\
w =& w_0 \cos( \sqrt{\alpha} t)  - (\eta_0 -\tau_0)\sqrt{\alpha}\sin( \sqrt{\alpha} t) 
\end{split}
\end{equation}
with $\alpha=\mu/\chi$. 

The operator $O_{\textsf{Hyp}}$  is nothing else than  the hyperbolic homogeneous part of  system \eqref{eq:hyp-form},
which is integrated using a multi-stage method (with the same notation of \eqref{eq:EH-steps})
\begin{equation}\label{eq:Hyp-steps}
\begin{split}
U_{\textsf{Hyp}}^{k+1}=   & \sum_{\ell\ge1}\alpha_{\ell} U_{\textsf{Hyp}}^{k+1-\ell} - \Delta t \sum_{\ell\ge1}\beta_{\ell} \Psi_{\textsf{H}}(U_{\textsf{Hyp}}^{k+1-\ell}),
\end{split}
\end{equation}
 In practice here we use    a two stage method.

\subsection{Discretization of the spatial  differential operators}

Spatial differential operators are approximated on a uniform discretization of the 1D spatial domain, with size $\Delta x=1/N$,
$N$ representing the number of mesh cells. 

The elliptic operator in  \eqref{eq:EH-steps} is discretized with linear finite elements, using the variational
formulation \eqref{eq:EHS-variational}. The invertibility of the  resulting problem is characterized  by proposition 
\ref{prop:EHS-coercive}. In practice, the system being tri-diagonal, we have used Thomas method which allows
to obtain the solution in two explicit seeps. The third derivative of $h$ appearing in the right hand side is evaluated with a second order formula.

The hyperbolic operators, are  discretized using a second order   explicit Residual Distribution method
 \cite{hdr-M,ar17,ar22}.      To first order of accuracy in space one stage of the method can be written as 
\begin{equation}\label{eq:RD-predictor}
 U^{k+1}_i =  U^{n}_i - \dfrac{\Delta t}{\Delta x} \phi_i^{i-1/2}(U^{n})- \dfrac{\Delta t}{\Delta x} \phi_i^{i+1/2}(U^{n})
\end{equation}
where the fluctuations    $\Phi_i^{i\mp 1/2}(U^{n})$ are defined by some upwind biased distribution of 
 the full residuals in mesh cells $[x_{i-1}, x_i]$ and $[x_{i}, x_{i+1}]$. In particular, 
 for a hyperbolic system reading
 $$
 U_t + F_x(U) = S 
 $$
 let $A(U)=\partial F/\partial U$ denote the flux Jacobian. We define
 $$
\varphi^{i-1/2} = \int_{x_{i-1}}^{x_i}( F_x -S ) = F_i - F_{i-1} - \Delta x (S_{i-1} + S_i)/2
 $$
and similarly for $ \varphi^{i+1/2}$. With this definition we set here
\begin{equation}\label{eq:RD-split}
\varphi_i^{i-1/2}= \varphi^{i-1/2}/2 + \delta A(U_{i-1/2}) \varphi^{i-1/2}\;,\quad 
\varphi_i^{i+1/2}=  \varphi^{i+1/2}/2 - \delta A(U_{i+1/2}) \varphi^{i+1/2}
\end{equation}
where the term multiplied by the matrix $\delta$ is a stabilizing upwind bias, and $U_{i\pm 1/2}$ denote simple the cell averages
of the unknown.  Classical upwinding is  obtained for $\tau = |A|^{-1}$, Lax-Wendroff like schemes are obtained for $\delta=\Delta t /2\Delta x$.
This is the choice used here.

Instead of using  a reconstruction, or more complex Lax-Wendroff like procedures  (see e.g. \cite{Dumbser08,Cauquis}),
 a second order extension is built via   a two-step procedure  exploiting a  first update
to construct a strongly consistent residual. Given the first order predictor \eqref{eq:RD-predictor} we set 
$(\cdot)^{n+1/2} = ( (\cdot)^n +(\cdot)^{k+1})/2 $,  and we  compute in the second stage
$$
\begin{aligned}
 \Phi^{i-1/2} =& \dfrac{1}{\Delta t} \int_{t^n}^{t^{n+1}}\int_{x_{i-1}}^{x_i}( U_t + F_x -S ) \\=& 
\Delta x \dfrac{ U^{k+1}_{i-1/2} - U^n_{i-1/2}}{\Delta t}  + \dfrac{1}{2} \varphi_i^{i-1/2}(U^n) + \dfrac{1}{2} \varphi_i^{i-1/2}(U^{k+1})
\end{aligned}
 $$
 and similarly  for $ \Phi^{i+1/2}$. The  update is more simply written now in the form 
$$
 U^{n+1}_i =  U^{k+1}_i - \dfrac{\Delta t}{\Delta x} \Phi_i^{i-1/2} - \dfrac{\Delta t}{\Delta x} \Phi_i^{i+1/2} 
$$
where similarly to   \eqref{eq:RD-split} we set
\begin{equation}\label{eq:RD-split1}
\Phi_i^{i-1/2} =  \Phi^{i-1/2}/2 + \delta A(U_{i-1/2}) \Phi^{i-1/2}\;,\quad 
\Phi_i^{i+1/2} =   \Phi^{i+1/2}/2 - \delta A(U_{i+1/2})  \Phi^{i+1/2}
\end{equation}
The interested reader can refer to  \cite{hdr-M,ar17,ar22}  and references therein for additional information, and theoretical aspects.

Finally, in the case of the elliptic-hyperbolic approximation the time step  is computed from the shallow water part as  
\begin{equation}\label{eq:cfl-eh}
\Delta t = \nu \dfrac{\Delta x}{\max_{i}( |u_i| + \sqrt{gh_i} ) }
\end{equation} 
The interested reader can refer to \cite{Cauquis,Filippini,Kazolea} for  Fourier analyses justifying this choice.
For the fully hyperbolic approximation the source terms are exactly integrated in the splitting method,
so  we use    the classical  condition based on the largest eigenvalue
\begin{equation}\label{eq:cfl-hyp}
\Delta t = \nu \dfrac{\Delta x}{\max_{i}( |u_i| + \sqrt{gh_i + \mu \tau_i^2} ) },
\end{equation} 
which shows a dependence on the value of the relaxation constant $\mu$. 

 \bibliographystyle{plain}
 \bibliography{biblio} 
\end{document}